\documentclass[sigconf, nonacm]{acmart}

\newcommand{\stitle}[1]{\vspace{1ex}\noindent\textbf{#1}}
\usepackage{xspace}
\newcommand{\tree}{skd-tree\xspace}
\newcommand{\bstree}{B$^S$-tree\xspace}

\newcommand{\maxkey}{MAXVAL\xspace}

\newcommand{\hide}[1]{}

\usepackage{fancyvrb}
\usepackage[ruled, linesnumbered, vlined]{algorithm2e}
\usepackage[noend]{algorithmic}
\usepackage{subcaption}     % For subfigures

\SetKwInOut{Input}{Input}
\SetKwInOut{Output}{Output}
\SetKwInOut{Variables}{Variables}
\SetKwComment{comm}{\hfill$\triangleright$\ }{}

\usepackage{titlesec}

\titlespacing*{\section}
  {0pt}   % left margin
  {1.5ex} % space before
  {0.8ex} % space after
\titlespacing*{\subsection}
  {0pt}
  {1.2ex}
  {0.6ex}

\newcommand{\vect}[1]{\mathbf{#1}}

\usepackage{tikz}
\usepackage{xcolor}
\newcommand*\circled[1]{\tikz[baseline=(char.base)]{
           \node[shape=circle,fill,inner sep=1pt] (char) {\textcolor{white}{#1}};}}

\usepackage{cleveref}

\begin{document}

%%
%% The "title" command has an optional parameter,
%% allowing the author to define a "short title" to be used in page headers.
%\title{SIMD-powered Multidimensional Indexing}
\title{In-memory Multidimensional Indexing Using the \tree}

%%
%% The "author" command and its associated commands are used to define the authors and their affiliations.
\author{Achilleas Michalopoulos}
\orcid{0009-0002-8424-2359}
\affiliation{%
  \institution{Dept. of Comp. Sci. and Engineering}
  \streetaddress{University of Ioannina}
  \city{Ioannina}
  \state{Greece}
}
\email{amichalopoulos@cse.uoi.gr}

\author{Dimitrios Tsitsigkos}
\orcid{0009-0003-9929-962X}
\affiliation{%
  \institution{Archimedes, Athena RC}
  \city{Athens}
  \country{Greece}
}
\email{dtsitsigkos@athenarc.gr}

\author{Nikos Mamoulis}
\orcid{0000-0003-3423-4895}
\affiliation{%
  \institution{Dept. of Comp. Sci. and Engineering}
  \streetaddress{University of Ioannina}
  \city{Ioannina}
  \state{Greece}
}
\email{nikos@cs.uoi.gr}

\begin{abstract}
In this paper, we revisit the problem of indexing multi-dimensional
data in memory for the efficient support of multi-dimensional range
queries and nearest neighbor queries.
%\nikos{and kNN?}.
This is a classic problem in main-memory databases, where there is a
need for indexing multiple columns simultaneously.
Established data structures include the R-tree, kd-tree, quad-tree, and
grid-based partitioning.
More recently, multi-dimensional learned indexes have also been proposed to address this
problem.
%In this paper,
We propose {\em slicing kd-tree} (\tree), a
%memory-based
%multiway
variant of the kd-tree,
%called
%slicing kd-tree (\tree),
where each node partitions the space of its
subtree into {\em multiple
slices} across a {\em single} splitting dimension.
By compressing the splitters of the partitions and with the
help of data-parallelism, we (i)
radically reduce the number of levels
of the tree and (ii) limit the number of computations required for
multi-dimensional range and proximity queries. 
%embed the use of data-parallel SIMD instructions in
%the design of an hierarchical index, which extends the kd-tree to
%define and use multiple partitions per node.
The  nodes of the \tree
resemble the nodes of a main-memory B$^+$-tree, however,
a different dimension is used at each level.
Our novel range and $k$NN algorithms on the \tree apply only a small constant
number of SIMD instructions at each node during tree traversal.
Our contributions also include
% proposal includes novel data-parallel multi-dimensional search
% algorithms,
a novel top-down construction algorithm,
different types of inner and leaf nodes that warrant tree balancing,
and a novel update algorithm.
Our \tree achieves strong performance
compared to existing methods, according to our experimental evaluation
on real and synthetic datasets. 
%We show how to achieve balanced partitioning and theoretical
%performance guarantees (?).
%+++
\end{abstract}

\maketitle

%%% do not modify the following VLDB block %%
%%% VLDB block start %%%
%\pagestyle{\vldbpagestyle}
%\begingroup\small\noindent\raggedright\textbf{PVLDB Reference Format:}\\
%\vldbauthors. \vldbtitle. PVLDB, \vldbvolume(\vldbissue): \vldbpages, \vldbyear.\\
%\href{https://doi.org/\vldbdoi}{doi:\vldbdoi}
%\endgroup
%\begingroup
% \renewcommand\thefootnote{}\footnote{\noindent
% This work is licensed under the Creative Commons BY-NC-ND 4.0 International License. Visit \url{https://creativecommons.org/licenses/by-nc-nd/4.0/} to view a copy of this license. For any use beyond those covered by this license, obtain permission by emailing \href{mailto:info@vldb.org}{info@vldb.org}. Copyright is held by the owner/author(s). Publication rights licensed to the VLDB Endowment. \\
% \raggedright Proceedings of the VLDB Endowment, Vol. \vldbvolume, No. \vldbissue\ %
% ISSN 2150-8097. \\
% \href{https://doi.org/\vldbdoi}{doi:\vldbdoi} \\
% }\addtocounter{footnote}{-1}\endgroup
% %%% VLDB block end %%%

%%% do not modify the following VLDB block %%
%%% VLDB block start %%%
% \ifdefempty{\vldbavailabilityurl}{}{
% \vspace{.3cm}
% \begingroup\small\noindent\raggedright\textbf{PVLDB Artifact Availability:}\\
% The source code, data, and/or other artifacts have been made available at \url{\vldbavailabilityurl}.
% \endgroup
% }
%%% VLDB block end %%%

\section*{Code Available at:}
\url{https://github.com/achmichalop/skd-tree_2027}

\section{Introduction}
\looseness=-1
Multi-dimensional range filtering is a core operation in 
analytical database engines. The goal is to retrieve the tuples from a
table that simultaneously satisfy range predicates in
multiple attributes. Typically, the filtering attributes are not many, so multi-dimensional indices for
low-dimensional spaces (e.g., kd-trees, quadtrees, R-trees) are
considered the standard way to index and search across multiple
attributes. In spatial databases, proximity queries, such as
%spherical
%range search and
$k$ nearest neighbor ($k$NN) search are also
prominent; besides, $k$NN queries are used for similarity search and as a
module of data mining algorithms (clustering, classification).  

\looseness=-1
Given the growth in memory sizes of commodity machines and the
capacity of modern processors for data
parallelism, 
modern database systems offer in-memory indexing solutions.
Besides adapting B-trees or tries
% and search
for one-dimensional
indexing in memory \cite{RaoR00,ZhouR02,ZhouR03,SchlegelGL09,LeisK013,BinnaZPSL18,bstree}, there have also
been efforts in modernizing multi-dimensional index structures such as
R-trees and kd-trees for
data-parallel and/or multi-core processing in memory 
\cite{RayhanA23,MenSGS25}. There is also
work on workload-aware multi-dimensional indexing
\cite{nathan2020learning,SudhirCM21},
where index partitions are determined
% not only
% based on the data distribution, but also
considering the distribution
of the expected query workload.

\begin{figure}
        \centering
	\begin{tabular}{@{}ccc@{}}
          \includegraphics[width=0.35\columnwidth]{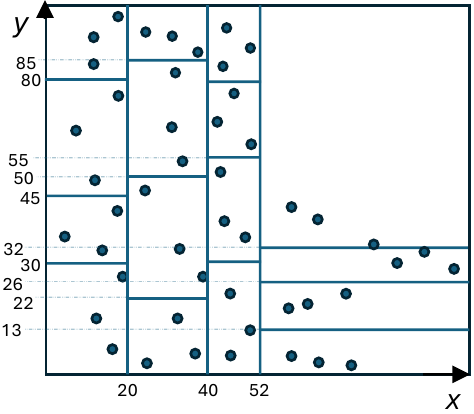}&
\includegraphics[width=0.25\columnwidth]{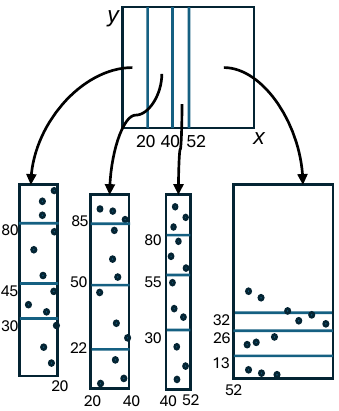}&
\includegraphics[width=0.3\columnwidth]{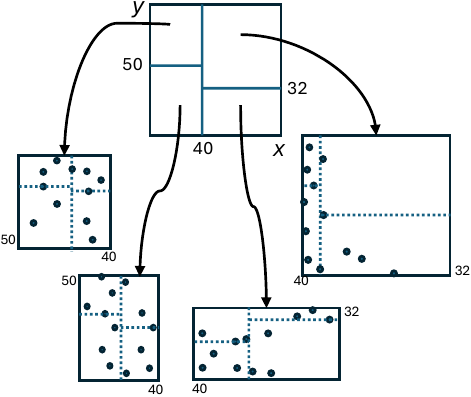}\\
        {\small (a) 2-d points}&
        {\small  (b) \tree (ours)}&
{\small  (c) m-way kd-trees \cite{Robinson81, AgarwalADH03}}\\
      \end{tabular}
        \vspace{-2mm}
      \caption{Multiway kd-trees}
      \label{fig:mkd}
\end{figure}

% Dividing the space into evenly populated partitions can be achie\-ved by
% the kd-tree; Figure \ref{fig:grids}(c)
% shows an example of a kd-tree that recursively
% divides the space alternating splitting dimensions
% at each recursion level ($x$,$y$,$x$,$y$),
% until leaves do not include more than $3$ points.
% Each time the current partition is split using the median value in the
% partition in the splitting dimension.
% While successful in
% achieving balanced partitions, the kd-tree is slow in search due to
% pointer chasing and dependencies between the accessed
% nodes. Specifically, during search, the accessed nodes depend on
% the relationship between their ancestors and the query; hence, search
% cannot easily be parallelized.

\stitle{Our proposal.}
Existing memory-based multi-dimensional data structures suffer from
poin\-ter chasing and branch mispredictions.
In this paper, we propose a {\em slicing} kd-tree, briefly \tree,
which achieves branchless search per accessed node
and balanced partitioning.
The structure of \tree is illustrated in Figure \ref{fig:mkd}(b) for a
set of 2-d points shown in Figure \ref{fig:mkd}(a).
We divide
the space recursively by alternating dimensions (as in the kd-tree),
but we perform $m$-way partitioning in a {\em single splitting
  dimension} at each node ($m$=4 in the figure).
In addition, we do not use points to define boundaries;
%(as this complicates updates);
all data are stored at the leaves of the tree (in this example, we
assume that each leaf gets 3 points).
Each node divides the data in its subtree to multiple
partitions that get (almost) the same number of points. This can be
done after computing quantiles along the split dimension for the space
to be partitioned and using these quantiles to define separators. For
example, in \Cref{fig:mkd}(b), the 0.25, 0.50, 0.75 quantiles of the
$x$-values are used to divide the space into four partitions
corresponding to the children of our \tree's root; for each partition,
the 0.25, 0.50, 0.75 quantiles of the
$y$-values that fall in them are used to recursively split the
partition.

\stitle{Novelty and contributions.}
\circled{1}
Each node of our \tree splits its subspace across a {\em single}
dimension. This facilitates search within a single
node by a small constant number of SIMD operations.
In previous multi-way kd-trees \cite{Robinson81, procopiuc2003bkd, AgarwalADH03},
each node is split in multiple dimensions, as illustrated in Figure
\ref{fig:mkd}(c), which limits data-parallelism.
\circled{2} We propose novel data-parallel
% and cache-conscious
algorithms for
multidimensional search and $k$NN queries,
which exploit the \tree structure.
\circled{3} We apply an adaptive quantization scheme to compress the splitters
    within each \tree node, effectively increasing node fanout and, in
    turn, reducing storage requirements, tree height, and search
    cost. Inner tree nodes can be of variable capacities, even at the
    same level.
    The novelty of our node compression is that it does
    not only adapt to the data (unique prefixes of splitters),
    but it also decides the level of compression based on a
target number of splits, aiming at a balanced tree.
\circled{4} To cope with ties and data skew, we introduce
three types of leaf nodes (light, heavy, and outlier) that help to maintain
balance and improve the robustness of the structure.
\circled{5} We present a novel top-down construction algorithm, which adaptively
    determines the number of splits at each node, to achieve a
    balanced, high-performance \tree.
    \circled{6} We propose a novel update algorithm,
    which postpones restructuring
    and reconstruction by exploiting heavy and outlier leaves.
To the best of our knowledge, this is the first work that integrates all these
functionalities in a high-performance cache-conscious multidimensional index
for main memory data.

\looseness=-1
We conduct an experimental evaluation with real and synthetic datasets,
    varying distribution and dimensionality, where we compare \tree against
    state-of-the-art conventional and learned
    multidimensional indices, including the boost.org R-tree
    implementation, which was shown to be the most robust in-memory
    multi-dimensional index in a recent experimental study
    \cite{LiuLZSC25}. The result show that \tree consistently
    outperforms the runner up  by 1.5x-2x in range and $k$NN queries,
    while being several times faster in updates. We also show that
    its search performance remains robust even after a large number of
    updates, as opposed to bulk-loaded R-trees and compressed
    multi-dimensional indices \cite{ZaschkeZN14}.

\stitle{Outline} Section \ref{sec:relatedwork} presents related
work.
%In Section \ref{sec:bstree}, we briefly review an in-memory \bt that
%uses SIMD instructions to search \nikos{to be moved to the related
% work and shortened }.
Section 
\ref{sec:method} presents the structure of our \tree, query
processing algorithms, and the \tree construction algorithm.
Updates on the \tree are discussed in Section \ref{sec:updates}.
Section \ref{sec:exp} presents our experimental evaluation and Section 
\ref{sec:conclusions} concludes the paper.

\section{Related Work} \label{sec:relatedwork}
%++

\subsection{Conventional multidimensional indices}
%Multidimensional indexing has been well studied for over 40 years  \cite{SellisRF97,GaedeG98}.
\looseness=-1
\stitle{The R-tree family.}
The popular multidimensional access method is the R-tree \cite{Guttman84,HadjieleftheriouMTT17}, a height-balanced structure similar to a B$^+$-tree where each node stores a set of minimum bounding rectangles (MBRs) that cover either actual objects in leaf nodes or in the corresponding child subtrees in internal nodes. Over the years, many variants, such as the R+-Tree \cite{SellisRF87}, the R*-Tree \cite{BeckmannKSS90}, and the Priority R-Tree \cite{ArgeBHY04},
bulk-loading methods \cite{KamelF94, LeuteneggerEL97, QiTCZ18, MotiSP22}, and optimizations \cite{KimCK01, beckmann2009revised}
have been proposed to improve its performance.
% eliminates overlap in internal nodes by allowing objects to appear in multiple leaves when necessary, resulting in more precise search paths at the cost of potential data duplication. The introduces advanced splitting and reinsertion strategies to minimize overlap, coverage, and perimeter, improving query performance.
%The Hilbert R-tree \cite{KamelF94} leverages Hilbert space-filling curves to impose a linear ordering on multidimensional objects prior to insertion, in R-tree bulk loading.
%STR-tree \cite{LeuteneggerEL97} uses a bulk-loading packing approach that sorts records along both axes to reduce the perimeters of leaf node MBRs and the overlap between them.
%The CR-tree \cite{KimCK01} compresses MBR coordinates relative to their parent and quantizes them to a fixed number of bits, reducing key sizes, widening nodes, and improving cache behavior. The Priority R-Tree \cite{ArgeBHY04} organizes data using priority-based bounding rectangles, achieving worst-case optimal query bounds while remaining efficient in practice. The Revised R*-Tree \cite{beckmann2009revised} further improves node splitting and reinsertion policies to aggressively minimize overlap and dead space. The WeR-Tree \cite{bozanis2007trees} maintains balance and reduces overlap using partial subtree rebuilds.
Recently, Rayhan and Aref \cite{RayhanA23} accelerate R-tree queries using SIMD instructions to vectorize intersection tests and traversal operations.
% while Qi et al. \cite{QiTCZ18} introduce a rank-space packing strategy based on Z-values that eliminates coordinate ties, enables highly parallel bulk loading, and achieves worst-case optimal query performance.

\looseness=-1
\stitle{kd-trees.}
Kd-trees represent another major class of multidimensional indices. Classic kd-trees \cite{Bentley75,Bentley79} recursively split space along alternating axes.
% , creating a hierarchical structure that allows efficient pruning of large portions of the search space. Balanced partitions and alternating-axis strategies ensure that all dimensions are evenly considered, supporting range queries, partial-match searches, and nearest-neighbor lookups.
The K-D-B-tree \cite{Robinson81} combines kd-tree space partitioning with the disk-based B$^+$-tree structure, storing bounding rectangles in region pages and data points in point pages.
% , efficiently supporting insertions, deletions, and range queries while maintaining tree balance.
%Orenstein and Merrett \cite{orenstein1984class} proposed associative searching structures that use Z-ordering (bit interleaving) to map multidimensional data into a single dimension, unifying kd-tree and B-tree methods.
The BKD-tree \cite{procopiuc2003bkd} improves the K-D-B-tree by maintaining multiple static trees and incrementally rebuilding selected trees to efficiently handle large datasets. Cache-oblivious kd-trees \cite{AgarwalADH03} use recursive layouts to preserve spatial locality and reduce memory transfers.
The iKD-tree \cite{abs-2102-10808} extends kd-trees to support incremental updates without full reconstruction.
% Multi-way kd-trees were proposed
% for fast rendering of large point clouds in \cite{GoswamiZPG10}.
% Although we also propose a multi-way kd-tree in this paper, as opposed to \cite{GoswamiZPG10} we support compression at inner nodes, data-parallel search,  and dynamic updates. In addition, our objective is to support range and $k$-NN search, whereas the goal in \cite{GoswamiZPG10} is multiresolution construction and visualization.
Parallel implementations of kd-trees have also been explored. ParGeo \cite{WangYYD0S22} provides scalable parallel kd-tree construction and spatial partitioning. The BDL-tree \cite{abs-2112-06188} is a parallel batch-dynamic kd-tree that combines ideas from BKD-trees and cache-oblivious kd-trees, maintaining multiple static trees of exponentially increasing sizes and rebuilding only the necessary trees for high-throughput batch updates. Reif and Neumann \cite{ReifN22} propose scalable multi-dimensional range joins using kd-trees with parallel space partitioning. The CGAL C++ library (www.cgal.org) implements kd-trees for fast nearest-neighbor and range searches with support for parallel construction, though dynamic updates require full rebuilds. The Pkd-tree \cite{MenSGS25} further improves cache efficiency and parallelism by supporting parallel construction, batch updates, and queries such as $k$ nearest neighbor, range, and range-count.
% , while maintaining a weight-balanced tree structure.
Parallel kd-tree construction algorithms typically fall into two categories: one class \cite{Brown14b,9237636,YamasakiNH18} sorts all points in every dimension before construction and splits using medians, while the other group \cite{Al-furaihAGR00,ChoiKLSBAH10,4061549,ReifN22,ShevtsovSK07} computes medians on the fly during each recursive split.

\looseness=-1
\stitle{Other multi-dimensional indices.}
Grid-based indexing
  is popular for
highly-dynamic low-dimensional points, such as
moving objects
\cite{MokbelXA04,SidlauskasSCJS09,RayBG14}.
On the other hand, grids cannot guarantee balanced partitioning and low worst-case cost.
Quadtrees \cite{finkel1974quad} are 2-d hierarchical space-partitioning structures that recursively subdivide space into quadrants for fast insertion and range queries, while octrees \cite{meagher1982geometric,0014978} extend this principle to three dimensions by subdividing nodes into eight regions. Unlike R-trees and kd-trees, quadtrees are not balanced, so searching in dense regions can be expensive.
Grids and quadtree-based structures do not scale well beyond 3 dimensions.
More recently,
the UB-tree \cite{ramsak2000integrating, bayer1997universal} generalizes B-trees for multidimensional indexing by mapping objects to a single ordered domain.
The PH-tree \cite{ZaschkeZN14} combines bitwise partitioning with Patricia-trie compression to provide predictable performance, low memory usage, and efficient access patterns. Sprenger et al. \cite{00010L18} show that hardware-conscious designs—such as cache-friendly layouts and SIMD-aware traversal—significantly outperform traditional indexes, and they propose the BB-tree \cite{00010L19}, a space-efficient main-memory index that combines a m-way search tree with a linearized, cache-optimized layout and bubble-bucket leaf architecture to buffer updates and reduce expensive reorganizations.

% \stitle{Indexing very high dimensional points.}
% \nikos{this paragraph may go}
% In very high dimensional spaces, the Vantage Point Tree (VP-tree) \cite{yianilos1993data} and it's multi-way extension
% \cite{bozkaya1999indexing}
% hierarchically partition data according to their distances from chosen vantage points.
% The VA-file \cite{WeberSB98} approximates high-dimensional vectors using compact bit-level quantization, enabling fast filtering before exact distance computations. iDistance \cite{jagadish2005idistance} maps high-dimensional points into a one-dimensional space using cluster reference points, allowing nearest-neighbor queries to be processed as efficient 1D range searches.

% The PH-tree \cite{ZaschkeZN14} combines bitwise partitioning with Patricia-trie compression to provide predictable performance, low memory usage, and efficient access patterns. Sprenger et al. \cite{00010L18} show that hardware-conscious designs—such as cache-friendly layouts and SIMD-aware traversal—significantly outperform traditional indexes, and they propose the BB-tree \cite{00010L19}, a space-efficient main-memory index that combines a $k$-ary search tree with a linearized, cache-optimized layout and bubble-bucket leaf architecture to buffer updates and reduce expensive reorganizations.

\stitle{Comparison to our \tree.}
The SIMD R-tree \cite{RayhanA23} compares entire boxes at each node, limiting the level of parallelism.
On the other hand, our \tree splits using only one dimension per node and {\em compares one dimension per level}, allowing larger fanout and considering more children per SIMD instruction.
Besides, the SIMD R-tree was implemented and tested only for 2-d data, while \tree supports points of arbitrary dimensionality. Finally, as opposed to the \tree, the SIMD R-tree it does not support $k$NN queries and adaptive node compression.
Previous kd-tree variants that use multiple splitters per node
\cite{Robinson81, procopiuc2003bkd, AgarwalADH03} do not split across the same dimension at each node; hence, the number of branches that could be examined by a single SIMD instruction is limited.  The Pkd-tree \cite{MenSGS25} focuses on milti-threaded processing of batches of queries and not on data-parallelism in the evaluation of a single query.

\subsection{Multi-Dimensional Learned Indexing}
% In recent years, learned indexes have emerged as a promising alternative to traditional multidimensional indexing structures by leveraging machine learning models to predict data locations and optimize query performance.
An early multidimensional learned index is the ZM-Index \cite{wang2019learned}, which simply maps multidimensional points to a Z-order curve and employs 1-d learned indexing \cite{KraskaBCDP18, KraskaABCKLMMN19}.
% a multi-level model to predict the storage location of each data item.
% SageDB \cite{KraskaABCKLMMN19} proposes a novel database framework that integrates machine learning and program synthesis to automatically generate system components, enabling learned indexing at the database level.
ML-Index \cite{davitkova2020ml} is a multidimensional learned index that extends iDistance \cite{jagadish2005idistance}.
% using reference-point scaling and a two-layer learned model. This approach captures data distribution and learned ordering, efficiently supporting point, range, and k-nearest-neighbor queries while maintaining low memory usage.
The Learned KD-tree \cite{PengZZD20} extends replaces classic kd-tree splitting with learned models.
Wang and Xu proposed the Hilbert Model (HM) index \cite{wang2020spatial}, which maps multidimensional data to one dimension using the Hilbert curve.
%and applies a two-stage model to predict object locations.
RSMI (Recursive Spatial Model Index) \cite{qi2020effectively} recursively partitions multidimensional space and trains models at each partition.
%to predict point locations, demonstrating the potential of hierarchical model-based indexing.
LISA \cite{li2020lisa} partitions space into grid cells and uses models to map multidimensional points into a linear order, ensuring monotonicity between cells.
% By assigning points to shards and managing them with local models on consecutive disk pages, LISA achieves efficient range and nearest-neighbor queries while supporting dynamic updates, including insertions and deletions.
Flood \cite{nathan2020learning} partitions multidimensional data by a grid which uses $D$-1 dimensions, where $D$ is the data dimensionality; it then trains a 1-d model on the left-out dimension in each grid cell.
%to form the grid and the last dimension as a sorting axis within each cell. Learned cumulative distribution function (CDF) models in each cell enable improved range query performance.
Tsunami \cite{DingNAK20} extends Flood  to optimize the index structure for a known query workload.
%’s idea by introducing two data structures, the Grid Tree and the Augmented Grid, which are optimized for high-performance range queries on correlated and skewed datasets.
SPRIG \cite{ZhangRLZ21} uses adaptive grids and spatial (two-dimensional) interpolation functions as learned models to directly predict the position of a spatial search key.
The IF-Index (IFI) \cite{D0001KH20} replaces leaf-node search in a standard
R-tree 
by a model
to predict the locations of points which are included in a query range.
RW-Tree \cite{DongCLLFZ22} and AI+R-Tree \cite{Al-MamunHWA22}
are learned, workload-aware R-trees.
%, adaptive indexing
%combine machine learning with classical R-tree structures, resulting in workload-aware, adaptive indexing.
% Recent developments continue to refine learned multidimensional indexing.
LMSFC \cite{Gao0YZ023} leverages learned monotonic space-filling curves to enhance query efficiency, while Wazi \cite{PaiM024} introduces workload-aware z-indexing that dynamically adapts to access patterns. SLBRIN \cite{wang2023slbrin} combines historical and current ranges, using learned models to support fast spatial queries and efficient updates. The RLR-Tree \cite{GuFCL0W23} incorporates reinforcement learning to optimize R-tree construction and query performance, and COAX \cite{HadianGWH23} emphasizes correlation-aware indexing by learning inter-dimensional relationships, achieving high efficiency on complex, correlated datasets.
Refs. \cite{LiWDLCGCCLLC24,LiuLZSC25}  provide comprehensive evaluations of multidimensional learned indices vs. traditional indices.
Based on these studies,
the most robust non-learned multidimensional indices are the boost R-tree implementation and the kd-tree, which we also include in our evaluation (Section \ref{sec:exp}), together with the best performing 
learned indices (IFI, Flood).
Learned indices perform well in range queries, but they lose to the kd-tree in $k$NN search; in addition, they have high construction cost and limited support of updates.
%outperform the best non-learned index (boost R-tree) in range queries but they lose to the best non-learned index in KNN search (kd-tree ); in addition, learned indices have higher construction cost and limited support of updates. In Section \ref{sec:exp}, we compare our \tree with IFI and Flood.

% \nikos{Add the main findings of these evaluations, emphasizing the robustness of non-learned indices like the boost R-tree, especially in arbitrary dimensionalities and in the presense of updates. We need to support the selection of the competitors as much as possible.}
%including many important findings, based on tests on several real-world datasets.

\subsection{Data-parallel, main-memory indexing}
There are several data-parallel implementations of B-trees \cite{Graefe24}. The first work in this direction was a SIMD-ified m-way search algorithm (find which out of m partitions contains a search key) \cite{SchlegelGL09}.
FAST \cite{KimCSSNKLBD10} optimizes m-way tree search by leveraging cache locality, SIMD parallelism on CPUs, and GPU-parallelism.
B$^+$-trees for GPU and SIMD architectures were proposed in \cite{ShahvaraniJ16, YanLPZ19, KwonLNNPCM23}.
Finally, the recently proposed \bstree \cite{bstree} aligns each node
to a small number of cache lines that can be processed in parallel.
Branching at each node is performed by a small
number of SIMD operations (one operation per cache line).
Unused slots at the end of nodes have as key a maximum value \maxkey, while
gaps in the middle take the next used key in the node, to facilitate fast search and updates.

\stitle{Comparison to \tree.}
Although both the \bstree \cite{bstree}  and our \tree use SIMD instructions in search, there are significant differences between them.
The \bstree supports equality and range queries on a {\em single} attribute, whereas \tree supports multidimensional range queries and $k$NN queries.
The \bstree splits the {\em same
    single} dimension hierarchically (many times), but the \tree
  splits a different dimension per level.
The search algorithms are essentially different, as
%\bstree search accesses one or two tree paths, whereas 
queries on the \tree traverse multiple paths and the pending accesses are organized in a (priority) queue. Two operations are needed per accessed node in the mkd-tree (for the upper and lower query bounds in that dimension), but only one is used in the \bstree.
% Besides, the \bstree splits the {\em same
%     single} dimension hierarchically (many times), but the \tree
%   splits a different dimension per level.
The \bstree applies
  frame-of-reference (FOR) compression at the leaves, which
  requires reconstruction of keys during search. On the other hand,
  the mkd-tree uses a prefix compression technique at inner nodes that {\em requires
    no reconstruction} of key values during search.
  The construction algorithms of the two structures are different, since \tree is constructed in a top-down manner and compression is not only based on unique prefixes but also on balancing demands. Updates are simple in the 1-d \bstree, however, in \tree, splits do not propagate upwards, so {\em we apply reconstruction of subtrees}.
The \tree uses three different types of leaf nodes that store uncompressed points, while the leaf nodes of the \bstree have the same structure as its inner nodes. 
Finally, The \bstree only supports {\em unique} keys, but the \tree handles
points, which may have ties in one or more
dimensions.

\section{\tree}\label{sec:method}
This section presents \tree, a multi-dimensional index designed to efficiently handle $D$-dimensional data. We first describe the basic structure of the \tree,
%review the m-way kd-tree \nikos{cite?}, highlighting its structural and performance limitations.
which
adapts to modern memory hierarchies, focusing on cache-aware node layouts and data-parallel query processing using SIMD intrinsics. We show how to reduce the precision of splitters to increase node fanout, producing more compact trees with shorter traversal paths.
%while maintaining balanced partitions and ensuring branchless search at each node.
Search algorithms for range and $k$NN queries are then presented.
Finally, we present a novel top-down bulk-loading algorithm for the \tree which achieves a balanced structure.
For the ease of presentation, we assume that all dimensions take as values 64-bit unsigned integers; however, our approach can be generalized for other attribute types (e.g., doubles).
%\todo{revise introduction of this section after the content is finalized}

\subsection{The \tree structure} \label{sec:tree}
\begin{figure}
          \includegraphics[width=0.99\columnwidth]{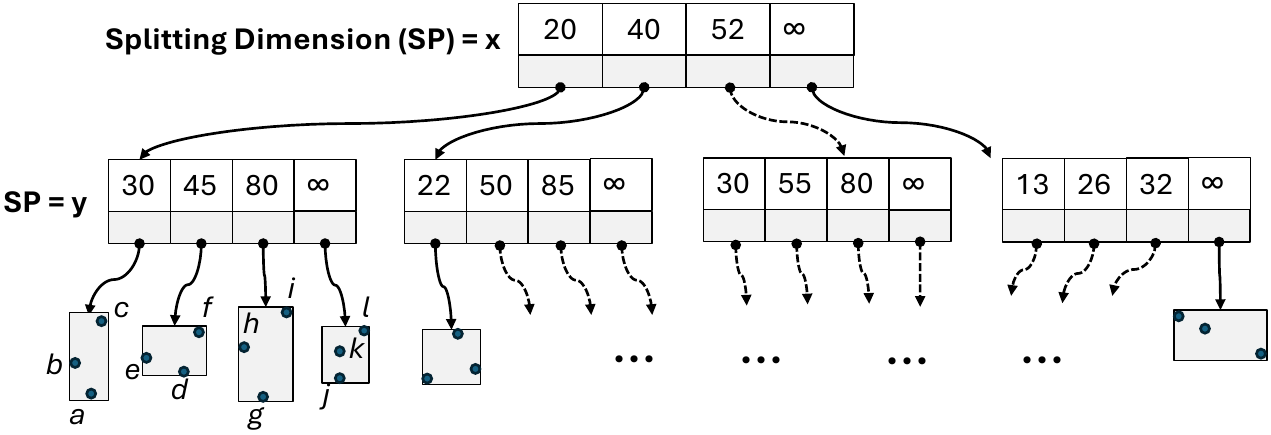}
      \caption{The \tree structure for the data in \Cref{fig:mkd}(d)}
      \label{fig:mkdt}
\end{figure}

%\subsection{Traditional m-way kd-tree}
The \tree generalizes the binary kd-tree by allowing each internal node to hold multiple splitters {\em in the same dimension} instead of a single splitter.
%Given a fanout $f$,
An internal node $v$ with maximum fanout $f$ stores up to $f-1$ splitters that partition the data space of the subtree rooted at $v$ into $f$ partitions across a dimension. The splitting dimension
%typically
alternates across consecutive tree levels, as in the classical kd-tree. The leaves of the tree store up to $C$ data points, where $C$ is a predefined maximum leaf capacity. \Cref{fig:mkdt} shows the structure of the \tree for the data plotted in \Cref{fig:mkd}(a) ($C=3$). The root splits the x dimension into $f=4$, using $f-1=3$ splitter values in the x domain, based on the 0.25, 0.5, and 0.75 quantiles of the total order of the points in the x dimension. Then, each child of the root, which spans a vertical stripe is split recursively based on the 0.25, 0.5, and 0.75 quantiles of the y-order of the points in the stripe. Finally, each leaf includes the points in the corresponding partition and their minimum bounding box (MBB). For example, the first leaf includes all the points in the partition at the lower-left part of the space in  \Cref{fig:mkd}(b). Note that non-leaf nodes  store splitters across a {\em single} dimension, whereas each leaf node stores the complete set of points in the corresponding partition. 

The \tree built from a set of $N$ D-dimensional points has a height $O(\log_f{(N/C)})$, as opposed to a binary kd-tree (with $C$ points at each leaf) that would have a $O(\log_2{(N/C)})$ height;
% , the \tree reduces the tree height from $O(\log_2{N})$ (as in the binary kd-tree) to $O(\log_f{N})$,
hence, a reduced number of nodes are visited along a search path.
Each node stores its $f-1$ splitters in a contiguous memory block (e.g., an array), and memory is fetched in cache lines (e.g., of 64 bytes). The choice of maximum fanout $f$ has a direct impact on performance: if $f$ is relatively small, all splitters fit within a single cache line, maximizing memory efficiency, but the tree becomes taller, requiring more node visits. If $f$ is larger, the tree height is reduced, but nodes may span multiple cache lines, increasing memory accesses and resulting in more cache misses.
%Therefore, the optimal fanout represents a tradeoff between minimizing tree height and maximizing cache-efficient access. This tradeoff depends on both data characteristics (volume, dimensionality, and distribution) and the hardware cache hierarchy, making a fixed $f$ rarely optimal in practice.

\stitle{Layout of inner nodes}
Our implementation of the \tree
%addresses
%follows the structure of an m-way kd-tree while addressing
%the tradeoff described above, using
uses a cache-friendly inner node layout;
node capacity (i.e., maximum fanout $f$) is set such that
all splitters fit within a single cache line. On 64-bit systems, if each coordinate of a data point is represented as 8 bytes (e.g., a double or 64-bit integer), %\nikos{why not 4 bytes? that depends on the attribute type}, so
a 64-byte cache line can store up to 8 elements. To fully utilize this space, each internal node stores an array of up to seven splitters.
Unused slots at the end of the array  are filled with 
an upper bound of the domain in the splitting dimension, denoted as \maxkey (and shown as $\infty$ in Figure \ref{fig:mkdt}).
%along with one record defining the upper bound of the domain in the splitting dimension, denoted as \maxkey.
We also store an array of 8 pointers, each holding the address of the corresponding child node, aligned to the splitters.  
This way, we can achieve a maximum node fanout $f = 8$.
%The role of \maxkey will be explain later, when we discuss search.
%\nikos{since \maxkey may vary among dimensions, we can say that it could be dimension-specific. Also, we have to explain at some point why \maxkey is necessary}

% Each splitter within an internal node $v$ defines a space partition for the data stored in the subtree of $v$, denoted by $T(v)$; these partitions are disjoint.
% Specifically,  the first splitter $s_0$ defines a partition that contains all points in $T(v)$ with values less than $s_0$ in the splitting dimension, each subsequent splitter $s_i$ defines the space that includes points with values at least $s_{i-1}$ and smaller than $s_i$, and the final partition extends from the last splitter up to \maxkey, ensuring complete coverage of the splitting dimension's domain.
% By structuring nodes in this way, the \tree guarantees that every internal node produces non-overlapping partitions of its subspace. Since all splitters are stored contiguously and aligned to a single cache line, the partitioning information can be fetched and evaluated with a single memory access, maintaining spatial locality \nikos{do you mean access locality? or avoidance of cache misses} and predictable traversal performance.
% \nikos{too abstract... needs rewriting...}

\looseness=-1
\stitle{Layout of leaf nodes}
Each leaf node stores $D$ coordinate vectors in a structure-of-arrays (SoA) layout (columnar representation).
That is, for each dimension $d$ there is a vector holding the values of all points in that dimension.
%Optionally,
There is also a vector with the record ids of all points in the leaf, to potentially access other attributes of query results. 
All these vectors are aligned; hence, values in the same position correspond to the same point (data object).
%Each leaf, also stores
Two more vectors $\vect{B}^{min}$ and $\vect{B}^{max}$ hold the bounding box of the points in the leaf. Node occupancy meta-data are kept in both inner and leaf nodes. 
%If the bounding box of the leaf is fully contained in the query (lines 2–4), all points are directly added to the result set $S$ using getPoint(i), which reconstructs the $i$-th point from the columnar vectors.
%\nikos{why don't you do this check for each dimension? if in one dimension the box is included in the query, no comparisons are needed for that dimension. An if-continue statement in the for loop would do the job.}
%This avoids per-dimension comparisons, since all points in the leaf lie entirely within the query hyper-rectangle. If the leaf is only partially covered by the query,

%\todo{describe SoA and give an example. One figure should exemplify the inner node layout and the leaf node layout.}

\begin{figure}
          \includegraphics[width=0.7\columnwidth]{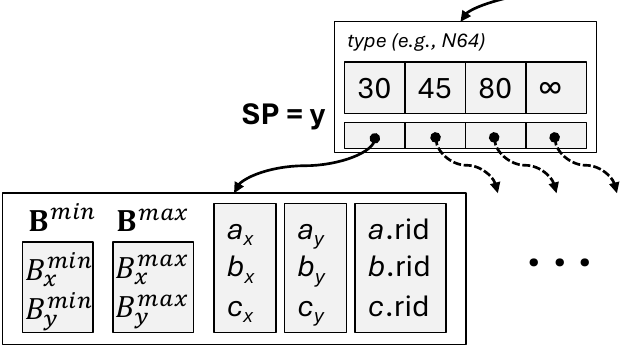}
      \caption{\tree node layout}
      \label{fig:layout}
\end{figure}

\Cref{fig:layout} exemplifies the \tree node layout for an excerpt of the tree shown in \Cref{fig:mkdt}, which includes the leftmost inner node above the leaves and the first leaf node, containing points $a$, $b$, and $c$. The inner node includes two vectors holding the splitters (upper array) and node pointers (lower array). It also includes a variable indicating the type of the node (N64, N32, or N16), to be explained in Section \ref{sec:compress_sep}. The leaf node includes two vectors with the coordinates of the points in the $x$ and $y$ coordinate (dimension 0 and 1, respectively), a record ids vector, and the two bounding box vectors $\vect{B}^{min}$ and $\vect{B}^{max}$.
% holding the minimum and maximum values per dimension in the leaf node, respectively.
% \achilleas{we can merge this paragraph with the previous one, since both describe the structure of \tree.}

\subsection{Compressed splitters and increased fanout} \label{sec:compress_sep}
%\todo{Describe here how to define the separators and use their prefix
%	for compression and increased fanout at each node}

% \dimitris{check this section}
% In any tree data structure, the fanout is important because it determines the height of the tree and, therefore, its performance. A smaller tree height leads to faster traversal and better overall efficiency. As described in Section \ref{sec:tree}, our cache-aware mkd-tree significantly reduces tree height compared to the kd-tree, achieving a height of $O(\log_f{N/C})$. However, this height can still be suboptimal in practice. In particular, when the data volume is large relative to the dimensionality, each dimension must be split many times to properly partition the data space. 

%\nikos{We must include a methodology for selecting between N64, N32, N16. One issue is feasibility, another issue is potential gain. I guess this is something you will describe in the Construction section, but feasibility should definitely be discussed here.} 

The fanout of our \tree is limited by the requirement that all splitters of an internal node must fit within a single cache line of 64 bytes.
%On 64-bit systems, this allows storing at most seven exact splitters plus one marker value (\maxkey), with each value represented using 64 bits. We refer to this layout as N64, which provides a fanout of $f$ = 8. This design is cache-efficient because all partitioning information can be loaded with a single memory access. However, the fixed fanout limits how much the tree height can be reduced, especially when the same dimension must be split repeatedly.
We suggest the use of {\em compressed splitters}, which trade numerical precision for a higher fanout while still fitting within a single cache line. The key idea is to
%is that splitters do not always need full 64-bit precision for dividing the domain of a dimension to disjoint ranges.
%We only need to
store the most significant bits of each splitter, up to the bit position where the set of kept prefixes of all splitters does not contain duplicates. 
This way, 
we can increase the number of splitters per node without increasing memory usage.
We support two additional node layouts based on this idea.
%In both layouts, exact splitters are either stored as 64-bit integers or obtained by converting double-precision splitters to a fixed-point integer format, and then compressed by keeping only their most significant bits.

In the {\em single-precision} layout (N32), each splitter becomes a 32-bit (unsigned) integer having the 32 most significant bits of the original splitter, considering only the bit positions from the first one where the largest original splitter in the node has a 1.
%of the largest of the original 64-bit splitters in the node.
% For a 64-bit splitter, we remove the lower 32 bits using a bit mask and then right-shift the result to store the compressed value.
With this layout, an internal node can store  up to 15 splitters plus one \maxkey entry in a single cache line, which doubles the maximum fanout compared to N64 ($f$ becomes 16).
In the {\em half-precision} layout (N16), only the 16 most significant bits of each splitter are kept, supporting maximum fanout $f$ up to 32.
%For a 64-bit splitter, we remove the lower bits using a bit mask and then right-shift the value. This allows each node to store up to 31 splitters plus one \maxkey entry in a single cache line, increasing the fanout by a factor of four.

\looseness=-1
The N32 and N16 layouts reduce tree height and traversal depth,
% especially when the same dimension needs to be split more than one times. Moreover,
while all layouts have the same search cost per node, since the SIMD operations that we use (to be discussed in the next section) support 64-bit, 32-bit, and 16-bit data types.
On the other hand, the use of compressed splitters does not guarantee even partitioning of the data, as when using  uncompressed 64-bit splitters, because padding 0's at their end does not result in the original 64-bit splitter. This potential issue of imbalance and how to address it is discussed in Section \ref{sec:construction},
where we choose between the N64, N32, and N16 layouts
during \tree construction.

% Although compression can slightly reduce partition accuracy, the construction process aims to keep the tree as balanced as possible. \nikos{I do not understand why there is a problem with precision or accuracy. If separators within each N32 or N16 node are unique, there should be no problem with precision.}
% If N32 and N16 are possible options (that depends on the data distribution), the choice between the N64, N32, and N16 layouts represents a tradeoff between precision and fanout and is determined during index construction.
% \nikos{again, the choice is not a matter of precision. It should be a matter of how many separators we need per dimension. If a dimension does not require 32 partitions but only 8 partitions, there is no reason to use N16 or N32.}

\subsection{Query processing}
\label{sec:query_processing}
The \tree supports
%the two fundamental multi-dimensional query types:
range queries and $k$ nearest neighbor ($k$NN) searches.
%Processing algorithms for these queries both exploit the hierarchical partitioning of the data space but differ in their traversal strategies.
%The \tree
%preserves the standard semantics of these operations while
%enhances traversal efficiency
We now present efficient processing algorithms for these queries, which 
exploit the cache-friendly node layout of the \tree and data-level parallelism enabled by SIMD-based query evaluation.

\subsubsection{Range Query}\label{sec:range}
Let $\mathcal{P} \subset \mathbb{R}^D$ be a set of data points in a $D$-dimensional space. Each point $\vect{p} \in \mathcal{P}$ is represented as $
\vect{p} = (p_0, p_1, \dots, p_{D-1})$. A range query is an axis-aligned hyper-rectangle, whose lower and upper bounds are defined by the following vectors:
$$
\vect{q}^{min} = [q_0^{min}, q_1^{min}, \dots, q_{D-1}^{min}], \quad
\vect{q}^{max} = [q_0^{max}, q^{max}_1, \dots, q_{D-1}^{max}]
$$
The result of range query $RQ(\vect{q}^{min}, \vect{q}^{max})$ is defined as follows:
$$
RQ(\vect{q}^{min}, \vect{q}^{max}) = \{\vect{p} \in \mathcal{P} | \vect{q}^{min} \le \vect{p} \le \vect{q}^{max} \},
$$
where $\vect{q} \le \vect{p}$ is vector-wise inequality, i.e., $q_i \le p_i, \forall i \in [0,D)$.

Traversal is performed in breadth-first order (BFS) using a queue $Q$ to store nodes to be examined. For each internal node, evaluation is accelerated using SIMD intrinsics, as summarized in Algorithm \ref{alg:simd-range-internal}. Initially, the node’s splitters are loaded into a SIMD register (splitters\_vec, line 1), and the bounds of the query along the current splitting dimension $d$ are broadcast to SIMD registers (qmin\_vec and qmax\_vec, lines 2–3). Data parallel comparisons (lines 4–5) find which splitters lie below or above the query bounds, producing masks (mask\_min and mask\_max). The popcount of these masks (lines 6–7) identifies the indices of the {\em first and last child whose subtree may contain points in the query range} (startPos and endPos, respectively).
% \achilleas{maybe it's useful to explain that popcount enumerates the sequential 1s in the masks in order to determine the desired position according to the comparison predicate. We don't have anywhere the formula $\sum_i{q\ge x_i}$}.
Finally, lines 8–9 add the corresponding child nodes to queue $Q$, to schedule visiting the relevant subtrees.

Leaf nodes are similarly evaluated using SIMD-based columnar operations (Algorithm \ref{alg:simd-range-leaf}).
%Each leaf stores $D$ coordinate vectors in a structure-of-arrays (SoA) layout.
%\nikos{the layout of the nodes should be described in section 4.1 and an example should be given. In addition, add a citation if this terminology has been used before in other papers} 
%Each leaf, also stores two vectors $\vect{B}^{min}$ and $\vect{B}^{max}$, which hold the bounding box of the points in the leaf.
%If the bounding box of the leaf is fully contained in the query (lines 2–4), all points are directly added to the result set $S$ using getPoint(i), which reconstructs the $i$-th point from the columnar vectors.
%\nikos{why don't you do this check for each dimension? if in one dimension the box is included in the query, no comparisons are needed for that dimension. An if-continue statement in the for loop would do the job.}
%This avoids per-dimension comparisons, since all points in the leaf lie entirely within the query hyper-rectangle. If the leaf is only partially covered by the query,
The points in a leaf $v$ are processed in SIMD-sized blocks.
We first initialize an array $masks$ with one bitvector per block to signify that each point is a result (all bits are set to 1).
Then,
for each dimension $d$,
we find which points are included in the projected query range to $d$.
If $v$'s bounding box projection to dimension $d$, i.e.,  $[v.B_d^{min}, v.B_d^{max}]$, is included in the query range, then we know that all points are results with respect to that dimension, so dimension $d$ is skipped (Lines 5-6).
Otherwise, the query bounds in $d$ are broadcast into SIMD registers (qmin\_vec and qmax\_vec, Lines 7-8), and each block of coordinates is loaded into a SIMD vector (x\_vec, line 11). Data parallel comparisons (Lines 12–13) produce masks identifying which points in the block lie within the query interval along that dimension. These masks are combined across dimensions using bitwise AND operations (line 14) to produce a final mask per block, marking points that are fully contained in the hyper-rectangle.
Finally, lines 15-21 iterate over the set bits in each mask. The global index of each point is computed as offset + idx, where $offset = b \cdot \text{SIMDWidth}$ and idx is the position of the bit in the mask. Each point is reconstructed from the columnar vectors using getPoint(offset + idx) and added to the result set $S$.
SIMD operations maximize throughput by processing multiple coordinates within the same block in parallel.

\begin{algorithm}[t]
	\caption{SIMD-based range search (inner nodes) }
	\label{alg:simd-range-internal}
	\footnotesize
        \Input{Internal node $v$, query bounds $[q_d^{min}, q_d^{max}]$ in $v$'s splitting dimension $d$, queue $Q$}
	\Output{Queue $Q$ updated with child nodes intersecting the query}
	
	$splitters\_vec \gets \text{load}(v.splitters)$\;  
	$qmin\_vec \gets \text{broadcast}(q_d^{min})$\;
	$qmax\_vec \gets \text{broadcast}(q_d^{max})$\;
	$mask\_min \gets \text{compare\_ge}(qmin\_vec, splitters\_vec)$\;
	$mask\_max \gets \text{compare\_ge}(qmax\_vec, splitters\_vec)$\;
	$startPos \gets \text{popcount}(mask\_min)$\;
	$endPos \gets \text{popcount}(mask\_max)$\;
	
	\For{$i = startPos$ \KwTo $endPos$}{
		$Q.\text{add}(v.childPtr[i])$\;
	}
\end{algorithm}

\begin{algorithm}[t]
	\caption{SIMD-based range search (leaf nodes)}
	\label{alg:simd-range-leaf}
        \footnotesize
	\Input{Leaf node $v$ with $D$ coordinate vectors $X_0, ..., X_{D-1}$, range query $RQ(\vect{q}^{min}, \vect{q}^{max})$, result set $S$}
	\Output{Result set $S$ updated with all points in the leaf satisfying the query}
	
	$leafCapacity \gets v.slotuse$\;
	% \If{leaf bounding box fully contained in $RQ(\vect{q}^{min}, \vect{q}^{max})$}{
	% 	\For{$i \gets 0$ \KwTo $leafCapacity-1$}{
	% 		$S.\text{add}(\text{getPoint}(i))$\;
	% 	}
	% }
	% \Else{
		$numBlocks \gets leafCapacity / SIMDWidth$\; 
		Initialize $masks[0..numBlocks-1] \gets$ all bits set\; 
		
		\For{$d = 0$ \KwTo $D-1$}{
                  \If{$q_d^{min}\le v.B_d^{min}$ and $q_d^{max}\ge v.B_d^{max}$}{
		{\bf continue}\comm*[f]{$v.\vect{B}$  contained in $\vect{q}$ in dimension $d$}\;
              }
			$qmin\_vec \gets \text{broadcast}(q_d^{min})$\; 
			$qmax\_vec \gets \text{broadcast}(q_d^{max})$\;
			
			\For{$b = 0$ \KwTo $numBlocks-1$}{
				$offset \gets b \cdot SIMDWidth$\; 
				$x\_vec \gets \text{load}(X_d, offset)$\;
				$mask\_le \gets \text{compare\_ge}(x\_vec, qmin\_vec)$\; 
				$mask\_ge \gets \text{compare\_le}(x\_vec, qmax\_vec)$\; 
				$masks[b] \gets masks[b] \textbf{ AND } (mask\_le \textbf{ AND } mask\_ge)$\; 
			}
		}
		
		\For{$b = 0$ \KwTo $numBlocks-1$}{
			$mask \gets masks[b]$\;
			$offset \gets b \cdot SIMDWidth$\;
			\While{$mask \neq 0$}{
				$idx \gets$ index of least-significant set bit in $mask$\;
				$S.\text{add}(\text{getPoint}(offset + idx))$\;
				Clear least-significant set bit in $mask$\;
			}
		}
	
\end{algorithm}

\subsubsection{$k$NN search}\label{sec:knn}
Let $\mathcal{P} \subset \mathbb{R}^D$ be a set of data points in a $D$-dimensional space, and let 
$\vect{q} \in \mathbb{R}^D$ be a query point. A $k$ nearest neighbor ($k$NN) query retrieves 
the subset of $k$ points in $\mathcal{P}$ that minimize the squared Euclidean distance to 
$\vect{q}$:
\[
kNN(\vect{q}, k) = \arg \min_{S \subset \mathcal{P}, \; |S| = k} \sum_{\vect{p} \in S} 
\|\vect{p} - \vect{q}\|^2,
\]
where the squared Euclidean distance between points $\vect{p}$ and $\vect{q}$ is
\[
\|\vect{p} - \vect{q}\|^2 = \sum_{i=0}^{D-1} (p_i - q_i)^2.
\]
\stitle{Search Strategy} The $k$NN search follows a best-first traversal (Best-Bin-First), guided by lower-bound distance estimates \cite{PapadiasMH05}. Two auxiliary heaps are used during the process: (i) a min-heap $L$ enabling access to tree entries by their minimum distance to the query and (ii) a fixed-size max-heap $H$ that stores the current $k$ nearest neighbors.
The maximum distance in $H$ (i.e., the distance of the top element in $H$) is used as
a global pruning threshold $bound$.
Initially, $bound$ is set to infinity until $k$ points have been found.
%Once $k$ points are reached, 
%$bound$ is updated to the distance of the farthest neighbor (the top element of the max-heap).
Any entry in $L$ whose distance exceeds $bound$ is skipped, effectively pruning regions that cannot contain closer neighbors.
%Algorithm \ref{alg:simd-knn} describes the search process.

\stitle{Heap Entries} Each entry in the min-heap $L$ is represented as a tuple $\langle v, idx, type, dist, \vect{\delta} \rangle$, where $v$ is a pointer to a tree node, $idx$ is a child or splitter index, $type \in \{\textsc{NODE}, \textsc{GROUP}\}$ denotes whether the entry represents a single child or a group of children. The value $dist$ is the accumulated squared Euclidean distance
%\nikos{unclear},
and $\vect{\delta}$ is a vector of per-dimension projection distances used to update distance bounds efficiently. Both $dist$ and $\vect{\delta}$ are computed incrementally during the query process.
For the max-heap $H$, entries are represented as tuples $\langle point, dist \rangle$, where $dist$ is the squared Euclidean distance from the query $\vect{q}$ for a point located inside a leaf node of the \tree.

\looseness=-1
\stitle{Internal Node Processing \& Grouping} When examining an internal node, search is guided by its splitting dimension $d$, and a point query is performed between the node's splitters and $q_d$ using SIMD intrinsics to identify the child containing the query coordinate. This child has zero projection distance along dimension $d$ (i.e., $\delta_d = 0$) and is immediately pushed to heap $L$ as a \textsc{NODE} entry.
The remaining subspaces, which lie on either side of coordinate $q_d$, are summarized into at most two  \textsc{GROUP} entries: one representing the subspaces to the left of $q_d$ and one for the subspaces to the right. Each group stores the minimum projection distance needed to reach that region, postponing unnecessary distance computations and heap operations. This approach ensures that only potentially relevant subspaces are explored, improving efficiency while maintaining correctness.
When a \textsc{GROUP} entry is extracted from heap $L$, the closest child in that group is enheaped first as a \textsc{NODE} entry. Any remaining children are reinserted into $L$ forming a new \textsc{GROUP} entry with updated $\vect{\delta}$ and $dist$ respectively.  

\stitle{Leaf Node Processing} When a leaf node is reached during the $k$NN search, the squared Euclidean distances between the query point $\vect{q}$ and 
all points in the leaf are computed. Since leaf contents are organized in a structure-of-arrays (SoA) layout,
% squared Euclidean distances between $\vect{q}$ and all contained points are
distances are computed in a dimension-wise manner using SIMD intrinsics. Specifically, each query coordinate is broadcast into a SIMD register, while blocks of point coordinates are loaded from the corresponding columnar arrays. For each block, squared differences are accumulated using combined multiply-add operations, producing a SIMD vector of squared distances. This process continues until distances for all points in the leaf have been computed.
Each point is then considered for insertion into the max-heap $H$, which maintains the current set of $k$ nearest neighbors.

% If $H$ exceeds its capacity, the farthest point is removed. Once $H$ becomes full, the pruning threshold $bound$ is updated to the $k$-th  distance stored in $H$. This threshold is subsequently used to prune entries in the min-heap $L$ whose lower-bound distances exceed $bound$, ensuring that no unnecessary leaf evaluations are performed.

\begin{algorithm}[t]
	\caption{SIMD-based $k$NN search}
	\label{alg:simd-knn}
        \footnotesize
	\Input{Tree root, query point $\vect{q}$, number of neighbors $k$}
	\Output{Set of $k$ nearest neighbors}
	
	Initialize empty min-heap $L$\;
	Initialize empty max-heap $H$ with capacity $k$\;
	$bound \gets \infty$\;
	Initialize projection vector $\vect{\delta} \gets \vect{0}$\;
	
	$L.\text{push}(\langle root, 0, \textsc{NODE}, 0, \vect{\delta} \rangle)$\;
	
	\While{$L$ is not empty}{
		$\langle v, idx, type, dist, \vect{\delta} \rangle \gets L.\text{pop}()$\;
		
		\If{$H \text{is full}$ \textbf{AND} $dist \ge bound$}{
			\textbf{break}\;
		}
		
		\If{$v$ is a leaf}{
			$\forall$ point $\in v$ compute sq. distance to $\vect{q}$ using SIMD\;
			Insert candidate points into $H$\;
			\If{$H$ is full}{
				$bound \gets H.\text{top}.dist$\;
			}
			\textbf{continue}\;
		}
		
		\If{$type = \textsc{NODE}$}{
			Let $d$ be the splitting dimension of $v$\;
			Use SIMD to locate the child subspace containing $q_d$\;
			Push this child as a \textsc{NODE} entry with $\delta_d = 0$\;
			
			Create up to two \textsc{GROUP} entries for subspaces left and right of $q_d$\;
			Compute their minimum projection distances and push them to $L$\;
		}
		\Else{
			Select the closest child within the group\;
			Push it as a \textsc{NODE} entry into $L$\;
			Reinsert remaining children as a reduced \textsc{GROUP} entry with updated
			$(dist, \vect{\delta})$\;
		}
	}
	
	\Return{$H$}\;
\end{algorithm}

Algorithm \ref{alg:simd-knn} summarizes $k$NN search. For simplicity, we present the high-level $k$NN traversal logic and omit the explicit SIMD intrinsics in this pseudocode. These operations either reuse the same SIMD comparison patterns illustrated in  Algorithms \ref{alg:simd-range-internal} and \ref{alg:simd-range-leaf} or are replaced by analogous SIMD arithmetic primitives (e.g., vectorized subtraction, multiplication, and accumulation).

\subsection{\tree construction}\label{sec:construction}
% \dimitris{we should mention the gaps here, because, we will need them to explain the insert and delete. When we have less ammount of splitters, that can store in a a cache line, the empty spaces are gaps}
% \nikos{gaps literally means that the splitters within each node are not stored continuously. Do we use exactly the same way as the BS-tree to handle gaps, or do we push all gaps at the end of the node?}
We propose a top-down construction algorithm for the \tree which is both effective and efficient.
%In this section, we describe how our mkd-tree is built using a top-down approach. Before presenting the main parts of the algorithm, we first discuss some general aspects. The leaf nodes are hyperrectangles, since this representation is well suited for handling agnostic queries.
%At each level, our mkd-tree uses nodes with multiple fanout, as described in Section 4.2. This differs from a traditional kd-tree, which uses only binary nodes.
Recall that, as in the kd-tree, 
%However, the choice of split dimensions
%follows the standard kd-tree approach, where
a different dimension is used to split each level
of the \tree.
However, nodes are multiway and may have different fanouts (i.e., N64, N32, and N16 nodes).
%The type of each node is chosen to maximize the number of separators while maintaining the required precision. As a result, nodes at the same level may use different fanout types, allowing
This allows the mkd-tree to adapt
%to the data distribution.
%Using different fanout sizes also helps
and keep balanced:
if a split results in imbalanced partitions (due to
ties in values in the splitting dimension or imprecision of the com-
pressed splitters), the partitions can use different node types. Partitions that contain more data can use nodes with larger fanout (e.g., N16), 
while partitions with less data can use nodes with smaller fanout (e.g., N64).
For leaf nodes, we empirically set the capacity $C$=128, a value that reduces number of tree levels while facilitating processing of multiple points in parallel. 

\looseness=-1
We first have to determine the order in which the dimensions are split.
More uniform dimensions with more unique values should be selected first, to facilitate compression of splitters and balanced subtrees .
%This step is particularly important when the data contains a large number of duplicate values in some dimensions.
%, as selecting the split dimension randomly at the upper levels may lead to an imbalanced \tree and poor query performance.
%To mitigate this issue,
Hence, we rank the dimensions according to their degree of uniformity (or spread) and based on the number of unique values in the data per dimension.
%More uniform dimensions with more unique values are selected first.
%, from most uniform to most skewed, as per the standard methodology applied for kd-trees \nikos{citation needed}.
%Further details of this model are provided in Section 4.4.1. \nikos{model to be removed?} \achilleas{I need time to check it. If we have an explanation for it, we will keep it}
Once the dimension order is established, we proceed with the construction. Our goal is to form leaf nodes whose bounding boxes are hypercubes, as this is well suited for query-agnostic workloads \cite{BeckmannKSS90,LeuteneggerEL97}.
To this end, we determine the number of splits to apply along each dimension.
Let $N$ be the total number of data points in the $D$-dimensional space, and let $C$
%=128
be the leaf capacity threshold. We first estimate the number of leaf nodes as $P=\lceil N/C \rceil$. The target number of splits for the first dimension is then estimated as $S=P^{1/D}$. This value represents the desired number of partitions along that dimension and serves as an upper bound on the fanout of nodes that split on it.
The estimated value $S$ guides the selection of the node type. If $S\le 8$, we exclusively use N64 nodes for this dimension, as their maximum fanout is 8. If $S>8$, we consider the possible options of N64, N32, and N16 nodes,
%if $S\le 16$ we evaluate the available node types
and select the option whose fanout most closely matches the desired number of partitions.
%for each dimension, we aim each node having $S$
%\dimitris{Maybe $S$, also in the next paragraphs}
%children.
However, if
%$S>8$ and this is not possible, either because
$S>8$ and it is not possible to achieve unique splitters by a N16 or N32 compression scheme, we split into $S/2$ child nodes and keep the dimension for a possible new split at a level below.
%(it is difficult to understand it)} \achilleas{it's the only part that it needs a clearer explanation}
%\nikos{unclear, we need a more definite description} \achilleas{If $S\le 8$, we exclusively use N64 nodes for this dimension, since the required splitters fit exactly into a cache line, and it is clear that we cannot benefit from using the alternative layouts. Now if $S > 8$, we select the appropriate combination of node layouts under the assumption that half-filled nodes are not allowed. For example, when $S \ge 16$, all three node layouts can be used. In contrast, if $S \in (8,16)$, only the N32 and N64 layouts are considered}

\stitle{Determining the Splitters}
To construct the splitters for a node, we generate candidate split boundaries using a {\em recursive median} selection approach. Specifically, we first compute the median $m$ of the points along the chosen dimension (this can be done fast using the median-of-medians algorithm or by computing the median of a small data sample).
Then, we quantize $m$ (if we aim for a N32 or N16 node), by right bit-shifting.
%it by 32 or 48, respectively.
Finally, we apply one step of quicksort to {\em crack} \cite{IdreosKM07} the subarray containing the points into two parts using $m$ (after padding it with an appropriate number of zeros in the end if $m$ is compressed).
After partitioning, we apply the same procedure recursively to the largest piece of the cracked points array, until the target number of splitters has been reached.
This procedure is applied iteratively to construct inner nodes.

During the median selection and cracking process, it is possible that we compute a new (quantized) splitter, which is identical to its next or previous splitter, especially when points have duplicate values in the splitting dimension. In this case, we skip the split of the corresponding cracked array segment and turn to the next largest segment.
%If we get to have multiple splitters with identical values, only one is retained.
This may create imbalance to the child subtrees. Another reason why we may have imbalance in the resulting partitions is that when we quantize a splitter, we essentially move the value of the original (uncompressed) splitter to the left. For example, suppose that we quantize value 156 (binary 10011100) by right-shifting it by 4 to become 9 (1001). The quantized splitter (9) will be used as $9 << 4 = 144$ in partitioning (and during search), which is much smaller than the original median value (156).

To deal with the possible imbalance of partitions at the next level, we {\em adapt} the number of splitters per node. Recall that target number of splits at each node is originally $S$.
%Instead of using $S$,
We adjust the target number of splits to $S'$ based how much  the number  $M$ of points to be partitioned at the node deviates from the expected number $M_{exp}$ of points that the subtree rooted at that node would get if the data were uniform and exact splitters (as in N64 nodes) were used.
Hence, $S'$ is set to $S\cdot (M/M_{exp})$.
%Hence, $S$ is dynamically and adaptively selecting, targeting at a 
%Hence, at each node, we first determine $S'$ and then attempt to split it accordingly.
If, after applying splitting we end up having fewer than $S'$ splits (due to splitter ties),
the imbalance propagates to (some of) the children and eventually addressed by either an additional round of splits for that dimension, or (in the case of numerous ties in that dimension) by creating {\em outlier} (i.e., overflown) leaf nodes, as we will explain now.

% During the construction process, we maintain a queue of nodes to process, where each node stores the range of points assigned to it, the dimension to split, and the target number of separators. At each step, we dequeue a node, compute its split separators, and partition the points according to these boundaries. Children whose point counts exceed the leaf capacity are added to the queue for constructing the corresponding subtree. This procedure continues until all nodes satisfy the leaf capacity constraint.

% For each child node, we consider
% only the subset of points assigned to the corresponding range of its parent to split the child, and recompute the target number of splits $S$ for the next dimension, and generate separators accordingly. As a result, in case of imbalance, the number of splits adapts to the size of each inner node, which helps to construct a balanced structure.

\stitle{Leaf types}
Due to the distribution of the data and the presence of duplicate values in the splitting dimensions, it may not be possible to produce leaves with exactly the same number of points. To accommodate this, we allow leaf nodes to store between $C/2=64$ and $C=128$ points. However, a larger than $C$ number of duplicates at the splitting dimension above the leaf level may result in larger leaves.
Due to this, we relax the basic structure of the  \tree to include three types of leaf nodes: {\em light}, {\em heavy} and {\em outlier}.
%Light, Heavy, and Outlier.\achilleas{Will we write them as Light, Heavy and Oulier or {\em light}, {\em heavy} and {\em outlier}?}
We define two thresholds to classify leaves. The heavy threshold, 
$T_{\text{h}}$, determines whether a leaf
% should be further subdivided
is subject to splitting
and is defined as $T_{\text{h}}$ $= 1.2\cdot max(\bar{C}, C)$, where $\bar{C}$ 
is the mean capacity of all leaves.
%and $C=128$ is the nominal leaf capacity.
The outlier threshold is defined as $T_{\text{o}}$ = $2\cdot T_{\text{h}}$. 
Leaves which contain between $C/2$ and  $T_{\text{h}}$ points are characterized as {\em light}.
Leaves having between $T_{\text{h}}$ and $T_{\text{o}}$ points are characterized as {\em heavy}. Finally, leaves with more points than $T_{\text{o}}$ are characterized as {\em outliers}. 
% Leaf types are assigned according to the following rules:
% 1. Light leaf: capacity in $[C/2, T_{\text{s}}]$.

% 2. Heavy leaf: capacity in $[T_{\text{s}},T_{\text{o}}]$, eligible for further splitting.

% 3. Outlier leaf: $capacity \geq T_{\text{o}}$.

Outlier leaves represent regions with a large number of points with duplicate values in the splitting dimension at the level above.
These nodes cannot be further subdivided, because all remaining points share identical coordinates along the splitting dimensions, resulting in zero-width partitions. Consequently, they are treated as terminal nodes in the \tree.

\section{Updates}\label{sec:updates}
% In this section, we describe how the \tree handles dynamic updates while keeping the tree balanced and maintaining its expected performance.
% Throughout update operations, the \tree exploits data parallelism using SIMD intrinsics in order to accelerate both traversal and leaf-level operations.

\subsection{Insertion}
To insert a new point $p$, we traverse the internal levels of the \tree as if we were searching for $p$ to locate the
leaf node where $p$ should be inserted.
%At each internal node, SIMD-based comparisons are employed, allowing the fast selection of the right child node with fewer instructions and reduced branching.
The traversal algorithm is similar SIMD-based range search described in
Section \ref{sec:range}, however, only one pointer per accessed node is selected (as in a point query) and we end up accessing just one path of the \tree.
Once the target leaf node is identified, the insertion procedure depends on the leaf type.

\stitle{Light leaf}
If the leaf is {\em light}, the point is simply inserted to the end of the leaf node; that is, for each dimension $d$, the value of $p$ in that dimension is appended at the end of the corresponding vector and the identifier of $p$ is also appended at the end of the id's vector, such that the vectors are fully aligned to each other. The bounding box (MBB) vector $\vect{B}$ of the node is also updated.
%if necessary.
After the insertion the leaf may become {\em heavy}.
%it has at least one empty slot at the end of the vectors (see Figure \ref{fig:mkdt}) that store the point values for each dimension. In this case, the insertion can be performed immediately by appending the new point, requiring only an update of the leaf's bounding box if the new entry expands it.

\stitle{Heavy leaf}
If $p$ is inserted to a {\em heavy} leaf, we also append the new point at the end of the corresponding vectors. However, we monitor whether multiple insertions cause the leaf size to exceed the threshold $T_{\text{o}}$.
%, in which case it is transformed into an {\em outlier} leaf.
When this happens, we attempt a {\em split} operation to the leaf.
For leaf node splitting, we follow two different strategies. If the parent node of the leaf is not full, we perform the split as in a B$^+$-tree;
we find the median value $m$ in the leaf based on the parent's splitting dimension $i$,
%sort the points according to the,
keep points smaller than $m$ in dimension $i$ the current leaf, create a new leaf with the points greater than or equal to $m$ in dimension $i$, and insert the splitter $m$ to the parent. If the parent node is of N32 or N16 type, we quantize the splitter before partitioning. Finally, the MBBs of the new leaf nodes are computed.
% similarly to traditional tree structures (e.g., \bt). This involves selecting a new splitter, allocating a new leaf node, and partitioning the points as evenly as possible. The bounding boxes of both leaves are updated accordingly.

If the parent node does not have an empty slot for an immediate split, we backtrack along the tree path to identify the closest ancestor node $v$ that has an empty slot.
Once the node is found, we collect all the data in the subtree pointed by the entry of $v$ that leads to the split leaf and construct two new subtrees that replace the original one, following the recursive construction algorithm described in Section \ref{sec:construction}.
A new splitter is created at node $v$ to separate the two new subtrees that replace the old one.
%a new subtree that approximately receives the half of the the points from the initial subtree. This reconstruction generalizes the previous splitting strategy, following the same concept described in Subsection \ref{sec:construction}.

There are two cases when a  {\em heavy} leaf node $l$ becomes an {\em outlier} node after an attempted split. The first case is when we fail to generate a new unique (quantized) splitter at the parent node $parent(l)$ of the leaf $l$ that we attempt to split. This may happen if the parent node is N32 or N16 and there is heavy skew, or when there are many duplicate values in the splitting dimension of $parent(l)$. Then, the new splitter may be identical to the next or the previous splitter in $parent(l)$.
The second case is when we fail to generate a new splitter at the closest ancestor node $v$ of $l$ that has an empty slot, for the same reason.

\stitle{Outlier leaf}
If the leaf node, where $p$ is to be inserted, is characterized as {\em outlier}, we do not attempt to split it. As discussed in Section \ref{sec:construction},
further splitting of such nodes is challenging due to the high replication of values in one or more dimensions. Only when the size of an {\em outlier} node becomes $2^1\cdot T_o$, we attempt a split; if the split fails, the leaf remains an outlier and we attempt again when its size becomes $2^2\cdot T_o$, and so on.
%\achilleas{If we have space, we can include a pseudoce with the described steps}
%\nikos{this is too abstract? what does significantly deviates mean in numbers? We have space for a pseudocode.}

\subsection{Deletion}
Like insertions, deletions first perform point search to locate the leaf node that contains the point $p$ to be deleted, using at each accessed node $v$ the value of $p$ in $v$'s splitting dimension.
% an examination strategy similar to that of insertion, where we traverse the \tree until reaching the appropriate leaf node that contains the point to be removed.
During traversal, SIMD-based comparisons are employed to efficiently find the corresponding tree path, as discussed in Section \ref{sec:range}.
Data parallelism enabled by SIMD intrinsics is exploited at the leaf node $l$ to perform equality search for $p.id$ in $l$'s vector which contains the identifiers of the points there.
%After accessing the leaf node, using $p.id$,
Assuming that the point $p$ to be removed exists, this search finds its position within the leaf.
Then, in all vectors (coordinate vectors, id vector), the last values are copied into the position of the deleted point, the node's occupancy is reduced by one, and its MBB is updated if necessary.
After the deletion, an outlier leaf may become heavy, a heavy leaf may become light, and a light leaf may {\em underflow}.
%points after it are then shifted to fill the resulting gap, and the leaf's bounding box is updated in a way analogous to the insertion. 
We do not handle underflows, relaxing the lower bound constraint of light leaves, following previous work (e.g., \cite{RaoR00}).
%\nikos{cite papers that tolerate underflows, I think we cite some in the bs-tree paper}.
The reason is that deletions are expected to be rare compared to insertions, so when a leaf underflows, we expect that it will accept new points in the future and that would address the occupancy requirement.
In addition, for in-memory indices, underflows have a much lower impact compared to disk-based access methods.
%decide to avoid leaf merging and allow underflow in leaves.
In the edge case where a leaf node becomes completely empty, we delete it and backtrack along the tree path and update the contents of the involved internal nodes.
% In a nutshell, starting from the parent node of the empty leaf, we remove the extraneous splitter and the corresponding pointer to the leaf node.
%, by left-shifting the remaining splitters and child pointers by one position, introducing an (additional) empty slot at the end of the node.
% Empty node deletion may propagate upwards as necessary.
% Node deletions effectively create empty slots to be exploited during node splits or subtree reconstructions.
%This backtracking continue upward along the traversal path as long as internal nodes become empty due to the deletion.

\section{Experiments}\label{sec:exp}
We experimentally compare \tree to alternative main-memory
multi-dimensional indices.
%(learned and non-learned).
All methods are implemented in C++ and compiled with \verb|gcc| (v13)
using the flags \verb|-O3| and \verb|-march=native|.
The experiments are conducted on a system with an 11th Gen Intel®
Core™ i7-11700K processor
running at 3.60 GHz, 128 GB of RAM, with AVX-512 support.
The operating system used is Ubuntu 22.04.
We extend the codebase
of Learnedbench \cite{LiuLZSC25} to include \tree.

\subsection{Setup}\label{sec:exp:setup}

\stitle{Datasets}
We conduct our experiments on standard benchmark datasets of varying dimensionality that have been widely used in previous studies \cite{LiuLZSC25, DingNAK20, nathan2020learning}. We consider only numerical dimensions, as in prior work \cite{LiuLZSC25,nathan2020learning,ZaschkeZN14,DingNAK20}, and normalize each dimension to the range [0,1]. The second and third columns of Table \ref{tab:mkdtree_stats} summarize the cardinality and dimensionality of the points in each dataset. EDGES \cite{spatialData} is a 2-d dataset containing polygon data from the continental United States; we use polygon centroids to obtain points. TORONTO \cite{tan2020toronto3d} is a large-scale urban outdoor point cloud dataset with 3-d points collected in Toronto, Canada, using a mobile laser scanning (MLS) system. OOKLA \cite{ookla} is a 5-d dataset that includes global fixed broadband and mobile network performance metrics. GAIA \cite{gaia, gaia_mission, gaia_data_release} is a 6-d dataset of nearly two billion Milky Way objects, providing precise astrometric and photometric measurements, along with spectroscopy and derived stellar properties. NYT \cite{nyt} is an 8-d dataset containing trip records from the New York City Taxi and Limousine Commission. TPC-H \cite{tpc_h} is a decision-support benchmark for evaluating database systems on complex analytical workloads. We use it to generate an 8-d dataset by joining the Lineitem, Orders, Partsupp, Customer, and Part tables using their standard foreign-key relationships and selecting numerical attributes from the result.

%\begin{table}
%	\caption{Dataset Characteristics}
%	\label{tab:dataset}
%	\centering
%	\vspace{-2mm}
%	\resizebox{\columnwidth}{!}{%
%		\begin{tabular}{|c|c|c|c|c|c|c|}
%			\hline
%			& \textbf{EDGES} & \textbf{TORONTO} & \textbf{OOKLA} & \textbf{GAIA} & \textbf{NYT} & \textbf{TPC-H} \\ \hline
%			\textbf{Records}    & 51M            & 78M              & 123M           & 216M              & 44M              & 240M           \\ \hline
%			\textbf{Dimensions} & 2              & 3                & 5              & 6                 & 8                 & 8              \\ \hline
%		\end{tabular}
%	}
%\end{table}
%

\stitle{Workloads}.
%We analyze the workloads used for range and $k$NN queries.
For range queries, we execute $1000$ hypercube queries on each
dataset, retrieving $0.001$\%, $0.01$\%, and $0.1$\% of the indexed
points. Query centers are randomly sampled from the data and each
dimension is then symmetrically expanded around the selected point
until the desired selectivity is achieved. We exclude queries with
selectivity of $1$\% or higher, as they return hundreds of thousands
of results in our datasets, which is not representative of a realistic
search scenario. For $k$NN queries, we randomly select 1000 points
from each dataset and use them as query points. We then perform $k$NN
searches while varying the number $k$ of nearest neighbors in  \{$1, 5, 10, 50,100$\}.

\stitle{Competitors}.
We compare our proposed skd-tree with seven conventional and learned
multi-dimensional index structures, most of which are implemented in
the Learnedbench codebase \cite{LiuLZSC25}, a benchmarking suite based
on the Boost C++ Libraries \cite{boost}. We exclude methods shown to
perform poorly or inconsistently in prior work \cite{LiuLZSC25}, as
well as those with unavailable or proprietary implementations.
% Among conventional indices, we omit the R*-tree \cite{BeckmannKSS90},
% as the STR bulk-loaded R-tree \cite{LeuteneggerEL97} consistently
% achieves better performance. The ANN index is also excluded since it
% supports only approximate $k$NN queries.
Regarding learned indices, we excluded those without publicly
available code
(e.g., SLBRIN \cite{wang2023slbrin}, LMSFC \cite{Gao0YZ023}, WAZI
\cite{PaiM024}, HMI \cite{wang2020spatial}) and those which
support only approximate queries (e.g., RSMI \cite{qi2020effectively})
or exhibit poor performance (e.g., ZMI \cite{wang2019learned}, LISA \cite{li2020lisa}).

%\stitle{Non-learned Indices}.
\looseness=-1
The first competitor is the strongest performing R-tree (with STR bulk-loading),
taken from the Boost C++ Libraries \cite{boost}, using a fanout of 128 (i.e., the
same as the leaf capacity $C$ threshold of \tree).
It supports $k$NN and range queries and also allows updates. The
kd-tree is based on the nanoflann library \cite{blanco2014nanoflann} and natively supports only $k$NN queries; we extend it to support range queries using the provided nanoflann interfaces.  Adaptive Grid (AG) and Uniform Grid (UG) use the same cell size (128 points per cell) and support only range queries. For AG, we apply the partitioning method described in our \tree (Section \ref{sec:construction}) to determine the boundaries of the partitions in each dimension using a sample of the dataset. During range queries, we perform binary search in each dimension to identify the relevant partitions and, ultimately, the grid cells covered by the query. The PH-tree implementation is taken from \cite{phtree_code} and integrated into Learnedbench; it supports $k$NN and range queries, as well as updates. Finally, we compare against Flood and IF-Index (IFI), two learned indices already included in Learnedbench. We use the default parameter settings of Learnedbench, and both support only range queries. For Flood, each cell contains 2000 points, while for IFI the fanout of inner nodes is 32 and that of leaf nodes is 2000.

\subsection{Effectiveness of the  \tree construction}
%\achilleas{I made changes to the text and the table in order to connect the statistics with the argument for why we select the corresponding datasets for the ablation study (or whatever it will be named)}
In the first experiment, we evaluate the effectiveness of the  \tree
construction algorithm described in \Cref{sec:construction}.
For this,
%To evaluate the impact of the construction process in terms of
%balancing,
we collected structural statistics of the \tree across each
dataset. In particular, we record which of the different internal node
layouts (N16, N32, N64) are used during construction. We also measure the average leaf capacity of the produced leaf nodes, along with their categorization into the three leaf types ({\em light}, {\em heavy}, {\em outlier}). All statistics are summarized in Table \ref{tab:mkdtree_stats}.
As shown in the table,
%the \tree indeed uses compressed internal node layouts according to
%the characteristics of each dataset.
EDGES, which is low-dimensional and duplicate values
in each dimension are rare, fully exploits compression and includes
exclusively N16 nodes. TORONTO and OOKLA use all three node types,
effectively compressing some regions with N16 and N32 nodes, while
GAIA does not require N16 compression due to its higher
dimensionality.
The 8-d datasets (NYT and TPC-H) require no compression at all, so
they just use N64 nodes.
%cannot exploit compresses some regions by only N32. For NYT and TPC-H, compression is not applied due to the higher dimensionality of these datasets, which results in a small target number of splitters per
%dimension (for which compression is not necessary).
%As shown in the table, \tree indeed compressed internal node layouts
%according to the characteristics of each dataset. EDGES and TORONTO,
%which are low-dimensional exploit both N16 and N32 layouts for
%compression, while OOKLA and GAIA use only N32. For NYT and TPC-H,
%compression is not applied due to the higher dimensionality of these
%datasets, which results in a small target number of splitters per
%dimension (for which compression is not necessary).
For all datasets, the \tree achieves average leaf occupancy consistently
close to the target value of 128.
In addition, the number of leaf nodes classified as heavy
or outliers is negligible across all datasets, indicating that the
tree achieves a well-balanced partitioning in practice.

\begin{table}[]
	\caption{\tree construction statistics across datasets (rows) and reported structural characteristics (columns)}
	\label{tab:mkdtree_stats}
	\centering
	\vspace{-2mm}
	\resizebox{\columnwidth}{!}{%
		\begin{tabular}{cccccccc}
			\toprule
			\textbf{Datasets} & \multicolumn{1}{l}{\textbf{\begin{tabular}[c]{@{}c@{}}Card. \\ ($N$)\end{tabular}}} & \multicolumn{1}{l}{\textbf{\begin{tabular}[c]{@{}c@{}}Dim. \\ ($D$)\end{tabular}}} & \textbf{\begin{tabular}[c]{@{}c@{}}Inner Node\\ Layouts\end{tabular}} & \textbf{\begin{tabular}[c]{@{}c@{}}Avg. Leaf \\ Capacity\end{tabular}} & \textbf{\begin{tabular}[c]{@{}c@{}}Light \\ Leaves {[}\%{]}\end{tabular}} & \textbf{\begin{tabular}[c]{@{}c@{}}Heavy \\ Leaves {[}\%{]}\end{tabular}} & \textbf{\begin{tabular}[c]{@{}c@{}}Outlier \\ Leaves {[}\%{]}\end{tabular}} \\ \midrule
			\textbf{EDGES} & 51M & 2 & N16 & 128.6 & 98.43 & 1.56 & 0.01 \\
			\textbf{TORONTO} & 78M & 3 & N16/32/64 & 123.6 & $\sim$99.53 & $\sim$0.47 & \textless 0.01 \\
			\textbf{OOKLA} & 123M & 5 & N16/32/64 & 125.5 & $\sim$100 & \textless 0.01 & \textless 0.01 \\
			\textbf{GAIA} & 216M & 6 & N32/64 & 133.9 & 100 & 0 & 0 \\
			\textbf{NYT} & 44M & 8 & N64 & 128.4 & $\sim$100 & \textless 0.01 & 0 \\
			\textbf{TPC-H} & 240M & 8 & N64 & 122.5 & 100 & 0 & 0 \\ \bottomrule
		\end{tabular}
	}
\end{table}

\subsection{Impact of compression and data-parallelism}
\begin{figure}
          \includegraphics[width=0.95\columnwidth]{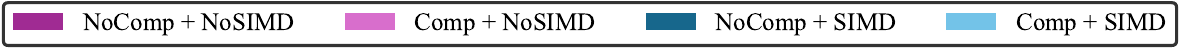}
	\begin{subfigure}{0.23\columnwidth}
         \centering
		\includegraphics[width=\columnwidth]{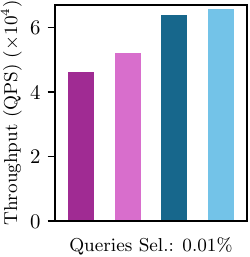}
		\caption{\footnotesize EDGES ($D=2$)}
	\end{subfigure}
	\begin{subfigure}{0.23\columnwidth}
		\centering
		\includegraphics[width=\columnwidth]{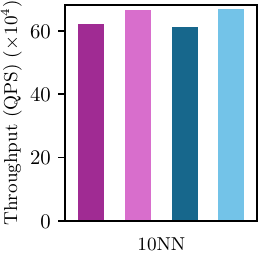}
		\caption{\footnotesize EDGES ($D=2$)}
	\end{subfigure}
		\begin{subfigure}{0.23\columnwidth}
		\centering
		\includegraphics[width=\columnwidth]{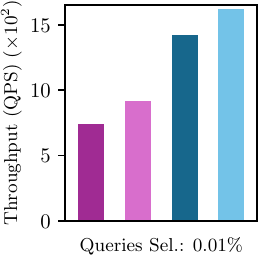}
		\caption{\footnotesize GAIA ($D=6$)}
	\end{subfigure}
	\begin{subfigure}{0.23\columnwidth}
		\centering
		\includegraphics[width=\columnwidth]{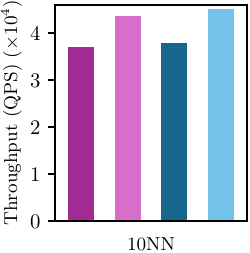}
		\caption{\footnotesize GAIA ($D=6$)}
	\end{subfigure}
	
	\caption{Impact of node layouts and SIMD in \tree}
       \label{fig:exp:ablation}
\end{figure}

Next, we test the impact of node compression and SIMD-based search in
the performance of the \tree.
\Cref{fig:exp:ablation} presents, for two different datasets, the
throughput in queries per second (QPS) of  the \tree, after turning off
compression (NoComp), i.e., enforcing all nodes to be N64, and when
using traditional for-loop search at each node (NoSIMD) instead of SIMD
intrinsics. Observe that using compression and SIMD improves the
throughput of \tree in range queries, but SIMD brings
insignificant benefit to $k$NN search. This is because
%e reason for this is that in
for $k$NN queries much fewer nodes are accessed and the computational
bottleneck is mainly due to the management of heaps which is not
parallelizable. For the rest of the experiments, we use the default
version of our \tree, which employs both compression and SIMD-based
search, as they are always beneficial.

%\begin{figure*}
	%	\hspace{-0.5ex}
%	\includegraphics[width=\textwidth]{exp/ablation_study/abla_legend.pdf}\\
%	\begin{subfigure}{0.24\textwidth}
%		\centering
%		\includegraphics[width=\textwidth]{exp/ablation_study/test.pdf}
%		\caption{Uniform ($D=4$)}
		%		\label{fig:udpates_edges}
%	\end{subfigure}
	%	\hspace{-0.5ex}
%	\begin{subfigure}{0.24\textwidth}
%		\centering
%		\includegraphics[width=\textwidth]{exp/ablation_study/test2.pdf}
%		\caption{Gaussian ($D=4$)}
		%		\label{fig:udpates_toronto}
%	\end{subfigure}	
%	\begin{subfigure}{0.24\textwidth}
%		\centering
%		\includegraphics[width=\textwidth]{exp/scalability/scal_uni_dim_knn.pdf}
%		\caption{Uniform ($N=400$M)}
		%		\label{fig:udpates_edges}
%	\end{subfigure}
	%	\hspace{-0.5ex}
%	\begin{subfigure}{0.24\textwidth}
%		\centering
%		\includegraphics[width=\textwidth]{exp/scalability/scal_norm_dim_knn.pdf}
%		\caption{Gaussian ($N=400$M)}
		%		\label{fig:udpates_toronto}
%	\end{subfigure}
%	\caption{Scalability test using uniform and gaussian data \dimitris{knn queries}}
%	\label{fig:exp:ablation}
%\end{figure*}

\subsection{Construction Time and Memory Footprint}

We compare all tested methods in terms of
construction time and memory footprint, when bulk-loaded with each of the
real datasets.
%using the full size of each dataset.
As shown in Table \ref{tab:construction_time}, UG and AG have the
lowest construction costs.
This is expected, as grids only
require simple and fast arithmetic operations to locate the cell to
insert a given point.
After the grid-based indices, the best methods are \tree
and the R-tree. \tree is marginally faster than the R-tree
on half of the datasets, and marginally slower on GAIA,
NYT, and TPC-H, which are the datasets with higher
dimensionality.
This demonstrates the efficiency of \tree's
construction algorithm, having comparable cost to an optimized R-tree
bulk-loading method, despite being top-down and
more complex.
%not using simple bottom-up construction after
%sorting
%bulk-loading like the R-tree.
%Instead, \tree employs a top-down construction algorithm, which is
%usually slower in many cases.

% IFI is the next competitive index, as it exploits the R-tree’s
% bulk-loading algorithm; however, \tree's construction is 1.27x–1.63x
% faster. All other competitors do not use bulk-loading, so higher
% construction times are expected. The PH-tree and Flood take
% 1.36x–4.90x and  1.96x–3.79x, respectively, more 
% time than \tree to construct.
% Even though Flood is grid-based, it requires additional training time
% for its model.
% Finally, the kd-tree is the slowest one to  construct (2.98x–15.32x
% slower than the \tree), which is expected for a binary tree.

\begin{table}[]
	\caption{Construction cost (seconds) across all datasets}
	\label{tab:construction_time}
	\centering
	\vspace{-2mm}
	\resizebox{\columnwidth}{!}{%
		\begin{tabular}{ccccccccc}
			\toprule
			\textbf{Datasets} & \textbf{\tree} & \textbf{kd-tree} & \textbf{R-tree} & \textbf{PH-tree} & \textbf{AG} & \textbf{UG} & \textbf{Flood} & \textbf{IFI} \\         \midrule
			
			\textbf{EDGES} & 5.07 & 12.54 & 5.46 & 8.28 & 2.36 & 0.64 & 11.24 & 7.47 \\
			\textbf{TORONTO} & 10.52 & 44.3 & 10.99 & 23.33 & 3.69 & 1.33 & 20.59 & 14.67 \\
			\textbf{OOKLA} & 23.76 & 356.21 & 28.04 & 116.5 & 18.65 & 4.15 & 65.35 & 38.94 \\
			\textbf{GAIA} & 50.07 & 444.26 & 46.6 & 68.15 & 22.41 & 8.17 & 189.59 & 63.63 \\
			\textbf{NYT} & 11.55 & 138.8 & 10.02 & 31.74 & 6.06 & 2.7 & 33.12 & 14.79 \\
			\textbf{TPC-H} & 71.05 & 1059.66 & 67.84 & 272.08 & 61.52 & 28.27 & 217.01 & 101.52 \\ 
			\bottomrule
		\end{tabular}%
	}
\end{table}

\looseness=-1
Table \ref{tab:memory_footprint} reports the memory footprint of \tree
compared with competing index structures across all datasets.
As expected, the
lowest memory consumption is achieved by grid-based indices (UG and AG), as grids
are simple data structures and require memory close to that of the raw
data.
%In our experiments, the most memory-efficient indices are UG and
%AG, which are grid-based and also perform well in terms of
%construction time.
Flood is a learned index that is built on top of a grid structure and also
exhibits low memory consumption.
Compared to UG, AG, and Flood, \tree uses up to 30\% more memory.
When compared to the kd-tree, \tree incurs a moderate memory overhead ranging from 17.3\% on EDGES to 27\% on TPC-H.
%While
%\tree uses slightly less memory on the EDGES dataset (a reduction of
%3.1\%), it incurs moderate memory overhead on datasets with higher
%dimensionality, ranging from 4.5\% on TORONTO to 20.7\% on TPC-H.
This overhead is mainly due to the additional storage required to
maintain a bounding box for each leaf and metadata for the whole nodes of \tree.
Finally, \tree consistently outperforms the R-tree and the PH-tree in
terms of memory usage, reducing memory consumption by 36–39\% compared
to the R-tree and by 73–92\% compared to the PH-tree across all
datasets.
% Overall, with the exception of the  PH-tree, all indices do not have
% high memory requirements on top of the base data.

\begin{table}[]
	\caption{Memory footprint (GB) across all datasets}
	\label{tab:memory_footprint}
	\centering
	 \vspace{-2mm}
	\resizebox{\columnwidth}{!}{%
		\begin{tabular}{ccccccccc}
			\toprule
			\textbf{Datasets} & \textbf{\tree} & \textbf{kd-tree} & \textbf{R-tree} & \textbf{PH-tree} & \textbf{AG} & \textbf{UG} & \textbf{Flood} & \textbf{IFI} \\ \midrule
			\textbf{EDGES} & 0.95 & 0.81 & 1.55 & 11.84 & 0.77 & 0.77 & 0.77 & 2.65 \\
			\textbf{TORONTO} & 2.30 & 1.86 & 3.58 & 26.14 & 1.76 & 1.76 & 1.76 & 4.68 \\
			\textbf{OOKLA} & 5.81 & 4.75 & 9.39 & 33.29 & 4.62 & 4.62 & 4.62 & 9.21 \\
			\textbf{GAIA} & 11.93 & 9.90 & 19.69 & 58.54 & 9.68 & 9.68 & 9.68 & 17.71 \\
			\textbf{NYT} & 3.27 & 2.71 & 5.40 & 13.45 & 2.65 & 2.65 & 2.65 & 4.30 \\
			\textbf{TPC-H} & 18.54 & 14.60 & 29.22 & 68.67 & 14.35 & 14.35 & 14.35 & 23.30 \\
			\bottomrule
		\end{tabular}%
	}
\end{table}

\subsection{Range queries}\label{sec:exp:range}

\begin{figure*}
	\includegraphics[width=0.8\textwidth]{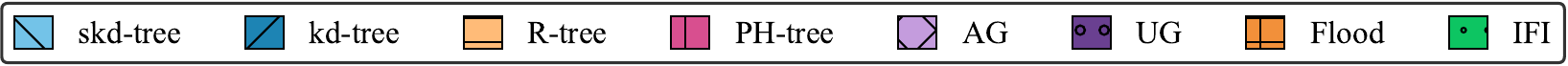}
	\begin{subfigure}{0.32\textwidth}
		\centering
		\includegraphics[width=\textwidth]{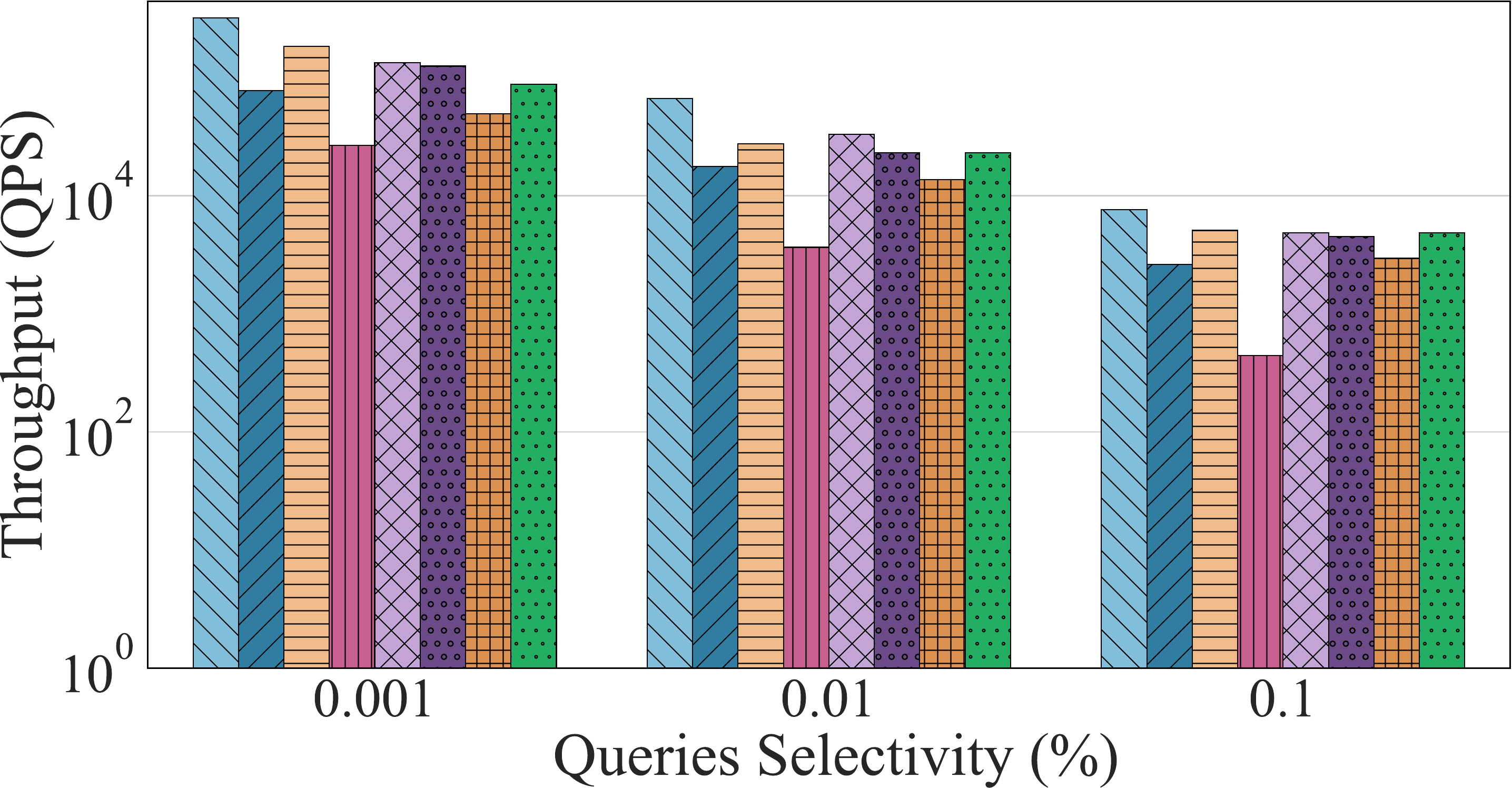}
		\caption{EDGES ($D=2$)}
		\label{fig:range_edges}
	\end{subfigure}
	\hspace{-0.5ex}
	\begin{subfigure}{0.32\textwidth}
		\centering
		\includegraphics[width=\textwidth]{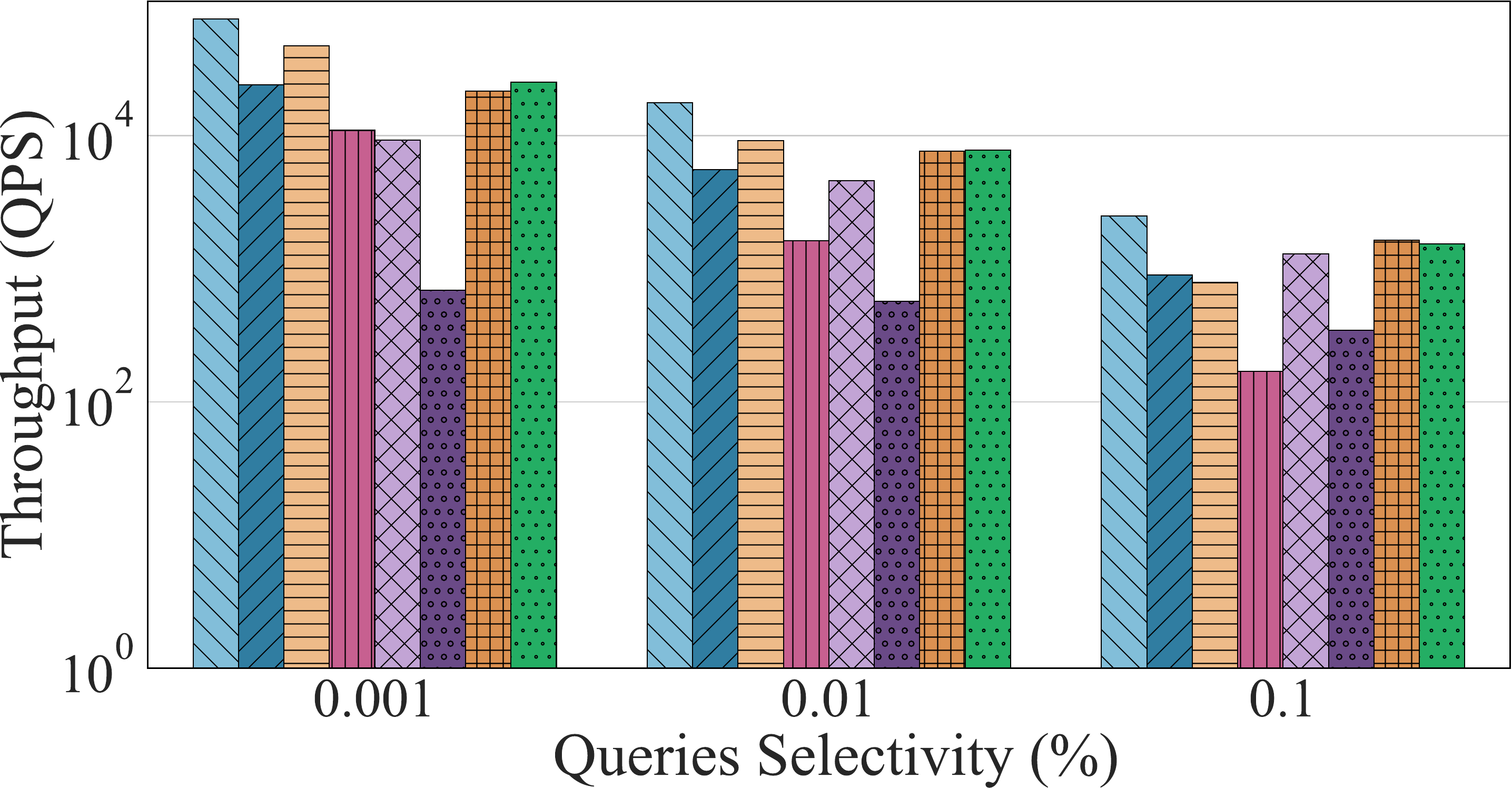}
		\caption{TORONTO ($D=3$)}
		\label{fig:range_toronto}
	\end{subfigure}
	\hspace{-0.5ex}
	\begin{subfigure}{0.32\textwidth}
		\centering
		\includegraphics[width=\textwidth]{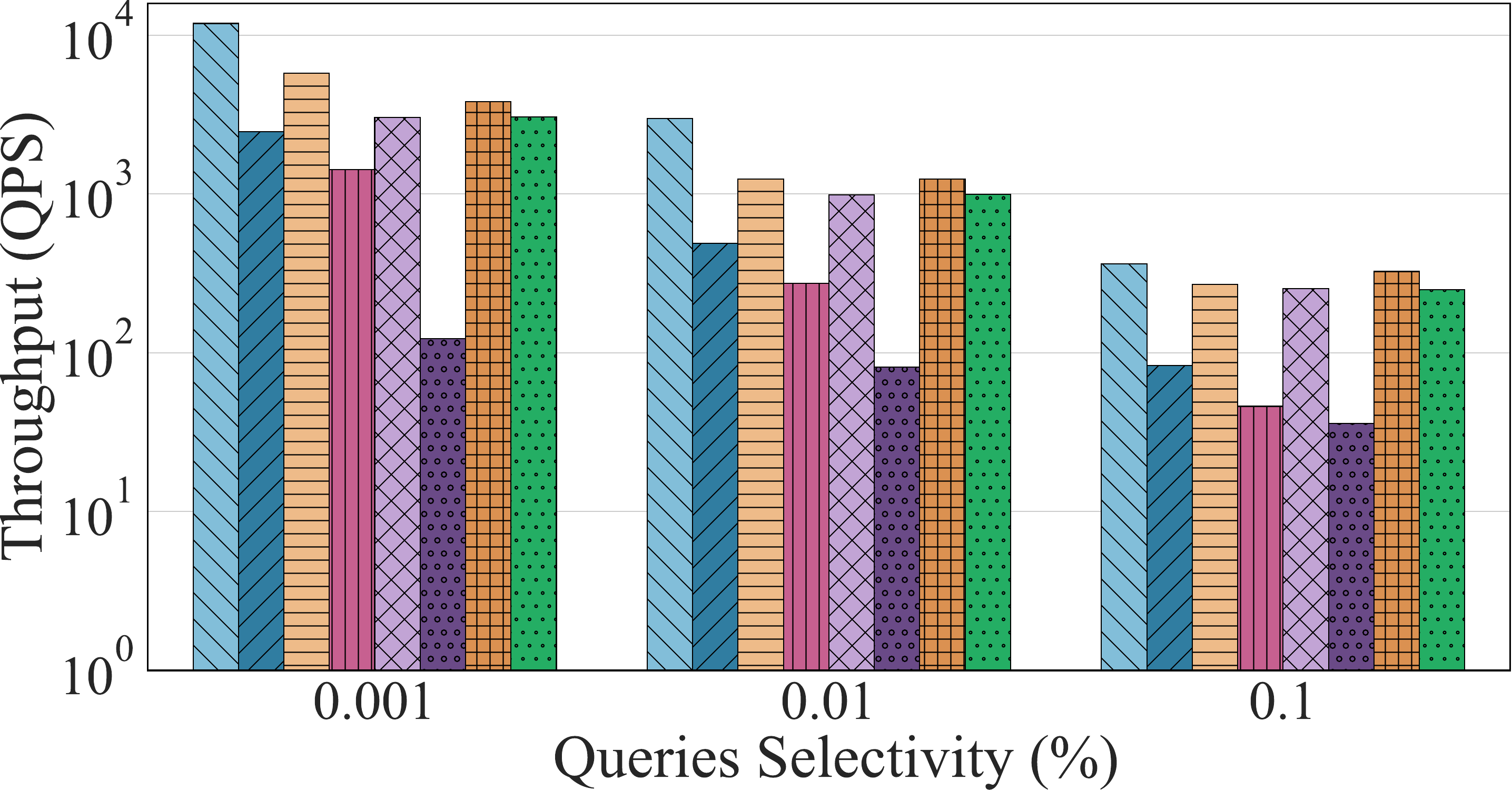}
		\caption{OOKLA ($D=5$)}
		\label{fig:range_ookla}
	\end{subfigure}
	\hspace{-0.5ex}
	\begin{subfigure}{0.32\textwidth}
		\centering
		\includegraphics[width=\textwidth]{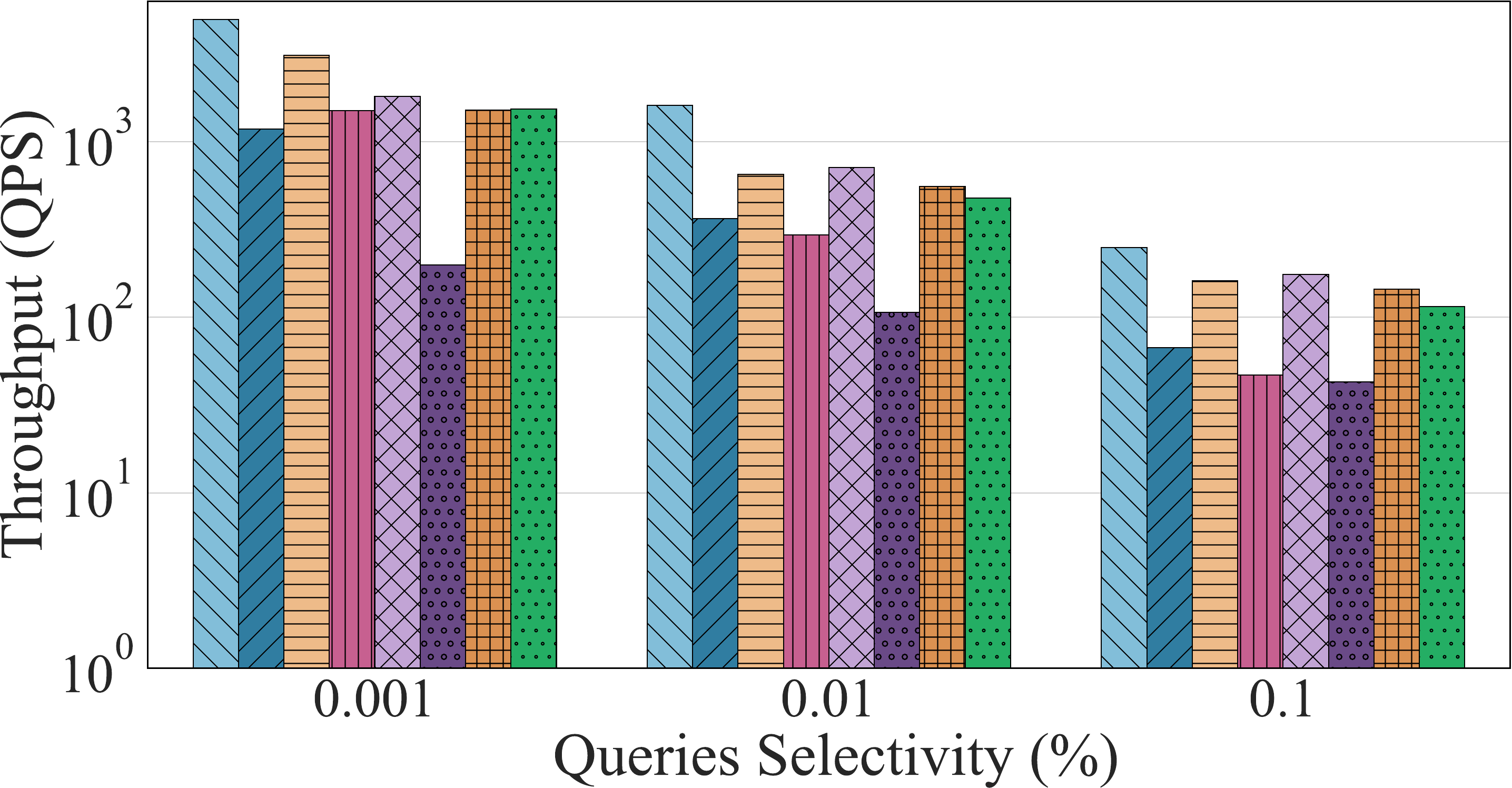}
		\caption{GAIA ($D=6$)}
		\label{fig:range_gaia}
	\end{subfigure}
	\hspace{-0.5ex}
	\begin{subfigure}{0.32\textwidth}
		\centering
		\includegraphics[width=\textwidth]{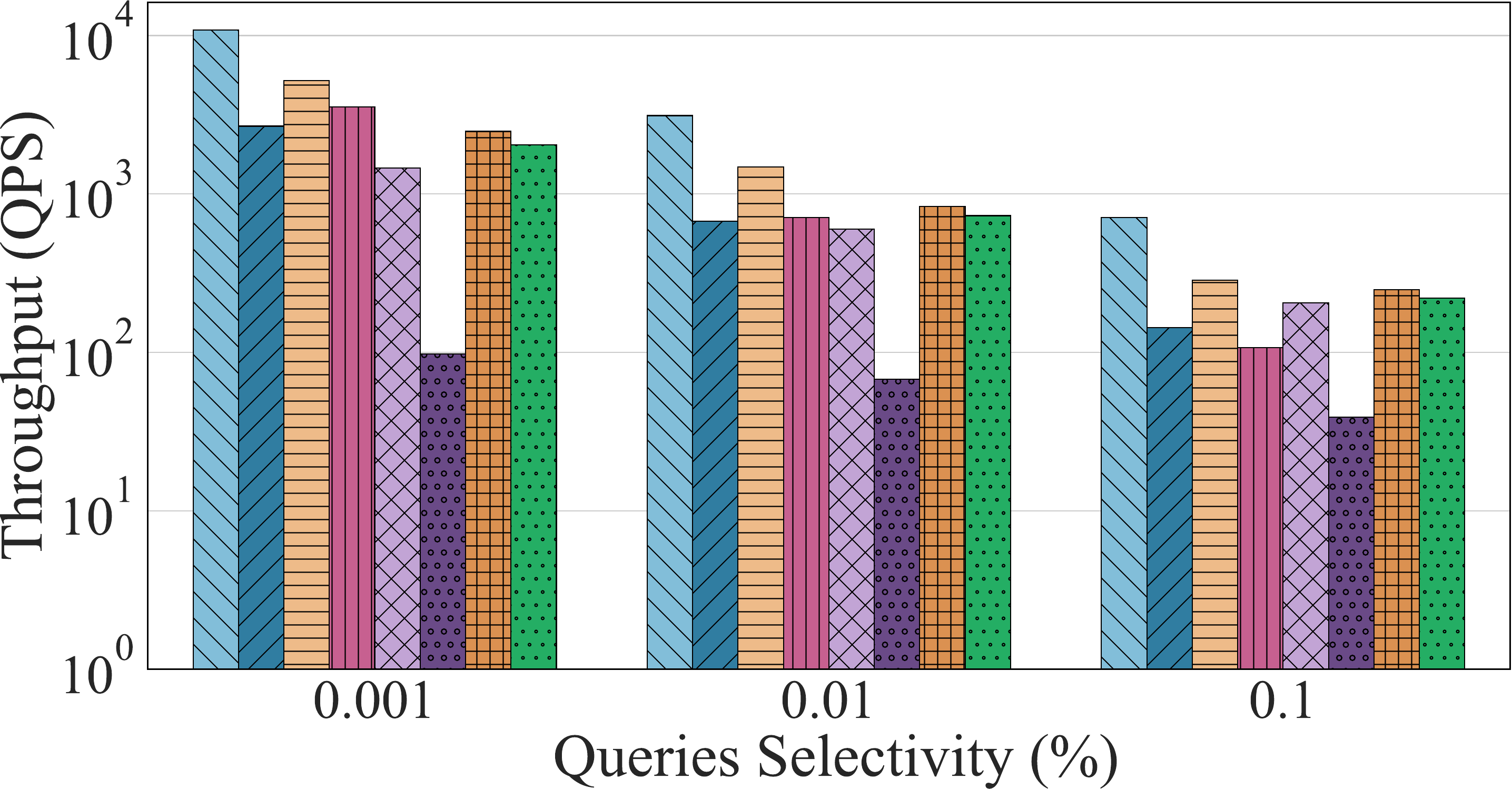}
		\caption{NYT ($D=8$)}
		\label{fig:range_nyt}
	\end{subfigure}
	\hspace{-0.5ex}
	\begin{subfigure}{0.32\textwidth}
		\centering
		\includegraphics[width=\textwidth]{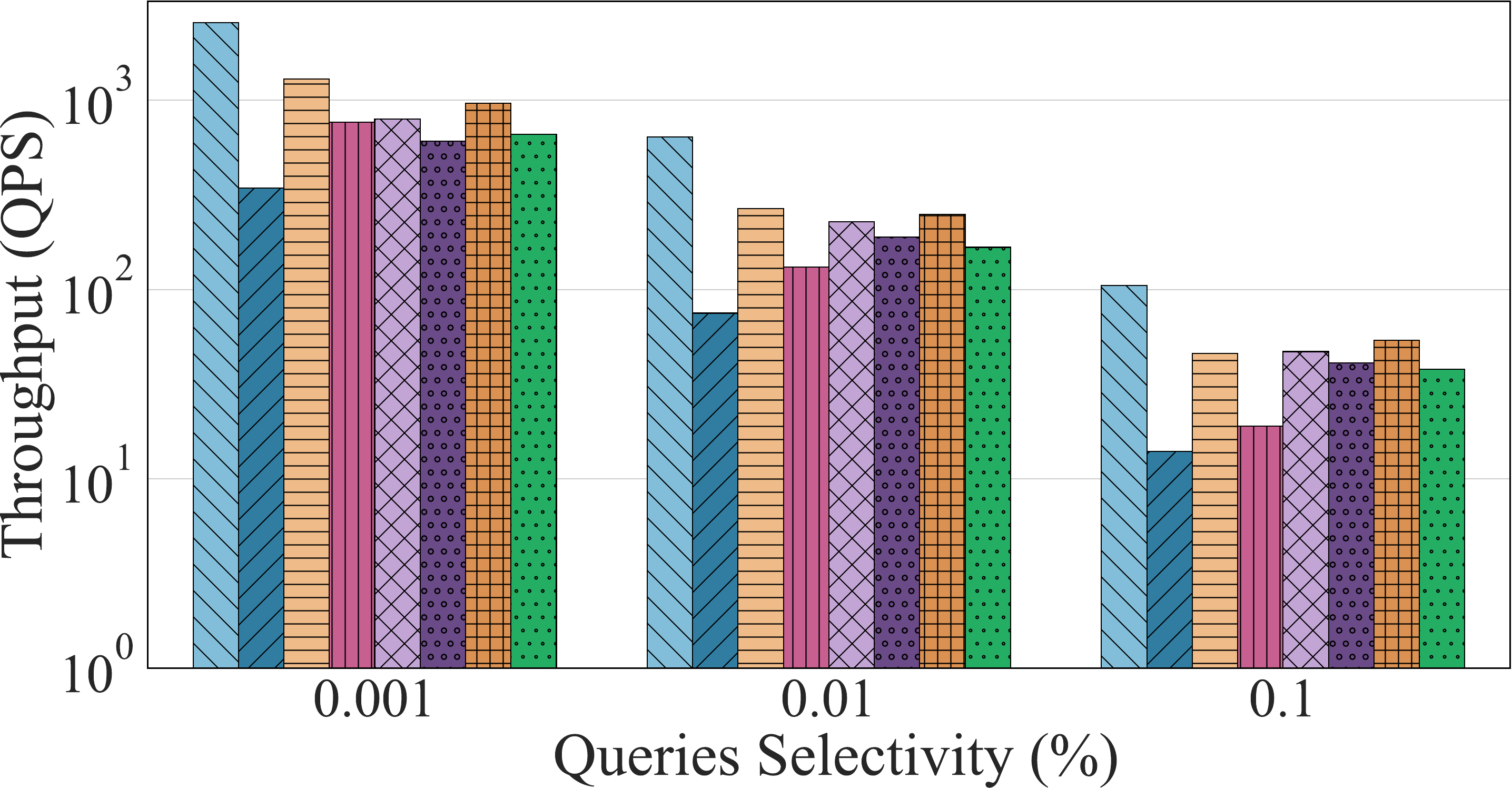}
		\caption{TPC-H ($D=8$)}
		\label{fig:range_tpch}
	\end{subfigure}
	\vspace{-2mm}
	\caption{Throughput in queries per second (QPS) of \tree vs. competitor
		indices}
	\label{fig:range_expr}
\end{figure*}

In this section, we compare all tested methods with respect to their
range query throughput. As shown in Figure \ref{fig:range_expr}, \tree exhibits remarkable robustness in range query performance, which is not affected by the data dimensionality and the
query selectivity.
%\dimitris{The description of the workloads has been moved to the Setup section.}
%For each dataset, we execute 1000 hypercube queries
%that
%retrieve 0.001\%, 0.01\% and 0.1\% of the indexed points. We exclude
%queries of  selectivity 1\% or above, as they produce hundreds
%of thousands of results on our datasets, which does not correspond to
%a realistic search scenario.

%as such queries are not realistic,
%producing hundreds of thousands of results.
%because for all datasets these settings produce results ranging from
%hundreds to tens of thousands of records, which are representative of
%realistic scenarios.
%Figure \ref{fig:range_expr} presents the throughput (queries per
%second) of all methods for range queries.
\looseness=-1
The best relative performance is observed on EDGES
(Fig. \ref{fig:range_edges}),
where the \tree best exploits compression, using only N16 inner nodes.
In most of the experiments,
the runner up method is the optimized boost R-tree,
which is 1.6x-3x times slower than the \tree.
Across all datasets, the performance of competing indexes varies notably with dimensionality and selectivity. UG exhibits acceptable throughput only on very low-dimensional data (EDGES) but its performance degrades sharply as dimensionality increases, making it unsuitable for higher-dimensional datasets.
AG has better relative performance than UG,
which drops with the data dimensionality.
%and has moderate performance as we increase the dimension of the
%datasets.
Flood, kd-tree, and IFI have  stable performance across all
datasets with small  fluctuations.
PH-tree shows limited throughput on low-dimensional datasets, with performance improving slightly as the number of dimensions increases.
% Overall, across all datasets and selectivity levels, \tree consistently delivers the highest throughput, demonstrating robust efficiency.
In TORONTO and OOKLA, for less selective queries (0.1\%),
the learned indices (Flood and IFI)
outperform the R-tree; this is due to the fact that their predictions
at the leaf nodes are more accurate compared to explicit search.
However, in most cases, learned indices underperform; adding this to the
facts that (i) to update them requires expensive model re-training,
(ii) they are not appropriate for $k$NN search, and (iii) they
outperform the R-tree only when the queries compute tens of thousands
of results or more, hints against using multidimensional learned indices.

Table \ref{tab:perf_counters_range}  presents hardware performance counters for range queries. \tree achieves the lowest values in Instructions, Cycles, and Branch Misses, highlighting its superior performance and the effectiveness of the SIMD-based search strategy.
Additionally, \tree exhibits the lowest number of L1 cache misses, demonstrating the efficiency of its memory access patterns and overall structure.

\begin{table}[thb]
	\centering
	\caption{Hardware performance counters per range query}
	\label{tab:perf_counters_range}
	\centering
	\vspace{-2mm}
	\resizebox{\columnwidth}{!}{%
		\begin{tabular}{lcccccccc}
			\toprule
			\multicolumn{9}{c}{\textbf{EDGES ($D=2$) - Queries Selectivity: 0.01\%}} \\
			\midrule \textbf{Events} & \textbf{\tree} & \textbf{kd-tree} & \textbf{R-tree} & \textbf{PH-tree} & \textbf{AG} & \textbf{UG} & \textbf{Flood} & \textbf{IFI} \\
			\midrule
			\textbf{Instr.} & 110K  & 361K  & 257K  & 756K  & 233K  & 308K  & 378K  & 360K \\
			\textbf{Cycles} & 95K   & 390K  & 200K  & 1.38M & 170K  & 233K  & 382K  & 237K \\
			\textbf{L1 Misses} & 6.8K  & 11.9K & 7.4K  & 28K   & 7.1K  & 7.5K  & 11K   & 7.7K \\
			\textbf{Br. Misses} & 614   & 1.1K  & 1.12K & 6.4K  & 1.4K  & 1.5K  & 1K    & 2.5K \\
			\bottomrule
		\end{tabular}
	}
	
	\vspace{1mm}
	\resizebox{\columnwidth}{!}{%
		
		\begin{tabular}{lcccccccc}
			\toprule
			\multicolumn{9}{c}{\textbf{GAIA ($D=6$) - Queries Selectivity: 0.01\%}} \\
			\midrule
			\textbf{Events} & \textbf{\tree} & \textbf{kd-tree} & \textbf{R-tree} & \textbf{PH-tree} & \textbf{AG} & \textbf{UG} & \textbf{Flood} & \textbf{IFI} \\
			\midrule
			\textbf{Instr.} & 2.57M & 4.15M & 7.31M & 7.94M & 9.04M & 71.5M & 6.69M & 11.5M \\
			\textbf{Cycles} & 3.08M & 14.9M & 7.74M & 17M   & 6.99M & 46.3M & 8.84M & 10.4M \\
			\textbf{L1 Misses} & 152K  & 362K  & 225K  & 337K  & 343K  & 2.77M & 302K  & 298K \\
			\textbf{Br. Misses} & 21K   & 49K   & 75K   & 66.6K & 43.8K & 206K  & 141K  & 132K \\
			\bottomrule
		\end{tabular}
	}
\end{table}

\subsection{$k$NN queries}

\begin{figure*}
	\includegraphics[width=0.44\linewidth]{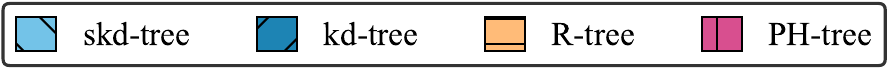}\\
	\begin{subfigure}{0.32\linewidth}
		\centering
		\includegraphics[width=\linewidth]{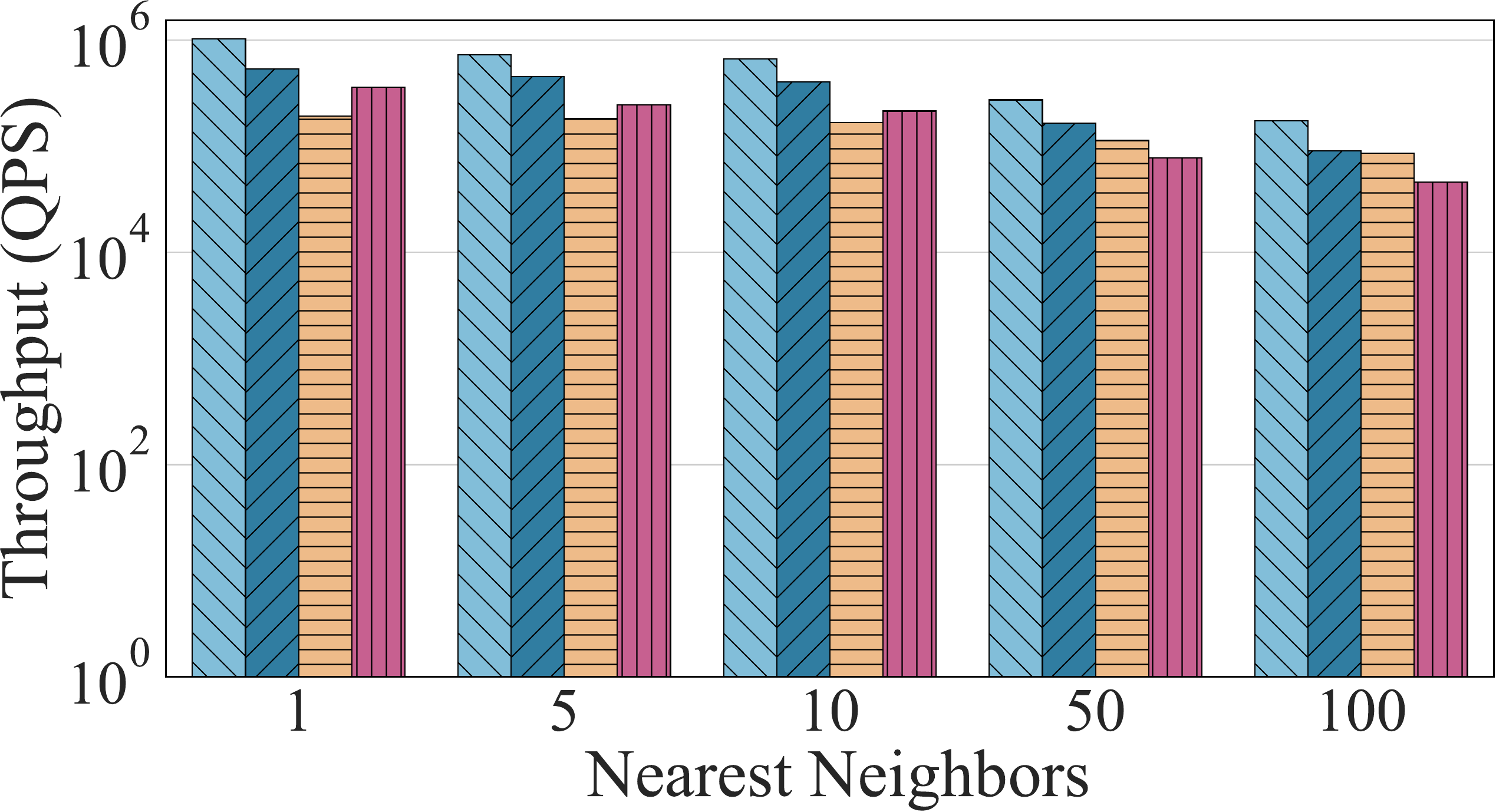}
		\caption{EDGES ($D=2$)}
		\label{fig:knn_edges}
	\end{subfigure}
	\hspace{-0.5ex}
	\begin{subfigure}{0.32\linewidth}
		\centering
		\includegraphics[width=\linewidth]{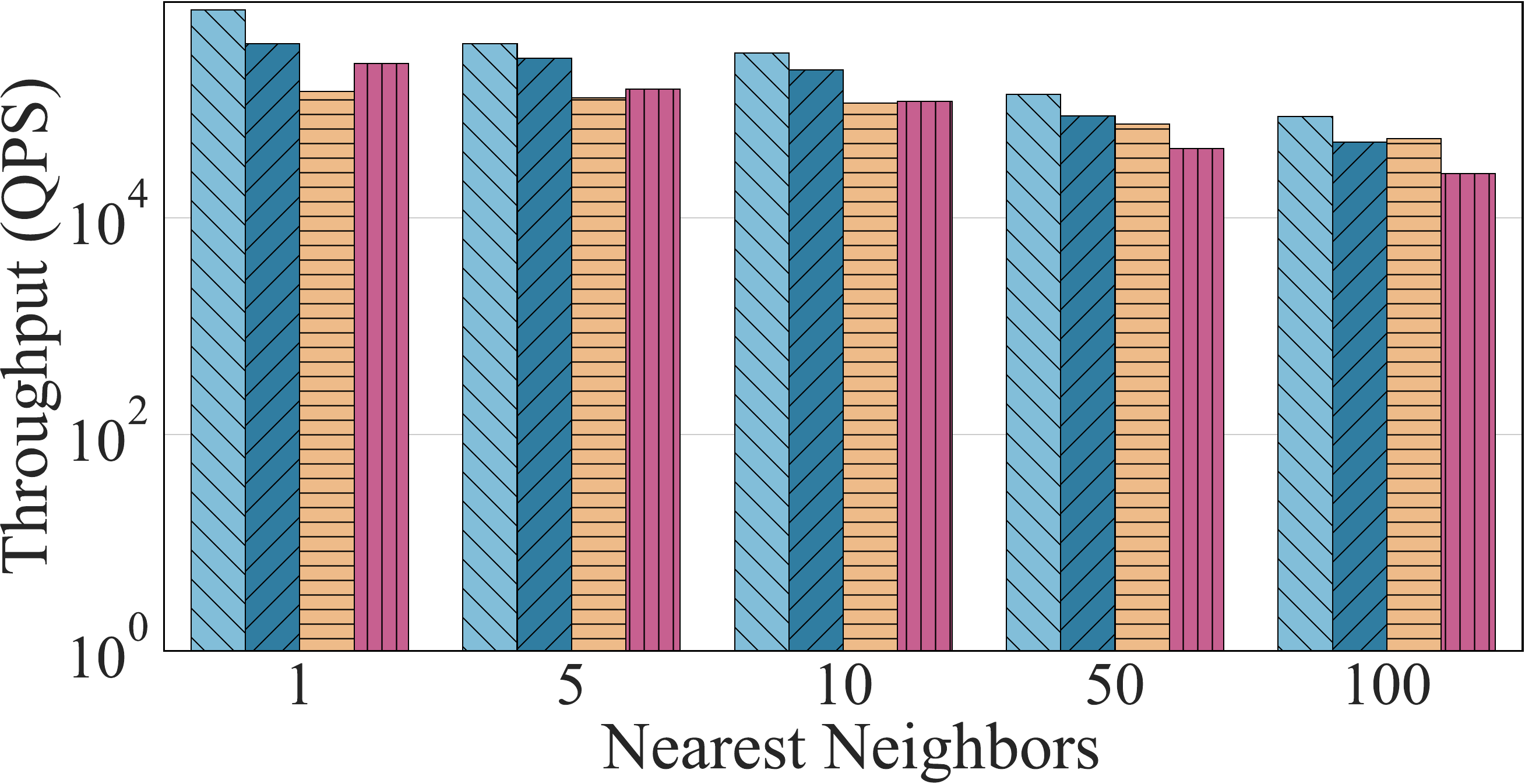}
		\caption{TORONTO ($D=3$)}
		\label{fig:knn_toront}
	\end{subfigure}
	\hspace{-0.5ex}
	\begin{subfigure}{0.32\linewidth}
		\centering
		\includegraphics[width=\linewidth]{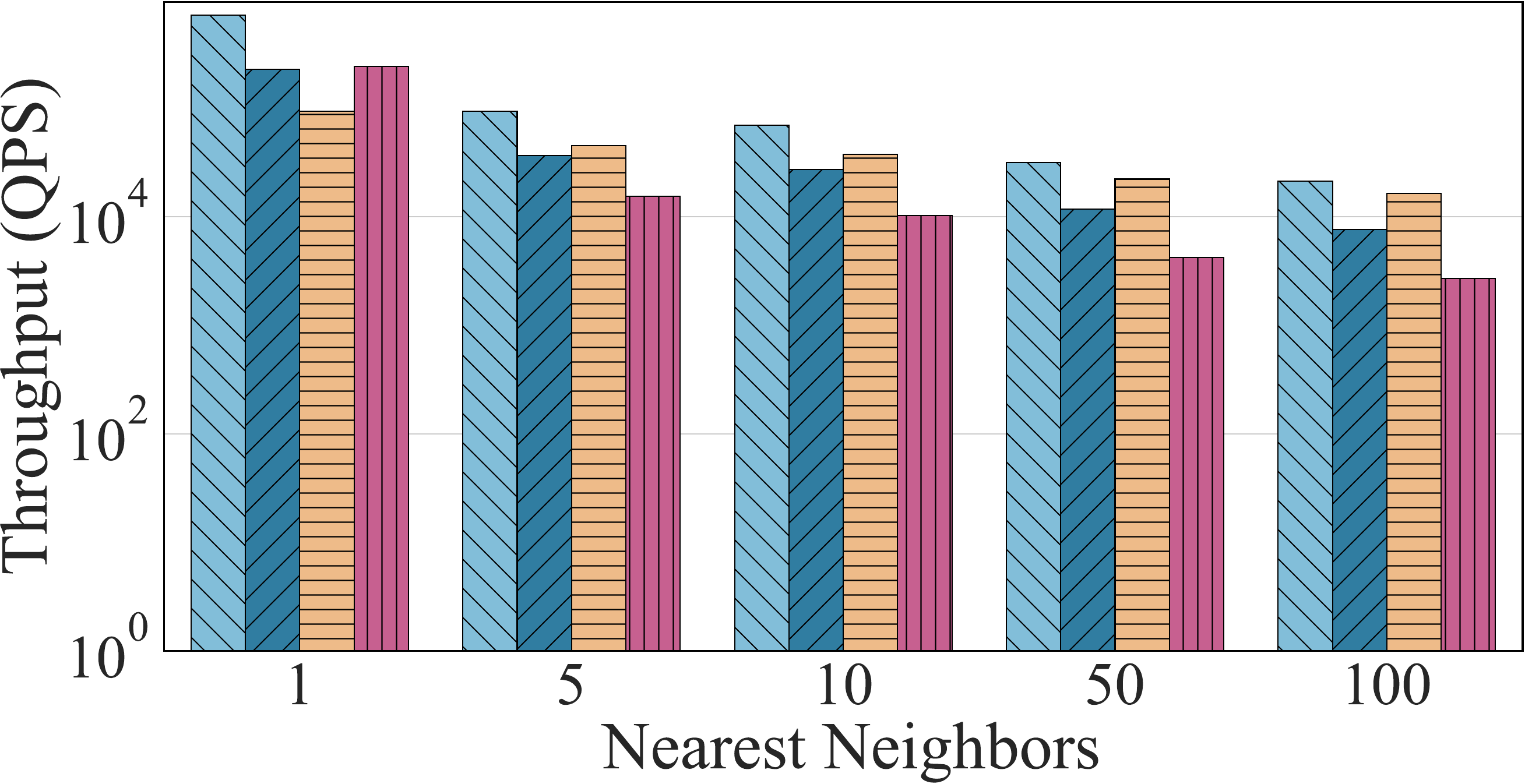}
		\caption{OOKLA ($D=5$)}
		\label{fig:knn_ookla}
	\end{subfigure}
	
	\hspace{-0.5ex}
	\begin{subfigure}{0.32\linewidth}
		\centering
		\includegraphics[width=\linewidth]{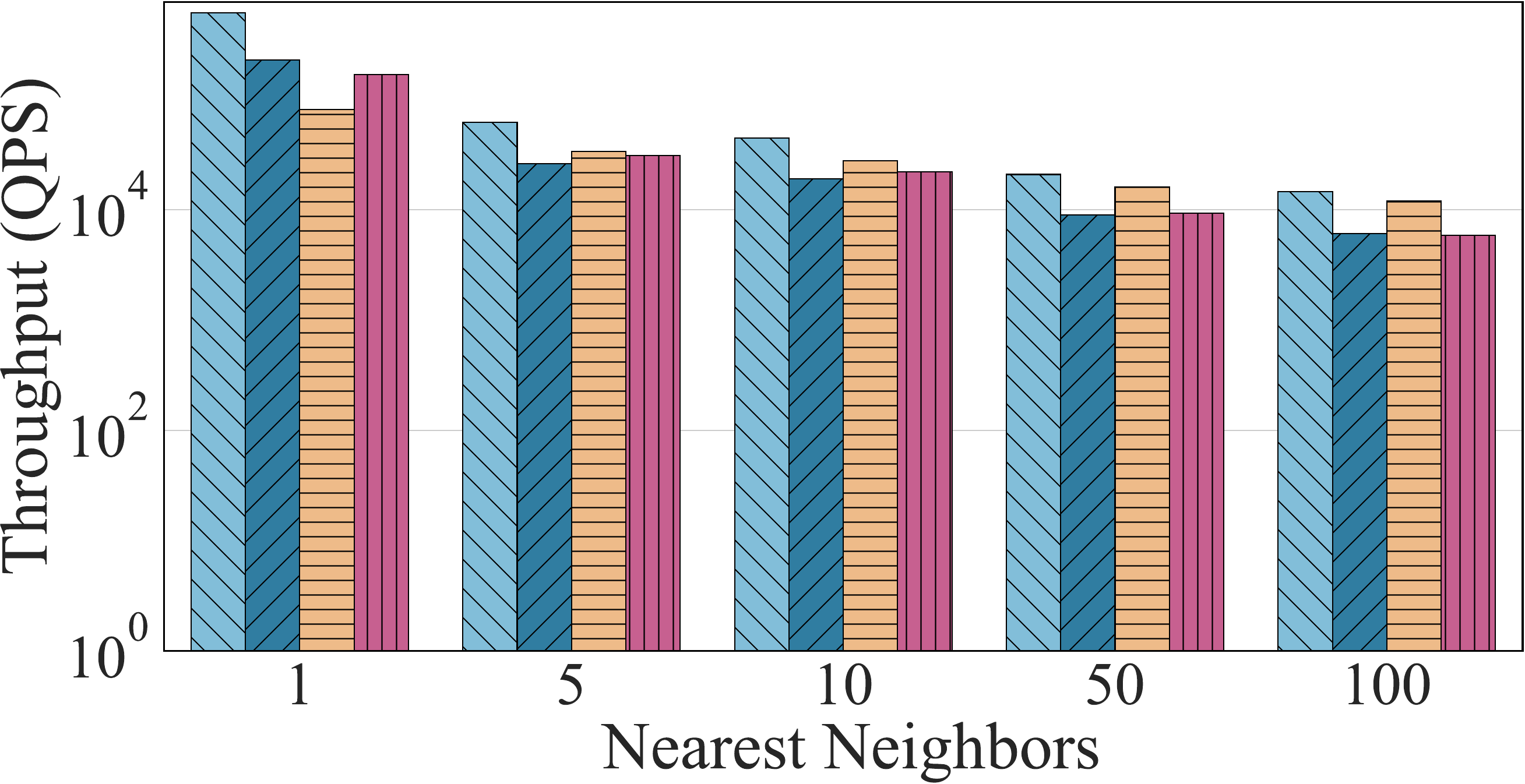}
		\caption{GAIA ($D=6$)}
		\label{fig:knn_gaia}
	\end{subfigure}
	\hspace{-0.5ex}
	\begin{subfigure}{0.32\linewidth}
		\centering
		\includegraphics[width=\linewidth]{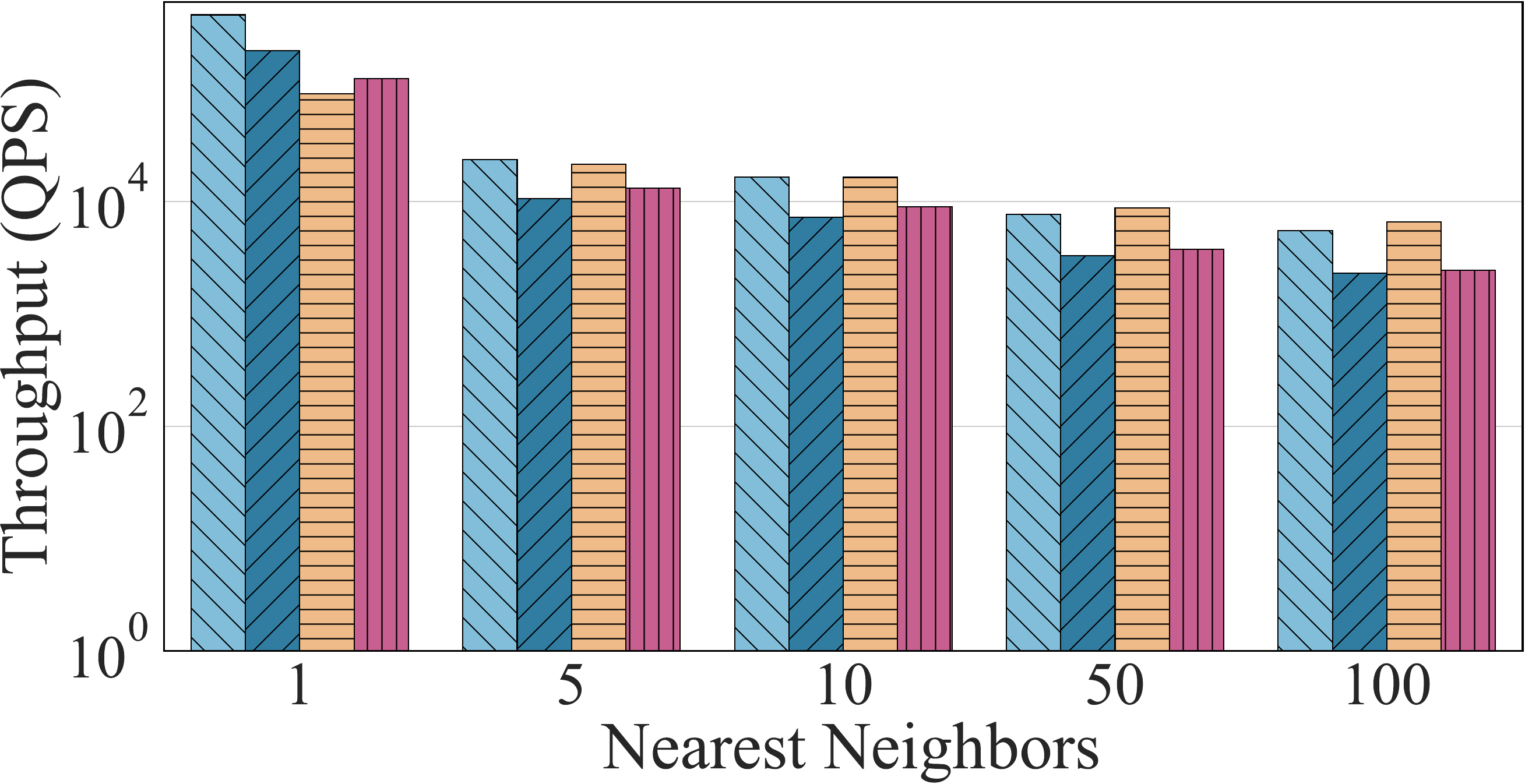}
		\caption{NYT  ($D=8$)}
		\label{fig:knn_nyt}
	\end{subfigure}
	\hspace{-0.5ex}
	\begin{subfigure}{0.32\linewidth}
		\centering
		\includegraphics[width=\linewidth]{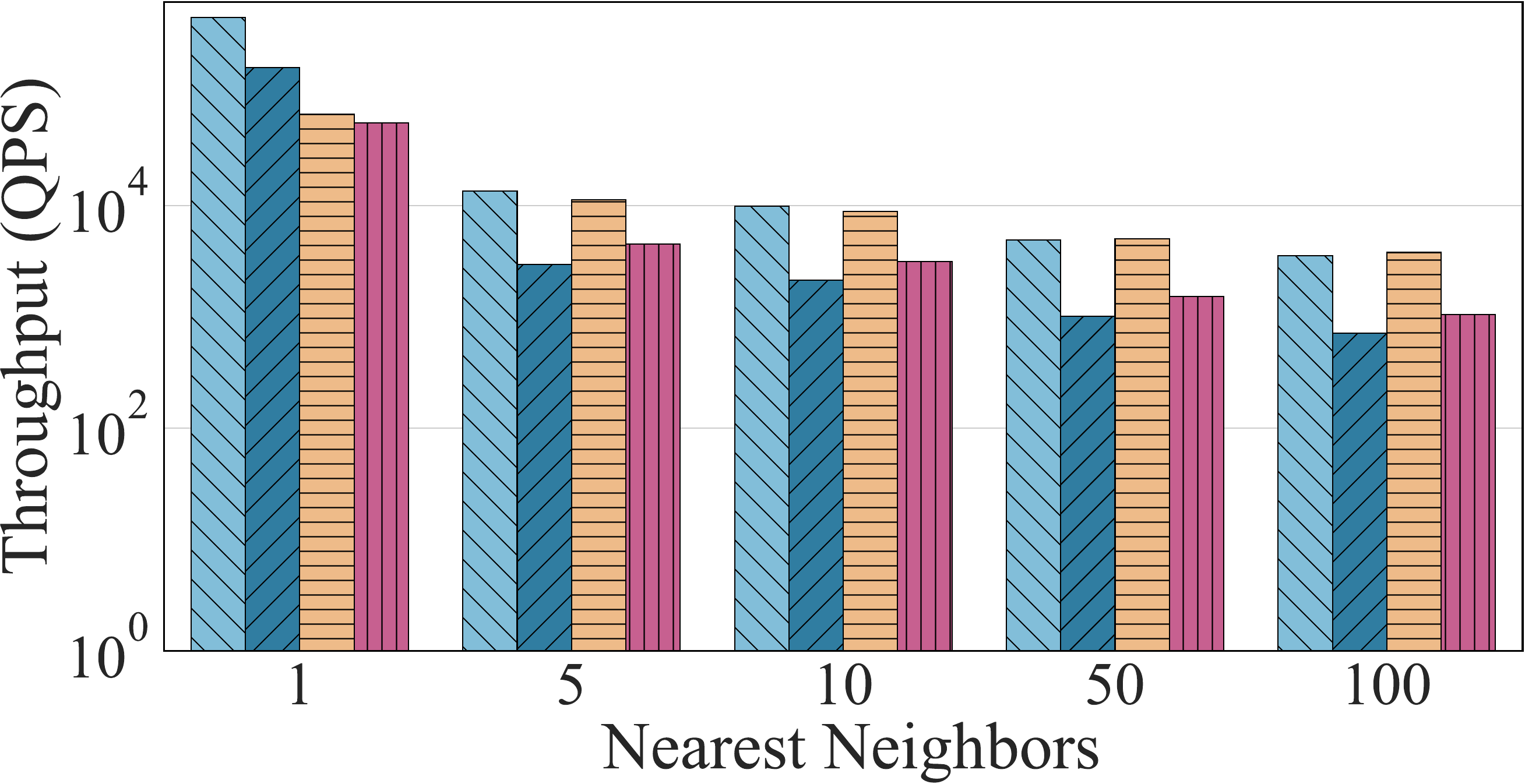}
		\caption{TPC-H ($D=8$)}
		\label{fig:knn_tpch}
	\end{subfigure}
	\vspace{-2mm}
	\caption{Performance comparison of \tree with kd-tree, R-tree, and PH-tree for $k$NN queries}
	\label{fig:knn_expr}
\end{figure*}

This section compares \tree against kd-tree, R-tree, and
PH-tree, with respect to their
$k$NN query throughput. These are the only competitors in our analysis that
support NN search.
%\dimitris{The description of the workloads has been moved to the Setup section.}
%At each experimental instance, throughput is computed after executing 1000 random queries.
%operations. Specifically,
%For each query we select a random point from the dataset and use it as
%the query point.
%We then perform $k$NN searches, varying the number $k$ of nearest
%neighbors.
%Figure \ref{fig:knn_expr} shows the performance gains of each index for different values of $k$, with $k\in \{1,5,10,50,100\}$.
As shown in Figure \ref{fig:knn_expr}, \tree consistently outperforms all competitors, except for a few
cases in the NYT dataset, where it marginally loses to the
R-tree. For the great majority of queries, the R-tree is the runner up;
however, the kd-tree outperforms the R-tree on datasets of low
dimensionality and when $k=1$. 
As we already saw from the previous experiment, the kd-tree is not
competitive in range queries, being a few times slower than our \tree.
The PH-tree is the worst performing method in almost all cases.
Overall, \tree consistently excels for a variety of data distributions
and dimensionalities and across all values of $k$.

In Table \ref{tab:perf_counters_knn}, we present the hardware performance counters for $k$NN queries. Overall, we observe a similar trend to that of range queries. However, the number of branch misses for \tree is slightly higher than for the kd-tree. This is expected, as \tree requires additional conditions to determine whether a popped heap entry corresponds to a leaf, an internal node, or a group of nodes during the $k$NN search (see Algorithm \ref{alg:simd-knn}, lines 10, 16, and 22).
Despite this, \tree maintains competitive performance across the remaining metrics, indicating that the additional comparisons introduce only a minor overhead and do not offset the benefits of its overall design.

\begin{table}[thb]
	\centering
	\caption{Hardware performance counters  per $k$NN query}
	\label{tab:perf_counters_knn}
	\centering
	\vspace{-2mm}
	\resizebox{\columnwidth}{!}{%
		\begin{tabular}{lcccccccc}
			\toprule
			&\multicolumn{4}{c}{\textbf{\begin{tabular}[c]{@{}c@{}}EDGES ($D=2$) - 10NN\end{tabular}}} 
			& \multicolumn{4}{c}{\textbf{\begin{tabular}[c]{@{}c@{}} GAIA ($D=6$) - 10NN\end{tabular}}} \\
			\cmidrule(lr){2-5} \cmidrule(lr){6-9}
			\textbf{Events}
			&\textbf{\tree} & \textbf{kd-tree} & \textbf{R-tree} & \textbf{PH-tree}
			& \textbf{\tree} & \textbf{kd-tree} & \textbf{R-tree} & \textbf{PH-tree} \\
			\midrule
			
			\textbf{Instr.} & 30.5K & 26.6K & 73.4K & 41.7K
			& 139.8K & 94.5K & 218.5K & 273.9K \\
			
			\textbf{Cycles} & 22.1K & 27.5K & 48.1K & 40.5K
			& 126.4K & 278.7K & 194.6K & 238.7K \\
			\textbf{L1 Misses} & 171   & 302   & 698   & 265
			& 3.42K  & 5.67K & 4.86K  & 3.55K \\
			
			\textbf{Br. Misses} & 130   & 106   & 451   & 239
			& 375    & 255   & 2.71K  & 1.59K \\
			
			\bottomrule
		\end{tabular}
	}
\end{table}

\subsection{Updates}
\looseness=-1

\begin{figure*}[thb]
	\includegraphics[width=0.34\linewidth]{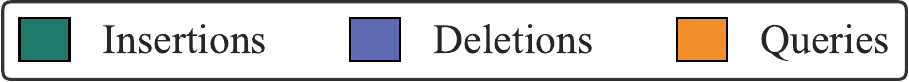}\\
	\begin{subfigure}{0.32\linewidth}
		\centering
		\includegraphics[width=\linewidth]{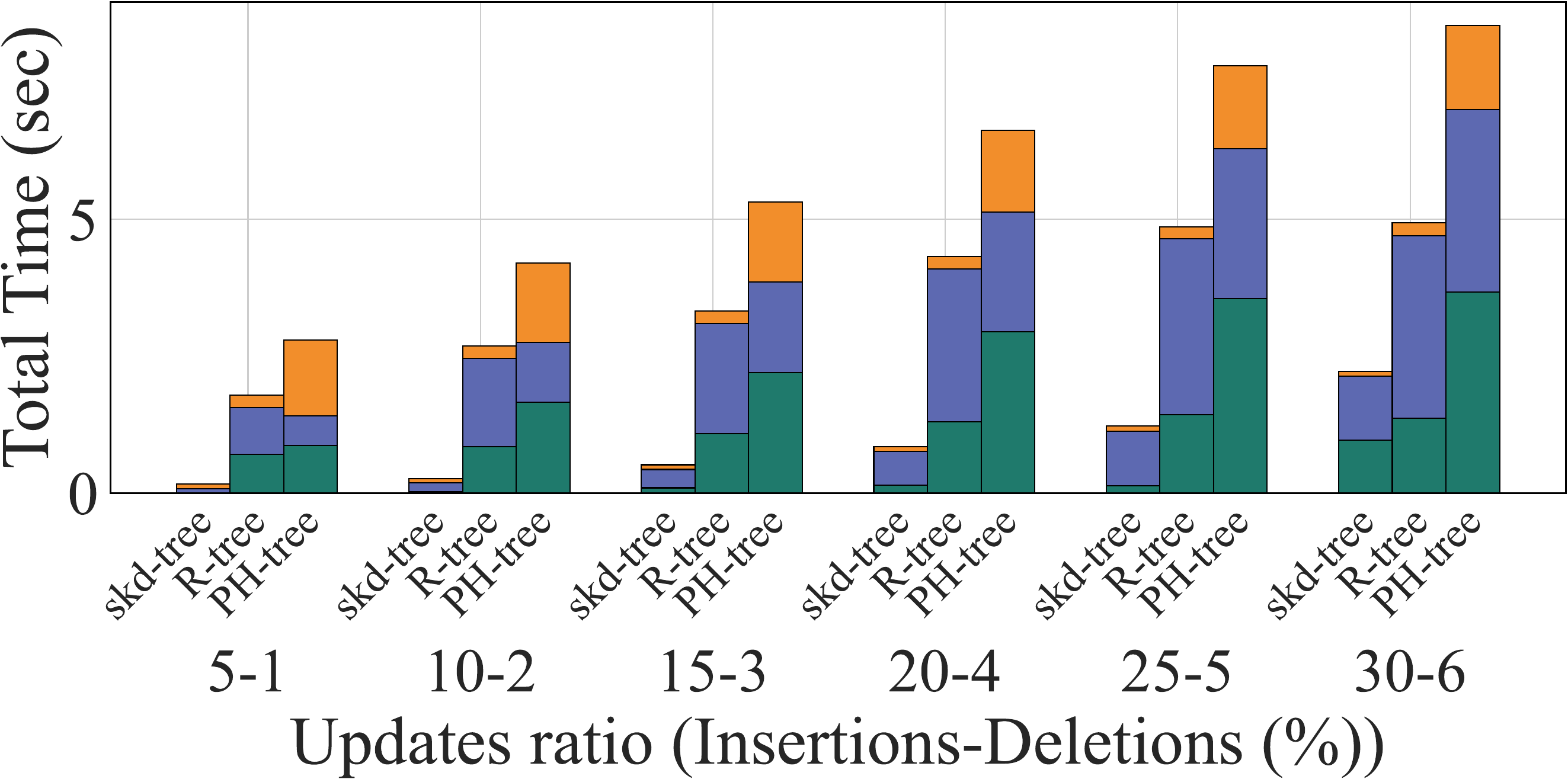}
		\caption{EDGES ($D=2$)}
		\label{fig:udpates_edges}
	\end{subfigure}
	\hspace{-0.5ex}
	\begin{subfigure}{0.32\linewidth}
		\centering
		\includegraphics[width=\linewidth]{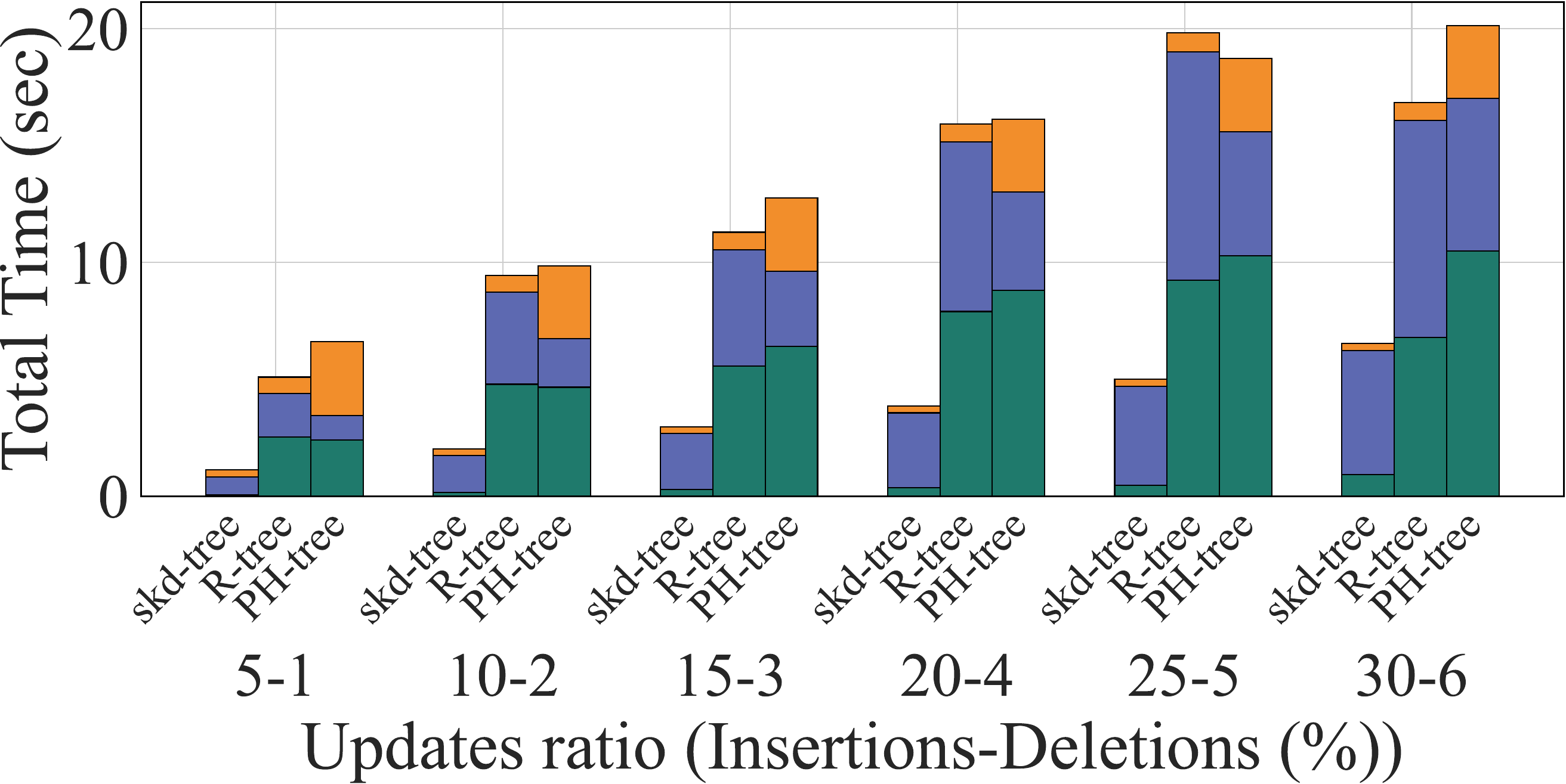}
		\caption{TORONTO ($D=3$)}
		\label{fig:udpates_toronto}
	\end{subfigure}
	\hspace{-0.5ex}
	\begin{subfigure}{0.32\linewidth}
		\centering
		\includegraphics[width=\linewidth]{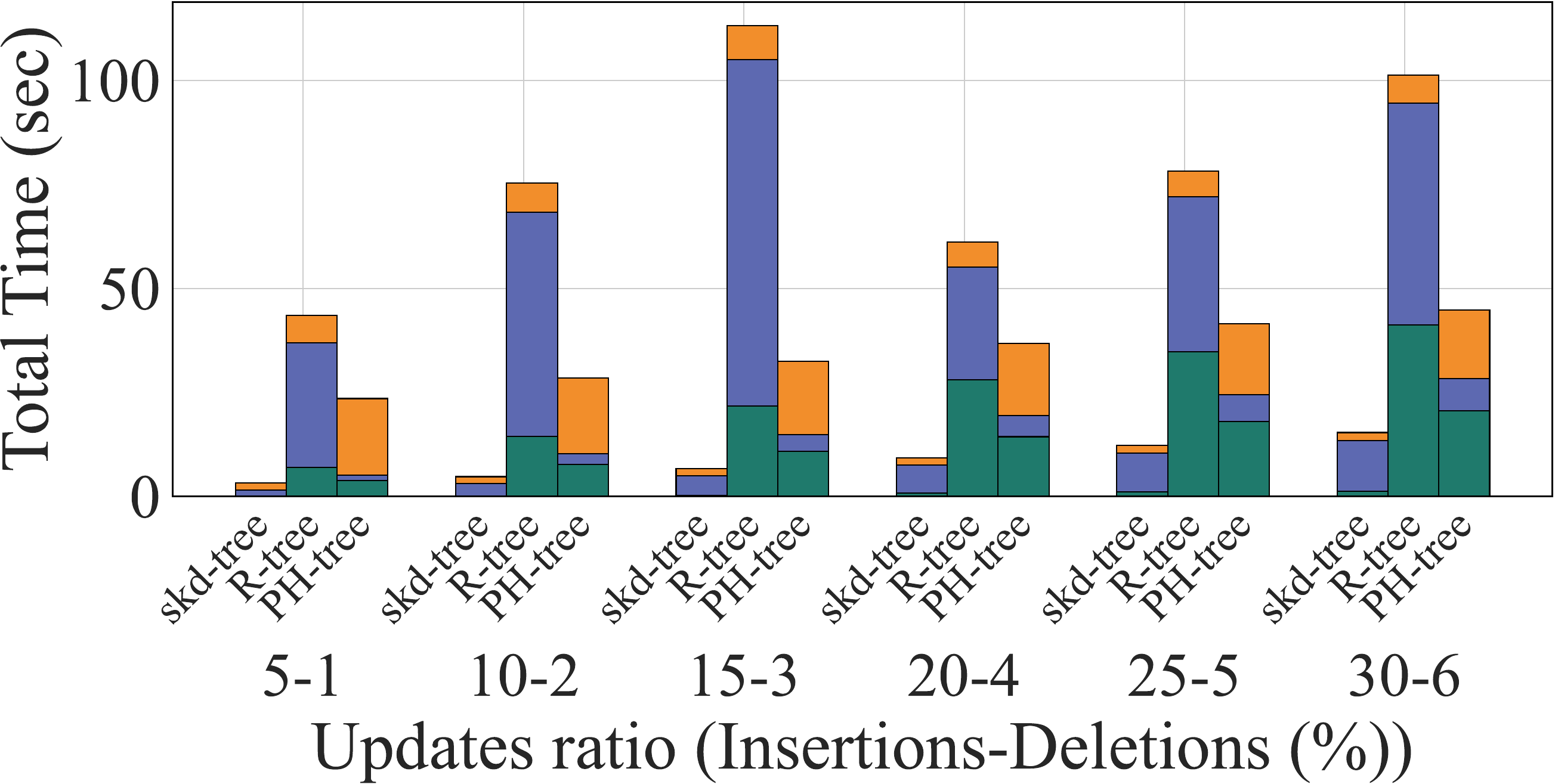}
		\caption{OOKLA ($D=5$)}
		\label{fig:udpates_ookla}	
	\end{subfigure}
	
	\hspace{-0.5ex}
	\begin{subfigure}{0.32\linewidth}
		\centering
		\includegraphics[width=\linewidth]{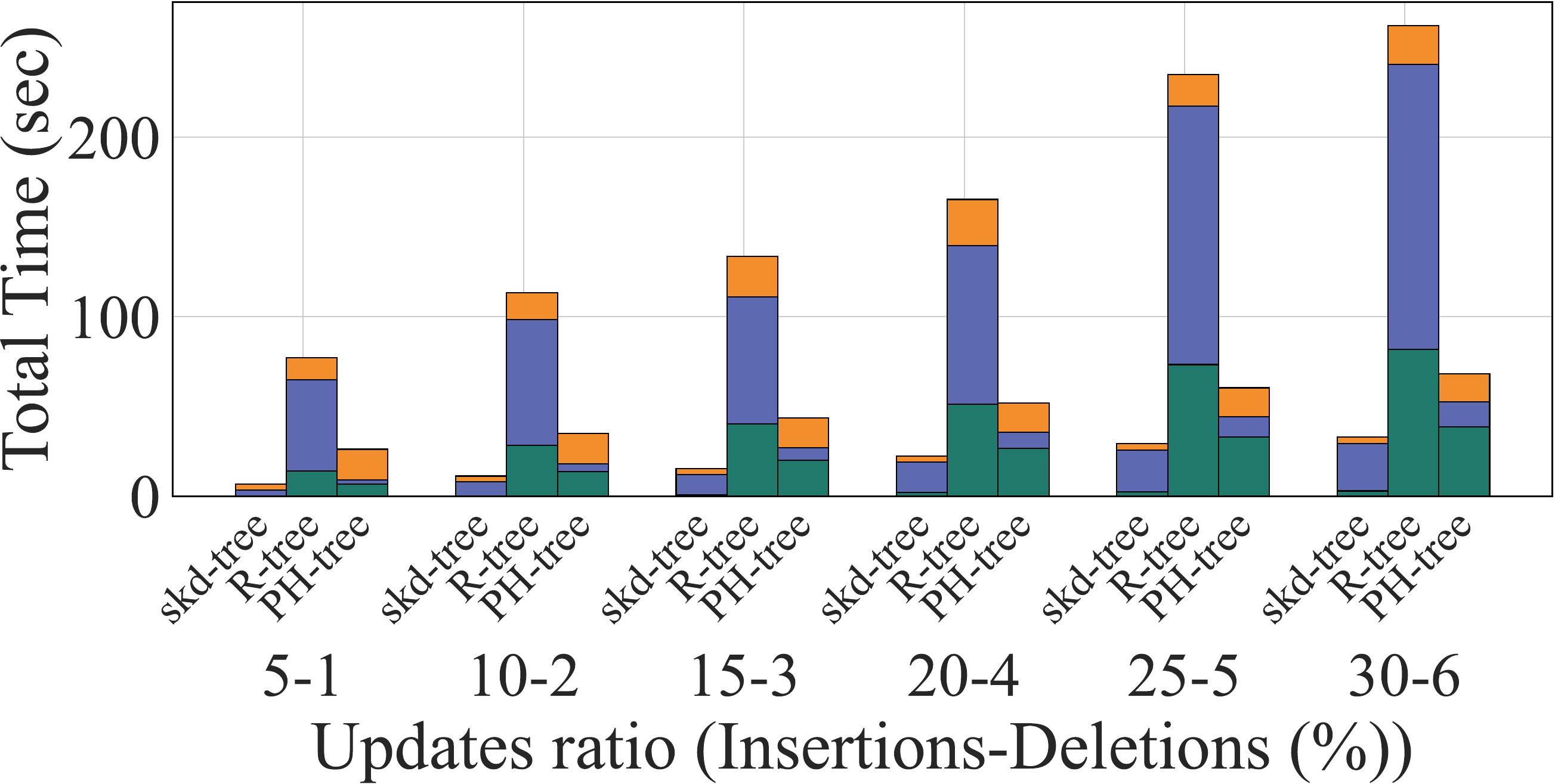}
		\caption{GAIA ($D=6$)}
		\label{fig:udpates_gaia}
	\end{subfigure}
	\hspace{-0.5ex}
	\begin{subfigure}{0.32\linewidth}
		\centering
		\includegraphics[width=\linewidth]{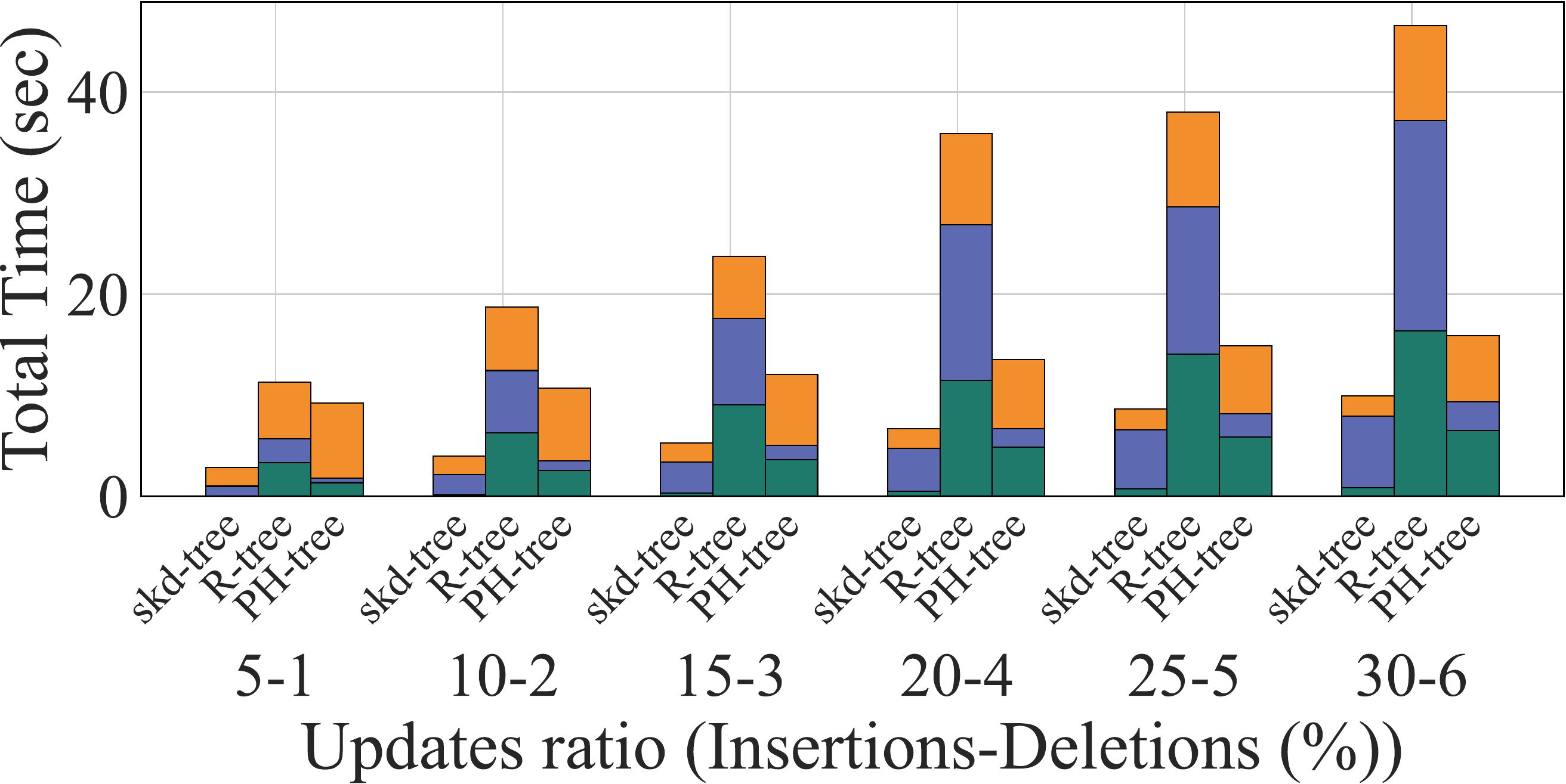}
		\caption{NYT ($D=8$)}
		\label{fig:udpates_nyt}
	\end{subfigure}
	\hspace{-0.5ex}
	\begin{subfigure}{0.32\linewidth}
		\centering
		\includegraphics[width=\linewidth]{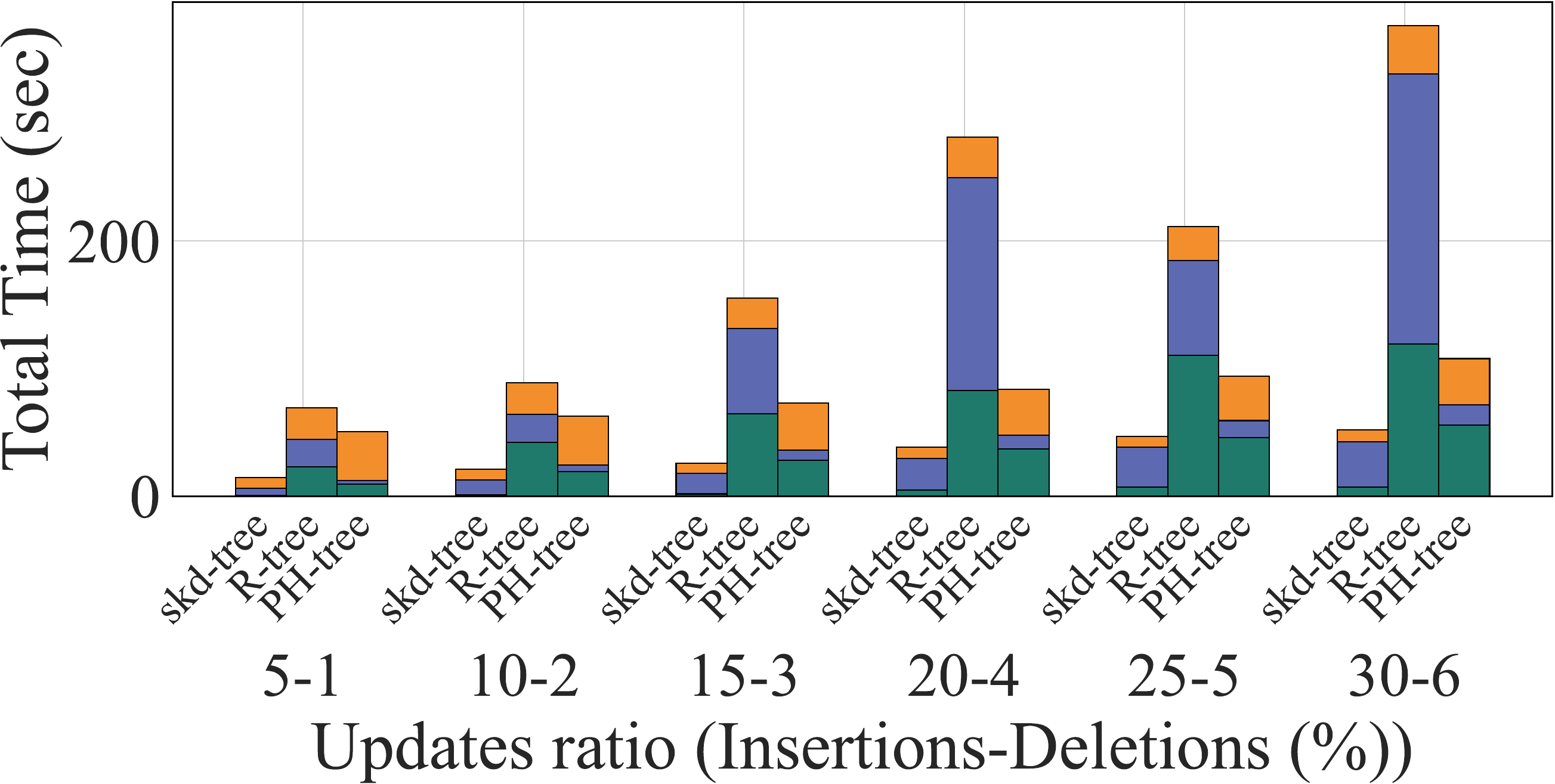}
		\caption{TPC-H ($D=8$)}
		\label{fig:udpates_tpch}
	\end{subfigure}
	\caption{Breakdown of execution time under a mixed workload of
		queries (mixed range and $k$NN) and updates. The x-axis shows the percentage of the initial dataset used for insertions and deletions}
	\label{fig:updates_expr}
\end{figure*}

We evaluate the behavior of \tree under {\em mixed workloads} that combine insertions, deletions, and queries, comparing it with R-tree and PH-tree, the only indices in our set supporting dynamic updates; the remaining indices are omitted due to lack of plug-and-play update support. 

\stitle{Workload setup.} Insertions and deletions are applied to a fraction of the initial dataset, varying from 5\% to 30\% for insertions and 1\% to 6\% for deletions. Queries consist of 1000 hypercube queries and 1000 query points, with a selectivity of 0.01\% and $k=10$ for $k$NN searches. Updates are split into five equal-sized batches, and queries are executed after each batch. For each scenario, we measure the total execution time and break it down into three components: insertion, deletion, and query time. This approach allows us to observe not only the overall performance but also the contribution of each operation to total cost.

\stitle{Performance comparison.}
Figure \ref{fig:updates_expr}
shows the performance of all tested methods across all datasets and
mixed workload scenarios. \tree consistently achieves the lowest total
execution time across all configurations. Its performance remains
stable as update intensity increases and continues to scale favorably
with higher dimensionality. Remarkably, the \tree is several times to
one order of magnitude faster than the R-tree on data with
dimensionality 5 or higher. This illustrates the deterioration of
the optimized boost R-tree index in workloads that include
updates. The PH-tree is more robust to updates on datasets of higher
dimensionality, but performs poorly when $D$ is 2 or 3.

\stitle{\tree structure under updates}.
%\achilleas{We will write less here. I keep an extented version to make it easier to follow me thought}
To interpret \tree's robustness,
% in mixed workloads that include
% numerous insertions and deletions,
we collect structural statistics,
including the average leaf capacity, number of leaf nodes, and
leaf-type categorization ({\em light}, {\em heavy}, {\em outlier})
before and after an update scenario.
Table \ref{tab:mkdtree_stats_updates} summarizes these statistics for
the EDGES and GAIA datasets under low-intensity (10\% insertions - 2\%
deletions) and high-intensity (30\% insertions - 6\% deletions)
scenarios. Overall, the \tree exhibits a stable and controlled structural adaptation, with only gradual changes in its organization as update intensity increases.

In the low-intensity scenario, the structure remains largely
unchanged. The number of leaf nodes is preserved, as a result of
the strategies we analyzed in Section \ref{sec:updates}. The average leaf
capacity increases slightly, showing that new insertions are
accommodated within existing leaves rather than causing extensive
splitting or redistribution (a small shift from {\em light} to {\em heavy} leaves, while {\em outlier} leaves remain negligible).
In the high-intensity scenario, structural changes are more visible
but remain controlled. GAIA maintains a nearly constant number of
leaves, while EDGES shows a moderate increase, indicating additional
splits. The average leaf capacity increases more noticeably in both
datasets; observe a clear shift from {\em light} to {\em heavy}
leaves. Despite this reorganization, {\em outlier} leaves remain rare, demonstrating that the structure avoids imbalance even under heavy updates.

\begin{table}[thb]
	\centering
	\caption{\tree structural characteristics across the EDGES and GAIA datasets under different update scenarios}
	\label{tab:mkdtree_stats_updates}
	\centering
	\vspace{-2mm}
	\resizebox{\columnwidth}{!}{%
		\begin{tabular}{lcccccccccc}
			\toprule
			\multicolumn{11}{c}{\bfseries 10\% Insertions - 2\% Deletions}\\
			\midrule
			\textbf{\begin{tabular}[c]{@{}c@{}} \\ Datasets\end{tabular}} & \multicolumn{2}{c}{\textbf{\begin{tabular}[c]{@{}c@{}}Avg. Leaf \\ Capacity\end{tabular}}} & \multicolumn{2}{c}{\textbf{\# Leaves}} & \multicolumn{2}{c}{\textbf{\begin{tabular}[c]{@{}c@{}}Light \\ Leaves {[}\%{]}\end{tabular}}} & \multicolumn{2}{c}{\textbf{\begin{tabular}[c]{@{}c@{}}Heavy \\ Leaves {[}\%{]}\end{tabular}}} & \multicolumn{2}{c}{\textbf{\begin{tabular}[c]{@{}c@{}}Outlier \\ Leaves {[}\%{]}\end{tabular}}} \\
			\cmidrule(lr){2-3} \cmidrule(lr){4-5} \cmidrule(lr){6-7} \cmidrule(lr){8-9} \cmidrule(lr){10-11}
			& Initial & Final & Initial & Final & Initial & Final & Initial & Final & Initial & Final \\
			\midrule
			\textbf{EDGES} & 128.7 & 140.13 & 355K & 355K & 98.68 & 92.47 & $\sim$1.32 & $\sim$7.53 & $<0.01$ & $<0.01$ \\
			\textbf{GAIA}  & 133.87 & 145.77 & 1.45M & 1.45M & 100 & 99.64 & 0 & 0.36 & 0 & 0 \\
			\bottomrule
		\end{tabular}
	}
%	\vspace{1mm}
%	
%	\centering
%	\resizebox{\columnwidth}{!}{%
%		\begin{tabular}{lcccccccccc}
%			\toprule
%			\multicolumn{11}{c}{\bfseries 20\% Insertions - 4\% Deletions}\\
%			\midrule
%			\textbf{\begin{tabular}[c]{@{}c@{}} \\ Datasets\end{tabular}} & \multicolumn{2}{c}{\textbf{\begin{tabular}[c]{@{}c@{}}Avg. Leaf \\ Capacity\end{tabular}}} & \multicolumn{2}{c}{\textbf{\# Leaves}} & \multicolumn{2}{c}{\textbf{\begin{tabular}[c]{@{}c@{}}Light \\ Leaves {[}\%{]}\end{tabular}}} & \multicolumn{2}{c}{\textbf{\begin{tabular}[c]{@{}c@{}}Heavy \\ Leaves {[}\%{]}\end{tabular}}} & \multicolumn{2}{c}{\textbf{\begin{tabular}[c]{@{}c@{}}Outlier \\ Leaves {[}\%{]}\end{tabular}}} \\
%			\cmidrule(lr){2-3} \cmidrule(lr){4-5} \cmidrule(lr){6-7} \cmidrule(lr){8-9} \cmidrule(lr){10-11}
%			& Initial & Final & Initial & Final & Initial & Final & Initial & Final & Initial & Final \\
%			\midrule
%			\textbf{EDGES} & 128.87 & 154.61 & 315K & 315K & 98.9 & $\sim$36.11 & $\sim$1.1 & 63.89 & $<0.01$ & $<0.01$ \\
%			\textbf{GAIA}  & 130.89 & 157.08 & 1.32M & 1.32M & 100 & 45.45 & 0 & 54.55 & 0 & 0 \\
%			\bottomrule
%		\end{tabular}
%	}
	\vspace{1mm}
	
	\centering
	\resizebox{\columnwidth}{!}{%
		\begin{tabular}{lcccccccccc}
			\toprule
			\multicolumn{11}{c}{\bfseries 30\% Insertions - 6\% Deletions}\\
			\midrule
			\textbf{\begin{tabular}[c]{@{}c@{}} \\ Datasets\end{tabular}} & \multicolumn{2}{c}{\textbf{\begin{tabular}[c]{@{}c@{}}Avg. Leaf \\ Capacity\end{tabular}}} & \multicolumn{2}{c}{\textbf{\# Leaves}} & \multicolumn{2}{c}{\textbf{\begin{tabular}[c]{@{}c@{}}Light \\ Leaves {[}\%{]}\end{tabular}}} & \multicolumn{2}{c}{\textbf{\begin{tabular}[c]{@{}c@{}}Heavy \\ Leaves {[}\%{]}\end{tabular}}} & \multicolumn{2}{c}{\textbf{\begin{tabular}[c]{@{}c@{}}Outlier \\ Leaves {[}\%{]}\end{tabular}}} \\
			\cmidrule(lr){2-3} \cmidrule(lr){4-5} \cmidrule(lr){6-7} \cmidrule(lr){8-9} \cmidrule(lr){10-11}
			& Initial & Final & Initial & Final & Initial & Final & Initial & Final & Initial & Final \\
			\midrule
			\textbf{EDGES} & 127.76 & 131.32 & 278K & 363K & 99.98 & 95.19 & 0.02 & 4.69 & 0 & 0.12 \\
			\textbf{GAIA}  & 124.73 & 165.07 & 1.21M & 1.23M & 100 & 46.37 & 0 & 53.63 & 0 & 0 \\
			\bottomrule
		\end{tabular}
	}
\end{table}

\stitle{Impact of updates on query performance}. Table \ref{tab:batch_query_time} compares query performance between the initial and final batches across all methods. A clear difference emerges between \tree and the competitors. In \tree, query times remain almost unchanged across both batches and datasets, indicating that updates have little impact on the structure relevant to query processing. In contrast, R-tree and PH-tree consistently show higher query times in the final batch, with the effect becoming stronger as update intensity increases. This reflects how updates affect structural quality. \tree preserves leaf organization and access efficiency over time, maintaining stable traversal costs. In contrast, R-tree and PH-tree gradually lose partitioning quality, reducing pruning efficiency and increasing query cost in later batches.

\begin{table}[]
	\caption{Queries  execution time under different update scenarios. We report the initial and final query batches for the EDGES and GAIA datasets}
	\label{tab:batch_query_time}
	\centering
	\vspace{-2mm}
	\resizebox{\columnwidth}{!}{%
		\begin{tabular}{lcccccc}
			\toprule
			\multicolumn{7}{c}{\textbf{10\% Insertions - 2\% Deletions}} \\
			\midrule
			
			& \multicolumn{3}{c}{\textbf{Initial Batch Query Time [sec]}}
			& \multicolumn{3}{c}{\textbf{Final Batch Query Time [sec]}} \\
			\cmidrule(lr){2-4} \cmidrule(lr){5-7}
			
			\textbf{Datasets}
			& \textbf{skd-tree} & \textbf{R-tree} & \textbf{PH-tree}
			& \textbf{skd-tree} & \textbf{R-tree} & \textbf{PH-tree}\\
			\midrule
			
			\textbf{EDGES} 
			& 0.016 & 0.05  & 0.265 
			& 0.017 & 0.05  & 0.32 \\
			
			\textbf{GAIA}  
			& 0.611 & 2.216 & 3.188 
			& 0.634 & 3.818 & 3.533 \\
			
			\bottomrule
		\end{tabular}
	}
	
%	\vspace{1mm}
%	
%	% -------------------- 20% --------------------
%	\resizebox{\columnwidth}{!}{%
%		\begin{tabular}{lcccccc}
%			\toprule
%			\multicolumn{7}{c}{\textbf{20\% Insertions - 4\% Deletions}} \\
%			\midrule
%			
%			& \multicolumn{3}{c}{\textbf{Initial Batch Query Time [sec]}}
%			& \multicolumn{3}{c}{\textbf{Final Batch Query Time [sec]}} \\
%			\cmidrule(lr){2-4} \cmidrule(lr){5-7}
%			
%			\textbf{Datasets}
%			& \textbf{skd-tree} & \textbf{R-tree} & \textbf{PH-tree}
%			& \textbf{skd-tree} & \textbf{R-tree} & \textbf{PH-tree}\\
%			\midrule
%			
%			\textbf{EDGES} 
%			& 0.015 & 0.044 & 0.253 
%			& 0.021 & 0.057 & 0.347 \\
%			
%			\textbf{GAIA}  
%			& 0.567 & 2.569 & 2.925 
%			& 0.801 & 8.607 & 3.58 \\
%			
%			\bottomrule
%		\end{tabular}
%	}
%	
	\vspace{1mm}
	
	% -------------------- 30% --------------------
	\resizebox{\columnwidth}{!}{%
		\begin{tabular}{lcccccc}
			\toprule
			\multicolumn{7}{c}{\textbf{30\% Insertions - 6\% Deletions}} \\
			\midrule
			
			& \multicolumn{3}{c}{\textbf{Initial Batch Query Time [sec]}}
			& \multicolumn{3}{c}{\textbf{Final Batch Query Time [sec]}} \\
			\cmidrule(lr){2-4} \cmidrule(lr){5-7}
			
			\textbf{Datasets}
			& \textbf{skd-tree} & \textbf{R-tree} & \textbf{PH-tree}
			& \textbf{skd-tree} & \textbf{R-tree} & \textbf{PH-tree}\\
			\midrule
			
			\textbf{EDGES} 
			& 0.014 & 0.038 & 0.237 
			& 0.019 & 0.058 & 0.378 \\
			
			\textbf{GAIA}  
			& 0.536 & 2.28  & 2.647 
			& 0.796 & 6.787 & 3.625 \\
			
			\bottomrule
		\end{tabular}
	}
	
\end{table}

\subsection{Scalability}\label{sec:exp:scalability}
\begin{figure}[]
	%	\hspace{-0.5ex}
	\includegraphics[width=\columnwidth]{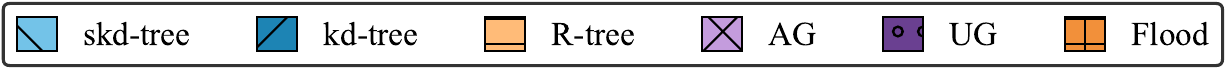}\\
	\begin{subfigure}{0.45\columnwidth}
		\centering
		\includegraphics[width=\columnwidth]{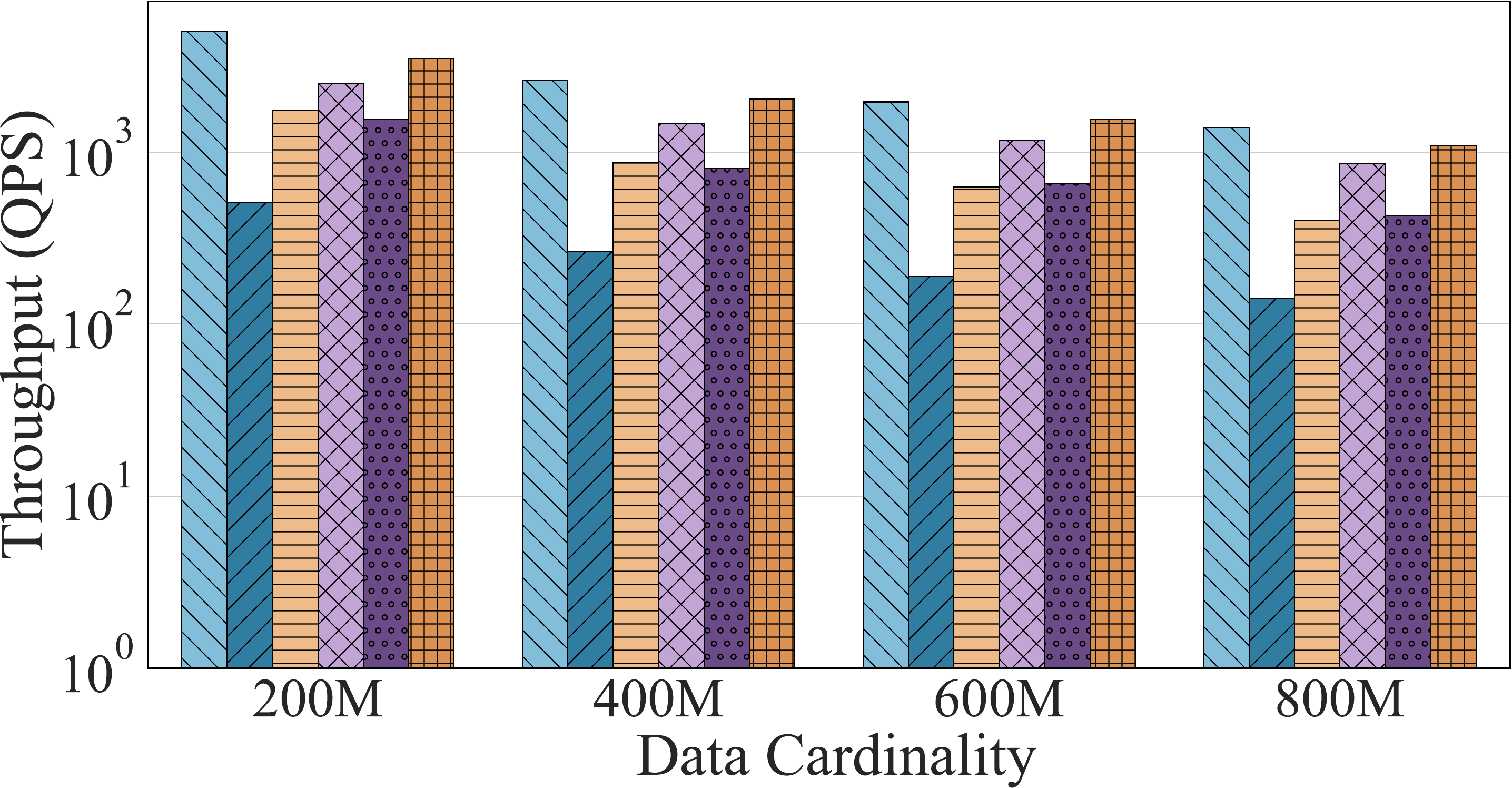}
		\caption{Uniform ($D=4$)}
		%		\label{fig:udpates_edges}
	\end{subfigure}
	%	\hspace{-0.5ex}
	\begin{subfigure}{0.45\columnwidth}
		\centering
		\includegraphics[width=\columnwidth]{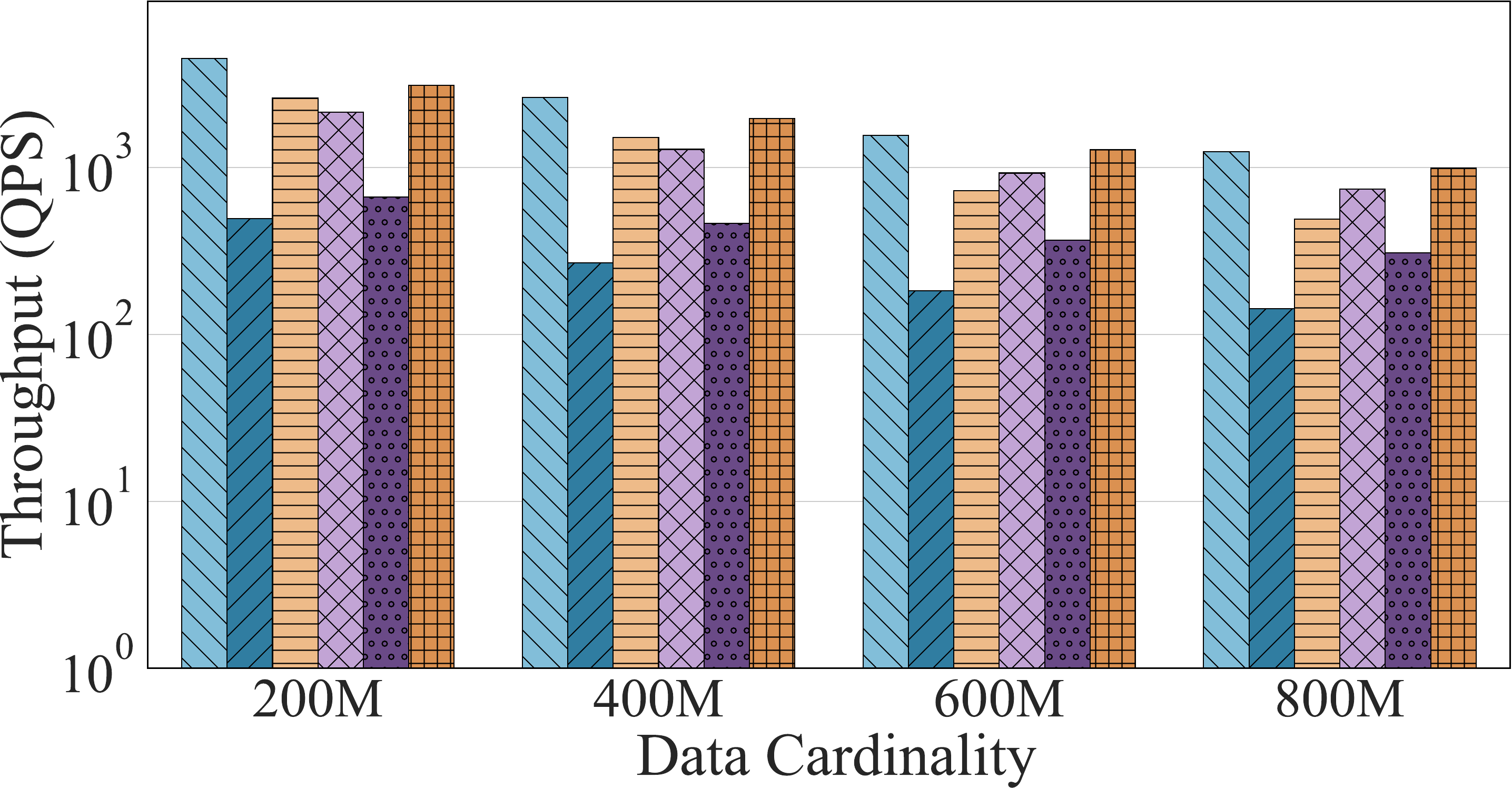}
		\caption{Gaussian ($D=4$)}
		%		\label{fig:udpates_toronto}
	\end{subfigure}	
	\begin{subfigure}{0.45\columnwidth}
		\centering
		\includegraphics[width=\columnwidth]{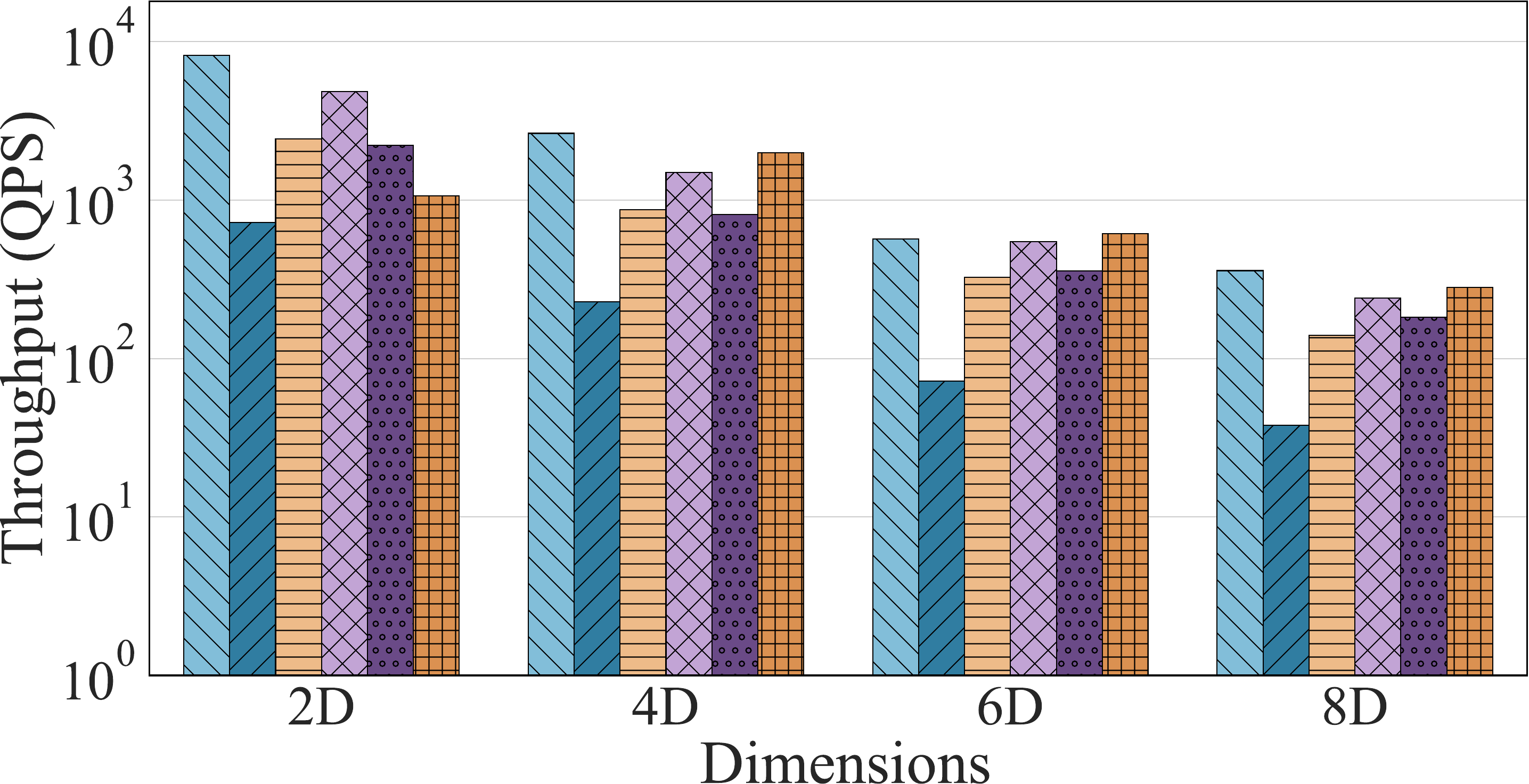}
		\caption{Uniform ($N=400$M)}
		%		\label{fig:udpates_edges}
	\end{subfigure}
	%	\hspace{-0.5ex}
	\begin{subfigure}{0.45\columnwidth}
		\centering
		\includegraphics[width=\columnwidth]{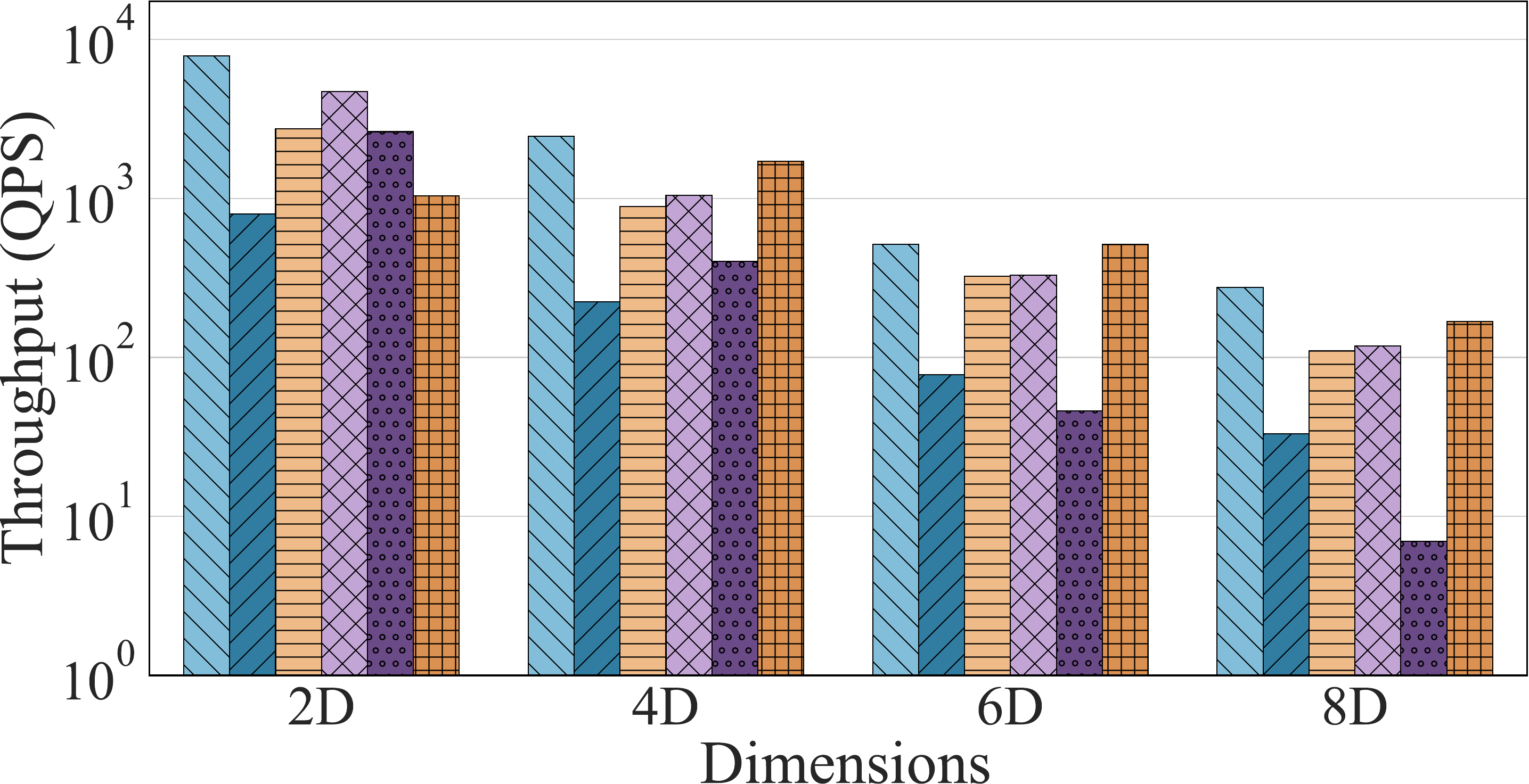}
		\caption{Gaussian ($N=400$M)}
		%		\label{fig:udpates_toronto}
	\end{subfigure}
	\caption{Scalability test using uniform and gaussian data for range queries}
	\label{fig:scalability_expr_range}
\end{figure}

\begin{figure}[]
	%	\hspace{-0.5ex}
	\includegraphics[width=0.5\columnwidth]{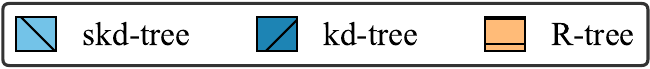}\\
	\begin{subfigure}{0.45\columnwidth}
		\centering
		\includegraphics[width=\columnwidth]{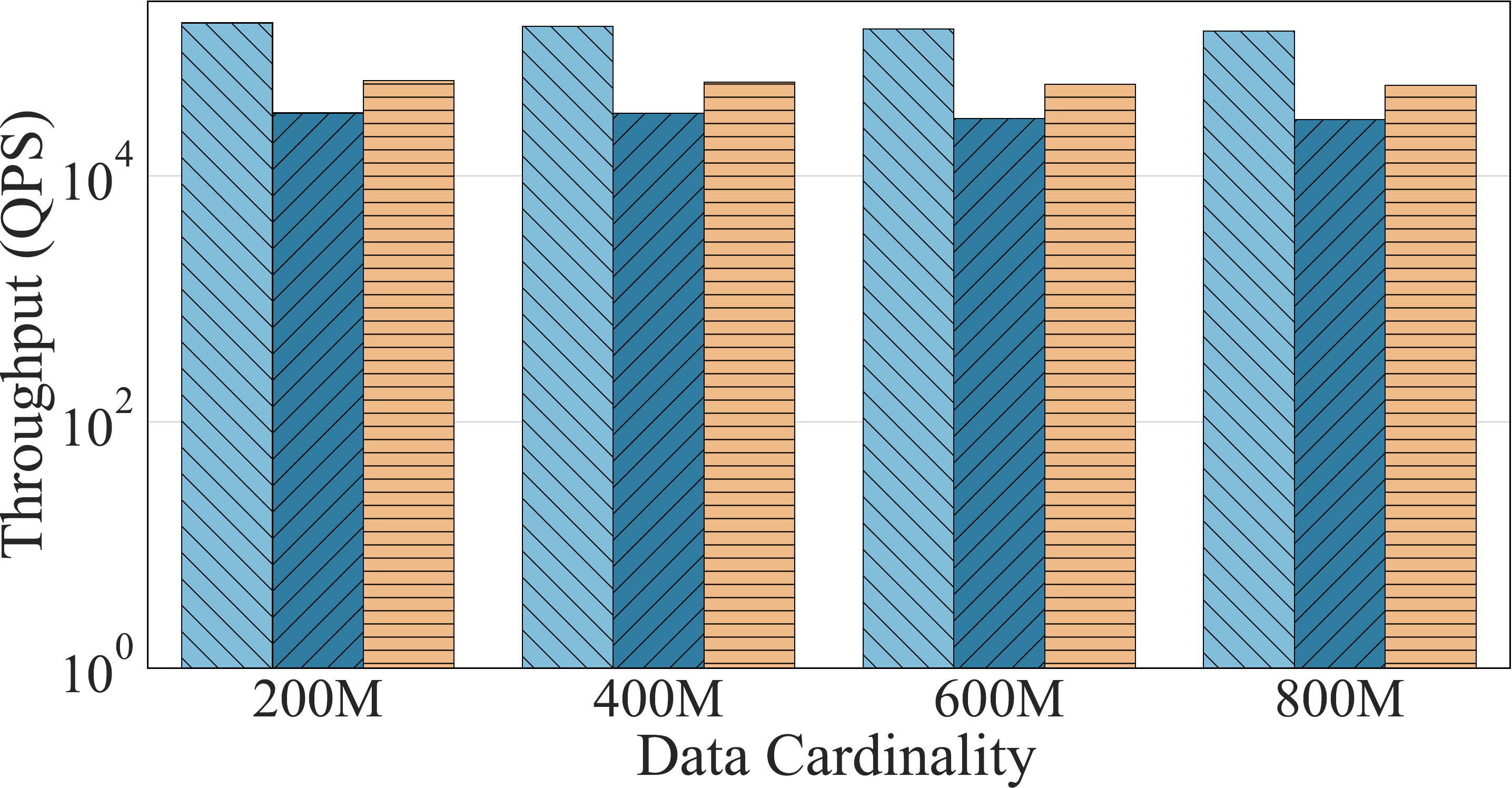}
		\caption{Uniform ($D=4$)}
		%		\label{fig:udpates_edges}
	\end{subfigure}
	%	\hspace{-0.5ex}
	\begin{subfigure}{0.45\columnwidth}
		\centering
		\includegraphics[width=\columnwidth]{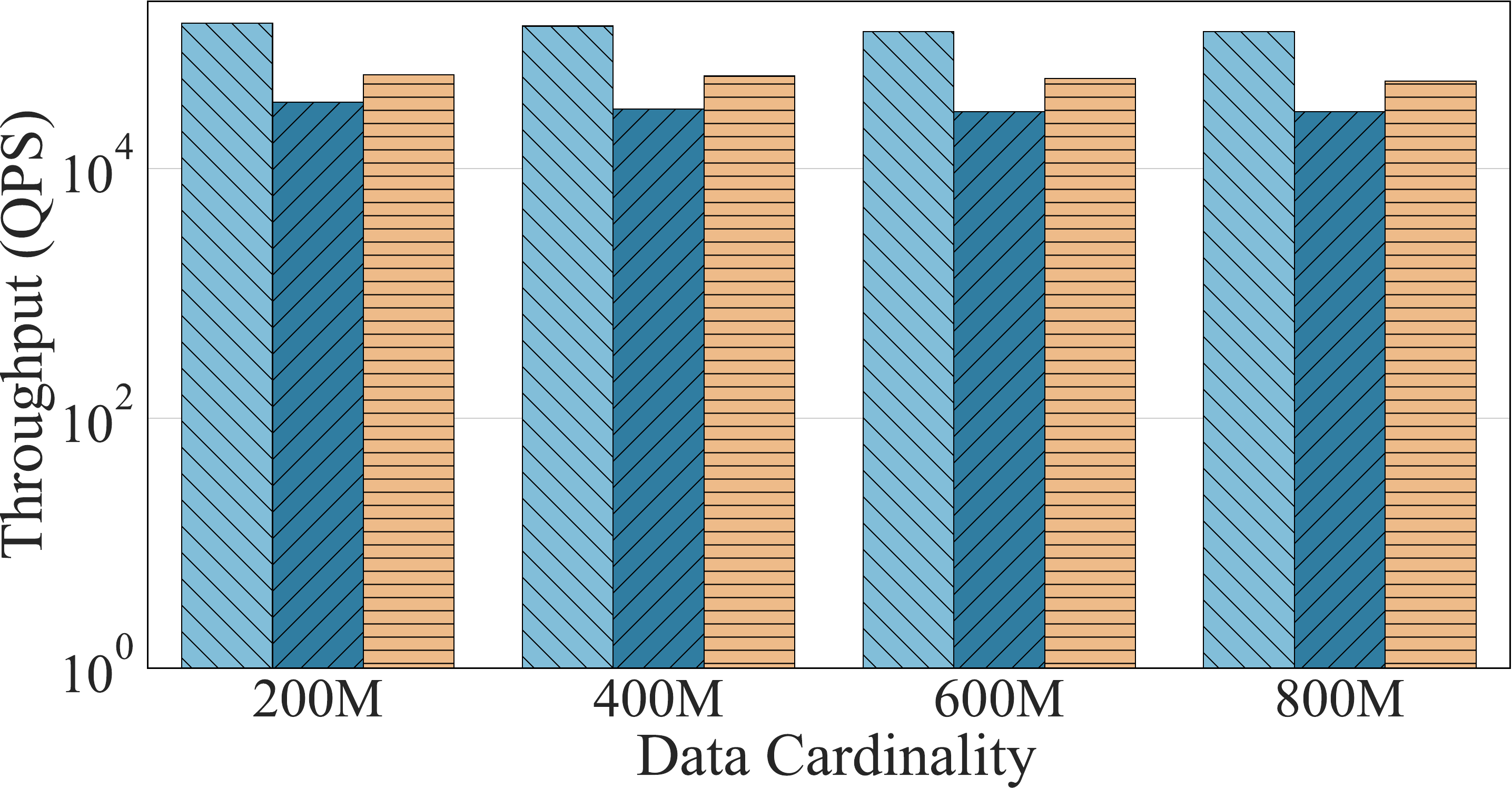}
		\caption{Gaussian ($D=4$)}
		%		\label{fig:udpates_toronto}
	\end{subfigure}	
	\begin{subfigure}{0.45\columnwidth}
		\centering
		\includegraphics[width=\columnwidth]{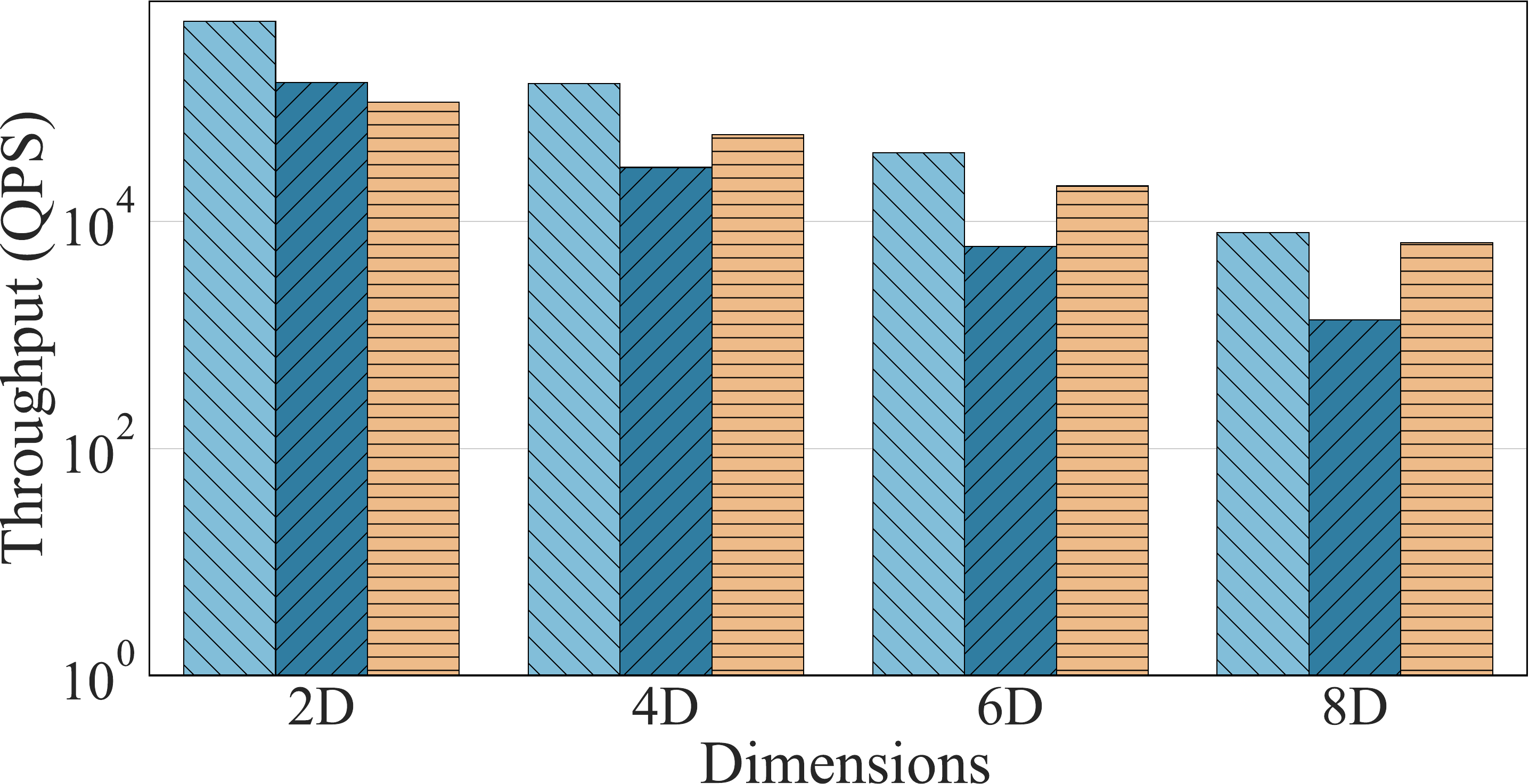}
		\caption{Uniform ($N=400$M)}
		%		\label{fig:udpates_edges}
	\end{subfigure}
	%	\hspace{-0.5ex}
	\begin{subfigure}{0.45\columnwidth}
		\centering
		\includegraphics[width=\columnwidth]{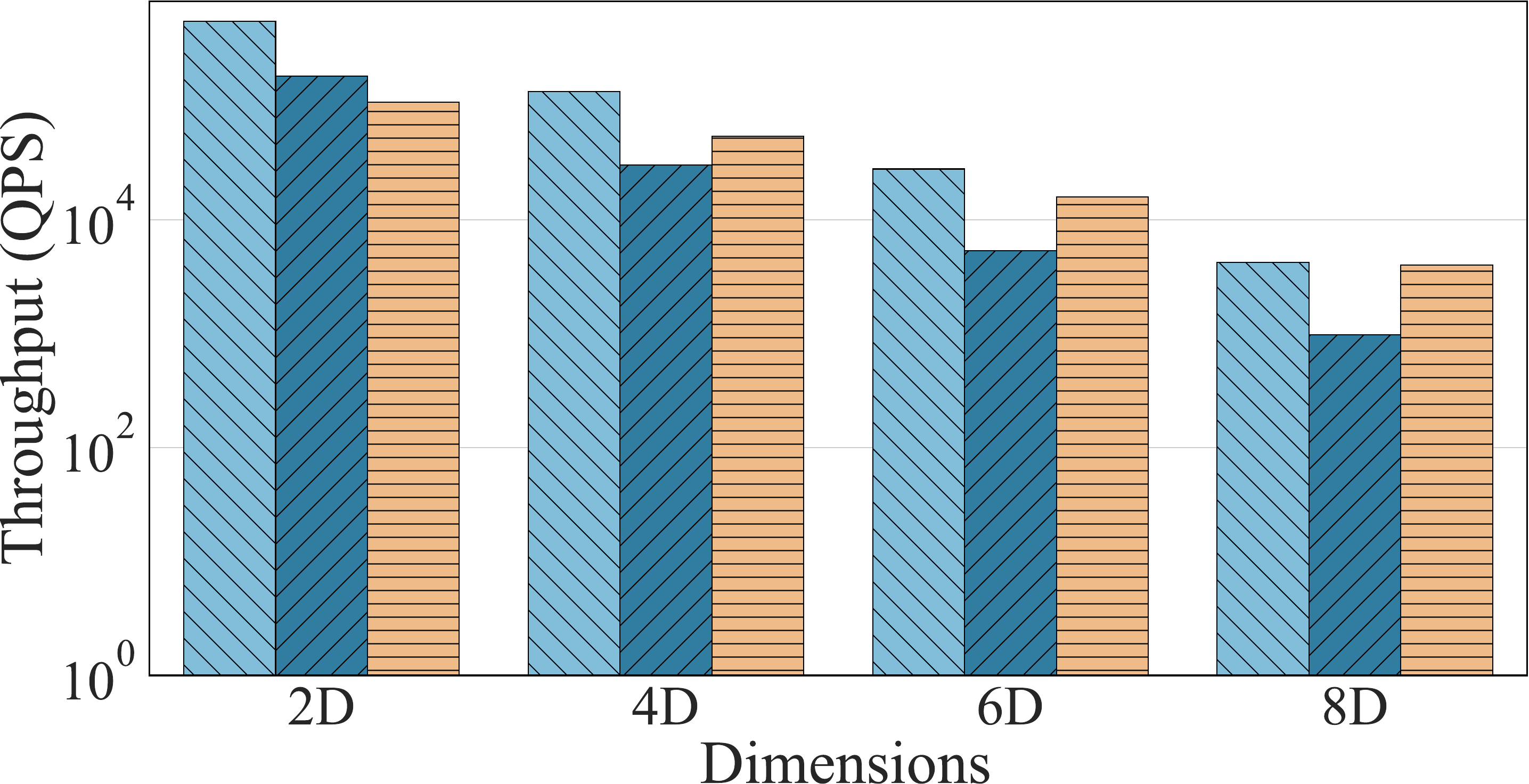}
		\caption{Gaussian ($N=400$M)}
		%		\label{fig:udpates_toronto}
	\end{subfigure}
	\caption{Scalability test using uniform and gaussian data for knn queries}
	\label{fig:scalability_expr_knn}
\end{figure}

\looseness=-1
In the last experiment, we study the scalability of the tested methods
to the data size and dimensionality,
using synthetic datasets (uniformly and normally distributed) for range queries and $k$NN queries.
We apply the same setup as the previous experimental sections, using range selectivity 0.01\% and $k = 10$.
Gaussian datasets were generated by choosing a random point as the
mean of the distribution and using as standard deviation in each
dimension a random number between 5\% and 30\% of the dimension's
domain.
We exclude IFI and PH-tree from these experiments, as they run out of
memory in some experimental settings.
We first fix the dimensionality to $D=4$ and vary the data size from
$N=200$M to $800$M.
Figures \ref{fig:scalability_expr_range}a and \ref{fig:scalability_expr_range}b
show that all methods scale well with the data size and their relative
differences are not affected by the data scale for range queries. The same behavior is observed in Figures \ref{fig:scalability_expr_knn}a and \ref{fig:scalability_expr_knn}b for the $k$NN queries.
In the second experiment, we fix $N=400$M and vary the data
dimensionality $D$. Figures \ref{fig:scalability_expr_range}c and
\ref{fig:scalability_expr_range}d show that \tree prevails in all cases and scales well
with $D$ for range queries. Note that Flood performs better compared to the case of real
data, as synthetic data distributions are easy to learn. However, 
Flood does not support $k$NN queries  and it is best for static data,
as already discussed. Regarding the $k$NN queries, Figures \ref{fig:scalability_expr_knn}c and \ref{fig:scalability_expr_knn}d show similar behavior, with the R-tree as the second-best performer.

\section{Conclusions}\label{sec:conclusions}
In this paper, we proposed \tree, a multi-way kd-tree for
indexing multidimensional points in memory, which applies at each node
(i) multiple partitions in the same dimension, (ii) compression, (iii)
data parallelism at search.
We proposed range and $k$NN search algorithms, an effective construction algorithm, and efficient update
methods for dynamic data.
We compared our \tree with the best performing, according to
\cite{LiuLZSC25},  multi-dimensional
classic and learned indices and shown that it consistently
outperforms all of them in range and $k$NN queries,
while being several times faster in mixed workloads with updates and queries.

One direction for future work is to make the \tree adaptive to a known
query workload before construction, in the same spirit as
\cite{DingNAK20, SudhirCM21}, by
adapting the number of splitters per dimension in each subtree during
construction, considering both the data and the queries distribution
locally.
% Previous work on query-workload
% adaptive multidimensional indexing \cite{DingNAK20, SudhirCM21} uses a
% coarse grid, which adapts to both the data and
% query distribution. Our \tree also adapts well to the data
% distribution, but it is agnostic to the query workload. If the query
% workload is also known, we can improve the design of our \tree by
% adapting the number of splitters per dimension in each subtree during
% construction, considering both the data and the queries distribution
% locally.
Another important direction is the implementation of a
multithreaded version of \tree that can process a workload
of queries and updates in parallel.
%Besides the straightforward idea of assigning query and update
%operations to available threads, we should apply concurrency control.
%For this, general-purpose locking mechanisms for concurrent traversal
%of search trees, such as
%Optimistic Lock Coupling (OLC) \cite{LeisH019} can be applied.

%\section{Discussion}\label{sec:future_work}

%\subsection{Adaptivity}\label{sec:equality}

% \subsection{Multi-core parallelism}
% \todo{Describe what to do when you have large batches of queries and
% you want to process them in parallel. The key idea is first
% partition the queries in parallel using the tree structure and then
% process sub-batches at each leaf. No rocket science but it should
% work out. Our goal is to beat the multi-threaded parallel kd-tree
% from sigmod25.} \dimitris{If we have the time we can have a parallel version only for range query and in the future work, we can explain how the olc can be implemented for mixed workload}
% \nikos{I do not see why it is so difficult to implement batch query processing (without locking). Even locking and mixed workloads should not be that hard.}

% \nikos{TODO: Add a discussion on how you will do multithreaded search and locking even if you did not implement it this time. Mention whether other competitor multidimensional indexing have dealt with multithreading and locking.}

%%
%% The next two lines define the bibliography style to be used, and
%% the bibliography file.
\bibliographystyle{ACM-Reference-Format}
\bibliography{references}

\end{document}